%% file: main.tex
\documentclass[a4paper,11pt]{article}
\usepackage{jcappub} 
\usepackage{aas_macros}
\usepackage{tabularx}
\usepackage{subfig} 
\usepackage{newtxtext,newtxmath}
\usepackage{amsmath}	
\usepackage{amssymb}	
\usepackage{cancel} 
\usepackage{hyperref}
\usepackage{xcolor} 

\input{bangla_commands}
\newcommand\ion[2]{\text{#1\,\textsc{\lowercase{#2}}}}	
\newcommand{\oii}{[\ion{O}{ii}] }
\newcommand{\gfib}{$g_{\rm fib}$ }
\newcommand{\rfib}{$r_{\rm fib}$ }
\newcommand{\zfib}{$z_{\rm fib}$ }
\newcommand{\fuji}{\emph{fuji} }

\arxivnumber{1234.56789} 
\title{Measuring $\sigma_8$ using DESI Legacy Imaging Surveys Emission-Line Galaxies and Planck CMB Lensing, and the Impact of Dust on Parameter Inference}

\author[1,2,3]{Tanveer Karim ({\bng tanviir kirm})}
\author[4]{Sukhdeep Singh}
\author[5]{Mehdi Rezaie}
\author[6]{Daniel Eisenstein}
\author[7,8,9]{Boryana Hadzhiyska}
\author[40,1,2,3]{Joshua S. Speagle}
\author[8]{Jessica Nicole Aguilar}
\author[10]{Steven Ahlen}
\author[11]{David Brooks}
\author[8]{Todd Claybaugh}
\author[12]{Axel de la Macorra}
\author[8,9]{Simone Ferraro}
\author[13,14]{Jaime E. Forero-Romero}
\author[15,16,17]{Enrique Gazta\~{n}aga}
\author[8]{Satya Gontcho A Gontcho}
\author[18]{Gaston Gutierrez}
\author[8]{Julien Guy}
\author[19,20]{Klaus Honscheid}
\author[21]{Stephanie Juneau}
\author[22]{David Kirkby}
\author[30,31,32]{Alex Krolewski}
\author[8]{Andrew Lambert}
\author[8]{Martin Landriau}
\author[8]{Michael Levi}
\author[21]{Aaron Meisner}
\author[23,24]{Ramon Miquel}
\author[25]{John Moustakas}
\author[12]{Andrea Mu\~{n}oz-Guti\'{e}rrez}
\author[26]{Adam Myers}
\author[27,28]{Gustavo Niz}
\author[29,8]{Nathalie Palanque-Delabrouille}
\author[30,31,32]{Will Percival}
\author[33]{Francisco Prada}
\author[34]{Graziano Rossi}
\author[35]{Eusebio Sanchez}
\author[36]{Edward Schlafly}
\author[8]{David Schlegel}
\author[37,38]{Michael Schubnell}
\author[21]{David Sprayberry}
\author[38]{Gregory Tarl\'e}
\author[21]{Benjamin Alan Weaver}
\author[39]{Hu Zou}

\affiliation[1]{David A. Dunlap Department of Astronomy \& Astrophysics, University of Toronto, 50 St George Street, Toronto, ON, M5S 3H4, Canada}
\affiliation[2]{Dunlap Institute for Astronomy \& Astrophysics, University of Toronto, 50 St George Street, Toronto, ON, M5S 3H4, Canada}
\affiliation[3]{Data Sciences Institute, University of Toronto, 17th Floor, Ontario Power Building, 700 University Avenue, Toronto, ON, M5G 1Z5, Canada}
\affiliation[4]{McWilliams Center for Cosmology \& Astrophysics, Department of Physics, Carnegie Mellon University, Pittsburgh, PA 15213,
USA}
\affiliation[5]{Department of Physics, Kansas State University, 116 Cardwell Hall, Manhattan, KS 66506, USA}
\affiliation[6]{Department of Astronomy, Harvard University, 60 Garden St, Cambridge, MA 02138, USA}
\affiliation[7]{Miller Institute for Basic Research in Science, University of California, Berkeley, CA, 94720, USA}
\affiliation[8]{Lawrence Berkeley National Laboratory, Berkeley, CA 94720, USA}
\affiliation[9]{Berkeley Center for Cosmological Physics, Department of Physics, University of California, Berkeley, CA 94720, USA}
\affiliation[10]{Department of Physics, Boston University, 590 Commonwealth Avenue, Boston, MA 02215, USA}
\affiliation[11]{Department of Physics \& Astronomy, University College London, Gower Street, London, WC1E 6BT, UK}
\affiliation[12]{Instituto de F\'{\i}sica, Universidad Nacional Aut\'{o}noma de M\'{e}xico,  Cd. de M\'{e}xico  C.P. 04510,  M\'{e}xico}
\affiliation[13]{Departamento de F\'isica, Universidad de los Andes, Cra. 1 No. 18A-10, Edificio Ip, CP 111711, Bogot\'a, Colombia}
\affiliation[14]{Observatorio Astron\'omico, Universidad de los Andes, Cra. 1 No. 18A-10, Edificio H, CP 111711 Bogot\'a, Colombia}
\affiliation[15]{Institut d'Estudis Espacials de Catalunya (IEEC), 08034 Barcelona, Spain}
\affiliation[16]{Institute of Cosmology and Gravitation, University of Portsmouth, Dennis Sciama Building, Portsmouth, PO1 3FX, UK}
\affiliation[17]{Institute of Space Sciences, ICE-CSIC, Campus UAB, Carrer de Can Magrans s/n, 08913 Bellaterra, Barcelona, Spain}
\affiliation[18]{Fermi National Accelerator Laboratory, PO Box 500, Batavia, IL 60510, USA}
\affiliation[19]{Center for Cosmology and AstroParticle Physics, The Ohio State University, 191 West Woodruff Avenue, Columbus, OH 43210, USA}
\affiliation[20]{Department of Physics, The Ohio State University, 191 West Woodruff Avenue, Columbus, OH 43210, USA}
\affiliation[21]{NSF NOIRLab, 950 N. Cherry Ave., Tucson, AZ 85719, USA}
\affiliation[22]{Department of Physics and Astronomy, University of California, Irvine, 92697, USA}
\affiliation[23]{Instituci\'{o} Catalana de Recerca i Estudis Avan\c{c}ats, Passeig de Llu\'{\i}s Companys, 23, 08010 Barcelona, Spain}
\affiliation[24]{Institut de F\'{i}sica d’Altes Energies (IFAE), The Barcelona Institute of Science and Technology, Campus UAB, 08193 Bellaterra Barcelona, Spain}
\affiliation[25]{Department of Physics and Astronomy, Siena College, 515 Loudon Road, Loudonville, NY 12211, USA}
\affiliation[26]{Department of Physics \& Astronomy, University  of Wyoming, 1000 E. University, Dept.~3905, Laramie, WY 82071, USA}
\affiliation[27]{Departamento de F\'{i}sica, Universidad de Guanajuato - DCI, C.P. 37150, Leon, Guanajuato, M\'{e}xico}
\affiliation[28]{Instituto Avanzado de Cosmolog\'{\i}a A.~C., San Marcos 11 - Atenas 202. Magdalena Contreras, 10720. Ciudad de M\'{e}xico, M\'{e}xico}
\affiliation[29]{IRFU, CEA, Universit\'{e} Paris-Saclay, F-91191 Gif-sur-Yvette, France}
\affiliation[30]{Department of Physics and Astronomy, University of Waterloo, 200 University Ave W, Waterloo, ON N2L 3G1, Canada
}
\affiliation[31]{Perimeter Institute for Theoretical Physics, 31 Caroline St. North, Waterloo, ON N2L 2Y5, Canada}
\affiliation[32]{Waterloo Centre for Astrophysics, University of Waterloo, 200 University Ave W, Waterloo, ON N2L 3G1, Canada}
\affiliation[33]{Instituto de Astrof\'{i}sica de Andaluc\'{i}a (CSIC), Glorieta de la Astronom\'{i}a, s/n, E-18008 Granada, Spain}
\affiliation[34]{Department of Physics and Astronomy, Sejong University, Seoul, 143-747, Korea}
\affiliation[35]{CIEMAT, Avenida Complutense 40, E-28040 Madrid, Spain}
\affiliation[36]{Space Telescope Science Institute, 3700 San Martin Drive, Baltimore, MD 21218, USA}
\affiliation[37]{Department of Physics, University of Michigan, Ann Arbor, MI 48109, USA}
\affiliation[38]{University of Michigan, Ann Arbor, MI 48109, USA}
\affiliation[39]{National Astronomical Observatories, Chinese Academy of Sciences, A20 Datun Rd., Chaoyang District, Beijing, 100012, P.R. China}
\affiliation[40]{Department of Statistical Sciences, University of Toronto, 9th Floor, Ontario Power Building, 700 University Avenue, Toronto, ON, M5G 1Z5, Canada}

\emailAdd{tanveer.karim@utoronto.ca}

\abstract{Measuring the growth of structure is a powerful probe for studying the dark sector, especially in light of the $\sigma_8$ tension between primary CMB anisotropy and low-redshift surveys. This paper provides a new measurement of the amplitude of the matter power spectrum, $\sigma_8$, using galaxy-galaxy and galaxy-CMB lensing power spectra of Dark Energy Spectroscopic Instrument Legacy Imaging Surveys Emission-Line Galaxies and the \emph{Planck} 2018 CMB lensing map. We create an ELG catalog composed of $27$ million galaxies and with a purity of $85\%$, covering a redshift range $0 < z < 3$, with $z_{\rm mean} = 1.09$. We implement several novel systematic corrections, such as jointly modeling the contribution of imaging systematics and photometric redshift uncertainties to the covariance matrix. We also study the impacts of various dust maps on cosmological parameter inference.
We measure the cross-power spectra over $f_{\rm sky} = 0.25$ with a signal-to-background ratio of up to $ 30\sigma$. We find that the choice of dust maps to account for imaging systematics in estimating the ELG overdensity field has a significant impact on the final estimated values of $\sigma_8$ and $\Omega_{\rm M}$, with far-infrared emission-based dust maps preferring $\sigma_8$ to be as low as $0.702 \pm 0.030$, and stellar-reddening-based dust maps preferring as high as $0.719 \pm 0.030$. The highest preferred value is at $\sim 3 \sigma$ tension with the \emph{Planck} primary anisotropy results. These findings indicate a need for tomographic analyses at high redshifts and joint modeling of systematics.}

\begin{document}
\maketitle
\flushbottom

\section{Introduction}
\label{sec4:intro}

The discovery of the accelerated expansion of the Universe \cite{Perlmutter98, Riess98} with the help of Type Ia supernovae, the cosmic microwave background (CMB) acoustic peaks \citep{Boomerang00, Maxima00}, and the Baryonic Acoustic Oscillation (BAO) in the large-scale structure \citep{Eisenstein05} established dark energy as the leading explanation for the observed cosmic acceleration. These cosmology experiments identify the existence of dark energy by measuring the Hubble constant, $H_0$. 

The negative pressure of dark energy inhibits the growth of large-scale structures, countering the force of gravitation \cite{rhodes13}, which has led to the measurement of structure growth as a power probe of dark energy complimentary to that from distance measurements. Suppression of the growth of large-scale structure is best measured with experiments that focus on weak gravitational lensing (galaxy-galaxy, cosmic shear, or CMB lensing), galaxy cluster counting, and redshift space distortions \citep{Huterer2015}. These experiments essentially measure the combination of the amplitude of the matter power spectrum, $\sigma_8$, and the cosmic matter density, $\Omega_{\rm M}$, by using matter tracers such as lensing fields, galaxy clusters, or spectroscopic surveys. 

These probes and the individual experiments have shown remarkable agreement with the $\Lambda$CDM paradigm. However, we have seen statistically significant disagreements between cosmological parameters inferred from CMB and late-time experiments in recent years. For example, the inferred measurement of $H_0$ from CMB experiments such as \emph{Planck} \citep{planck2020} and the Atacama Cosmology Telescope (ACT) \citep{aiola2020} are at about $5\sigma$ odds with Type Ia Supernovae probes \citep{riess2022}. Similarly, the inferred measurement of $\sigma_8$ from CMB experiments such as \emph{Planck} and ACT are at about $2-3\sigma$ odds with late-Universe probes such as weak lensing \citep{porredon21, pandey22, asgari2021, hikage2019}, spectroscopic galaxy clustering \citep{ivanov2021}, redshift space distortions \cite{eboss2021} and galaxy cluster counting \citep{abbott2020, mantz2015}. 

The presence of these so-called tensions between the early Universe and the late Universe is tantalizing; it may indicate the existence of new unknown physics, such as modification of gravity or dark energy (a detailed discussion can be found at \cite{abdalla2022}). However, a more down-to-earth explanation could be that previously unquantified systematic uncertainties induce these tensions. Cross-correlation of matter tracers with very different systematic properties could address this explanation \citep{rhodes13}. The aforementioned weak lensing experiments often take a $3 \times 2$ approach where they measure the auto-correlation of individual tracers and also their cross-correlation to mitigate the effect of systematics and to break parameter degeneracy. More recently, such techniques have also been used to combine galaxy clustering measurements with CMB lensing \cite{kim2024, sailer2024, Farren23, white22, krolewski21, Bianchini16}; some of these studies have shown mild tension in the $\sigma_8 - \Omega_M$ parameter space when compared to \emph{Planck}. The bulk of the growth of structure studies has been conducted with tracers at $z < 1$; thus, the tension we observe is between the $z < 1$ and $z = 1090$ Universe. This fact leads to the question of whether the answer lies in the intermediate redshift ($z > 1$) Universe because if $\Lambda$CDM is assumed to be accurate, then it is around $z \sim 1$ when the dark energy density, $\Omega_{\Lambda}$ became the same order of magnitude as the matter density, $\Omega_M$, for the first time in the cosmic history. Thus, there may be observable effects during this turnover period. 

Interestingly, some recent studies that use the Atacama Cosmology Telescope (ACT) data \cite{Farren23, act2024} do not find this tension. Thus, it has become even more critical to understand the physics of the $z > 1$ Universe to ascertain whether the deviations seen by earlier works are due to cosmology, astrophysics or systematics.

The Dark Energy Spectroscopic Instrument (DESI) Legacy Imaging Surveys \citep{Dey19} and the ongoing DESI spectroscopic survey\footnote{"DESI Collaboration Key Paper". Part of Key Project 1, convener: Paul Martini} \citep{desi2022} have provided a new window to exploring the $z > 1$ epoch of our Universe using galaxy clustering. DESI is a highly multiplexed spectroscopic surveyor that can obtain simultaneous spectra of almost $5000$ objects over a $3^{\circ}$ field \cite{desi2016b, silber2023, miller2023}, is currently conducting a five-year survey of about a third of the sky. This campaign will obtain spectra for approximately $40$ million galaxies and quasars \cite{desi2016a}.

Both the Legacy Imaging Surveys and the DESI spectroscopic survey use emission-line galaxies (ELGs) as matter tracers to probe the $z > 1$ Universe with the highest number density and sky coverage to-date. As the cosmic star formation rate peaks around $z \sim 2$ \citep{madau2014}, so does the number density of these star-forming ELGs. Thus, ELGs provide a unique window to the intermediate redshift Universe. Ongoing and future galaxy surveys such as \emph{Euclid}, \emph{Roman}, and the Rubin Observatory will use ELG-like galaxies to push beyond $z \sim 2$. The Legacy Surveys is the photometric catalog for DESI; it covers $0.4$ of the sky and contains $40$ million objects deemed as ELGs, thus making it the largest photometric ELG catalog to date.

In this paper, for the first time, we use the ELG targets from the Legacy Surveys and cross-correlate them with the \emph{Planck} CMB lensing map to constrain the growth of structure by measuring the ELG auto- and ELG-CMB lensing cross-power spectra. Given the size of the Legacy Surveys catalog, this is the most extensive analysis of ELGs as a cosmological tracer in a cross-correlation study. Additionally, we thoroughly analyze various systematics, especially how the choice of dust maps to correct foreground systematics affects cosmological parameter inference. 

We structure this paper as follows: Section~\ref{sec4:data} describes the Legacy Surveys, DESI Survey Validation, and the \emph{Planck} CMB lensing datasets. Section~\ref{sec4:samplesel} discusses how we selected our ELG sample and estimated the redshift distribution. In Section~\ref{sec4:method} we discuss the various modeling aspects of our analysis. Section~\ref{sec4:cov} discusses our choices for the likelihood inference framework as well as how we estimated the associated covariance matrix. In Section~\ref{sec4:sims}, we showcase the validation of our analysis pipeline. Section~\ref{sec4:results} presents the key findings of our paper, and in Section~\ref{sec:discussion} contextualizes the results. Finally in Section~\ref{sec4:conc} we summarize the results. 

\section{Data}
\label{sec4:data}
The overlapping footprint of the Planck CMB lensing and the Legacy Surveys maps is about $0.3$ of the entire sky, allowing us to take relatively large modes into account for our analysis, compared to previous studies.

For this analysis, we select galaxies that make it through the DESI Survey Validation 3 (SV3) color cuts in the g - r versus r - z parameter space (DESI-like ELGs). Here g, r, and z refer to green at $464$ nm, red at $658$ nm, and extended infrared at $900$ nm. DESI-like ELGs are a natural choice for this analysis because DESI has already made many observations during its Survey Validation phase and internally released reduced spectra and spectroscopic redshifts of about $4\times 10^{6}$ ELGs. These spectroscopic redshifts are invaluable in calibrating the photometric redshift distribution that we use in our analysis and for deciding which ELGs belong to which tomographic bins.

In the following sections, we briefly describe the three key datasets for our analysis -- the Planck CMB lensing map in Section~\ref{sec4:planckcmb}, the Legacy Surveys DR9 catalog in Section~\ref{sec4:dr9} and the DESI SV3 \fuji catalog in Section~\ref{sec4:sv3}.

\subsection{\emph{Planck} CMB lensing map}
\label{sec4:planckcmb}

The \emph{Planck} collaboration published the PR3 CMB lensing map in 2018 \citep{PlanckCMB18}. The CMB lensing map traces the distortion of the CMB photons along the line of sight as they encounter the gravitational potential of masses. Depending on the geometry between the surface of the last scattering and the observer, the gravitational potential can deflect the CMB photons, distorting the observed temperature and polarization anisotropies. By carefully studying the statistical patterns of these distortions, one can reconstruct the intervening projected gravitational potential kernel, which tells us the projected distribution of mass between us and the surface of the last scattering. Hence, the CMB lensing map traces the matter distribution exactly and is an unbiased tracer of matter density distribution in the Universe. 

For this analysis, we use the minimum-variance lensing estimates from the SMICA DX12 CMB maps\footnote{https://wiki.cosmos.esa.int/planck-legacy-archive/index.php/Lensing}. We specifically use the \textsc{com\_lensing\_4096\_r3.00} dataset that provides the baseline lensing convergence estimates in spherical harmonics basis, $a_{\ell m}$ up to $\ell_{\rm max} = 4096$. The observed lensing convergences are measured using temperature and polarization maps. In addition, the dataset contains the \emph{Planck} convergence reconstruction approximate noise, $N_{\ell}^{\kappa \kappa}$, and the survey mask. 

Note that the \emph{Planck} CMB lensing reconstruction is noise-dominated at almost every scale, especially at high-$\ell$. It becomes especially problematic when using a pseudo-$C_{\ell}$ estimator because the finite survey size causes mode-mode coupling and the noise from high-$\ell$ can leak into low-$\ell$ scales of interest (see Section~\ref{sec4:method}). Hence, we first process the data $a_{\ell m}$ with a low-pass filter. This function allows us to smoothly truncate the power to $0$ above a specific scale; we set $\ell_{\rm min}$ and $\ell_{\rm max}$ to be $400$ and $1200$ respectively as we constrain our analysis in the range $50 < \ell < 400$. The low-pass-filtered $a_{\ell m}$ is given by:

\[ a^{\rm filtered}_{\ell m} = \begin{cases} 
          a_{\ell m} & \ell \leq \ell_{\rm min} \\
          a_{\ell m} \cos{ \left[ \frac{\pi}{2} \frac{\ell - \ell_{\rm min}}{\ell_{\rm max} - \ell_{\rm min}} \right]} & \ell_{\rm min} < \ell < \ell_{\rm max} \\
          0 & \ell \geq \ell_{\rm max} x 
       \end{cases}
    \]

We then rotate the filtered $a_{\ell m}$ and the \emph{Planck} mask from the Galactic coordinate basis to the Equatorial coordinate basis since we perform our analysis in Equatorial coordinates. In the new basis, we apodize the provided mask to allow it to go smoothly to $0$ in real space. We specifically use the ``C2" apodization scheme from the package \textsc{NaMaster} \cite{namaster} with an apodization scale of $0.5$ deg which is optimal for the size of holes in the \emph{Planck} mask \cite{white22}.

\begin{figure}
    \centering
\includegraphics[width=\columnwidth]{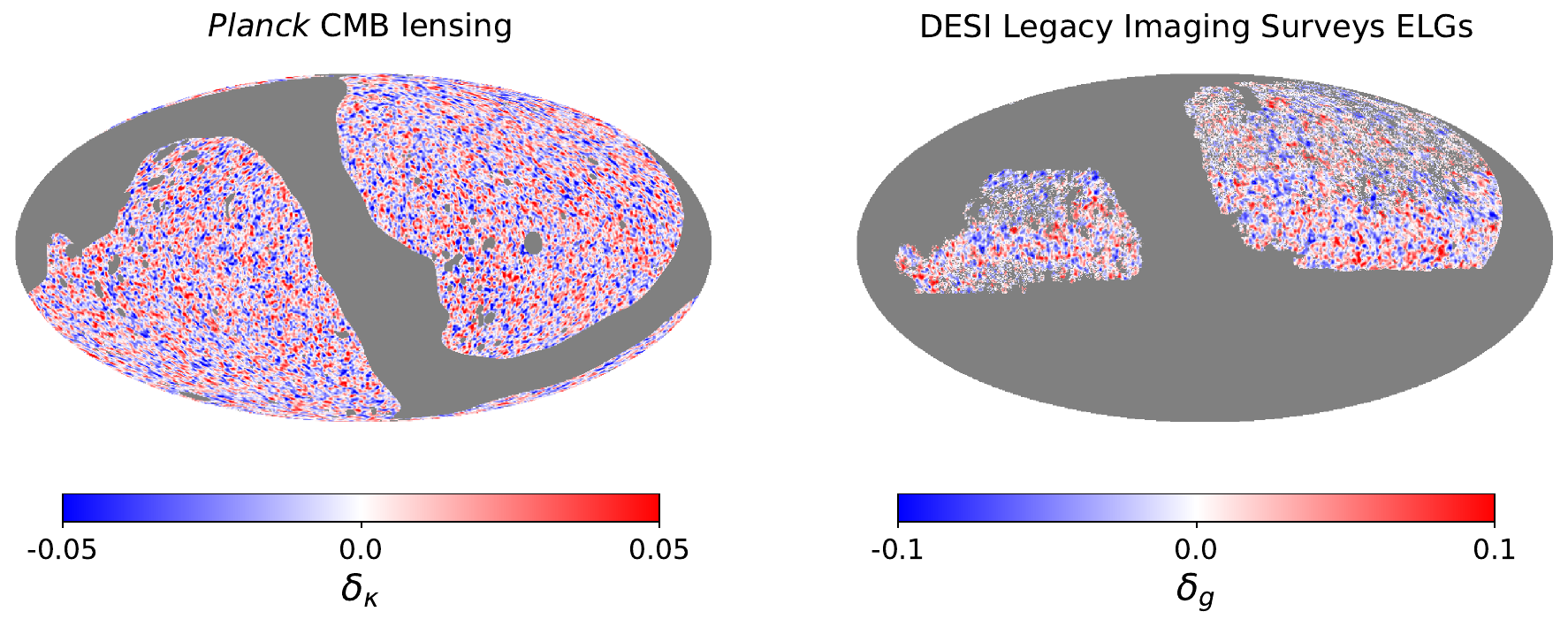}
    \caption{Processed \emph{Planck} CMB lensing and DESI Legacy Surveys ELG overdensity maps (in Equatorial projection) used in this analysis. We have smoothed the map with a $0.5$ degree Gaussian for visualization.}
    \label{fig:delta-maps}
\end{figure}

The final CMB lensing map is in Figure~\ref{fig:delta-maps}, where we have smoothed the map with a $0.5^{\circ}$ Gaussian for visualization. 

\subsection{Legacy Surveys DR9 catalogue}
\label{sec4:dr9}

The Legacy Surveys\footnote{https://www.legacysurvey.org/} is an inference model catalog of three different surveys -- the Dark Energy Camera Legacy Survey (DECaLS), the Beijing-Arizona Sky Survey (BASS) and the Mayall $z$-band Legacy Survey -- that jointly cover about $14000$ deg$^2$ of the sky visible from the Northern hemisphere in $g$-, $r$- and $z$-bands as well as four WISE infrared bands \citep{Dey19}. The main motivation for the  Legacy Surveys was to serve as the photometric input catalog of the DESI experiment; to achieve and maximize its scientific goals, DESI requires reliable photometry with sufficient depth to select target objects \textit{a priori}. The average galaxy depth for the $g$-, $r$- and $z$-bands are $23.6$, $23.07$ and $22.26$ AB magnitudes respectively. 

Because of its exact footprint overlap with the DESI Survey Validation, the DESI main survey, and the wide area coverage, the Legacy Surveys serve as the ideal photometric dataset for our cross-correlation analysis. Specifically, since the same cuts on the color and magnitude spaces that have been applied to select the DESI Survey Validation ELG targets can be applied to the full Legacy Surveys dataset, we can ensure that we are selecting galaxy populations with similar properties. As a result, we can robustly rely on using the spectroscopic redshift distribution, $p(z)$, as the redshift calibrator for our photometric sample and make tomographic bin assignments. 

The ninth data release\footnote{https://www.legacysurvey.org/dr9/description/} of the Legacy Surveys (hereafter DR9) is the final data release of the Legacy Surveys, which brings all three separate surveys under one reduction pipeline and produces uniform photometry across the three surveys, making it a robust galaxy catalog to probe large-scale modes. The DR9 catalog contains about $47$ million objects photometrically deemed ELGs. 

\subsection{DESI SV3 \fuji catalogue}
\label{sec4:sv3}

\begin{figure}
    \centering
\includegraphics[width=0.75\columnwidth]{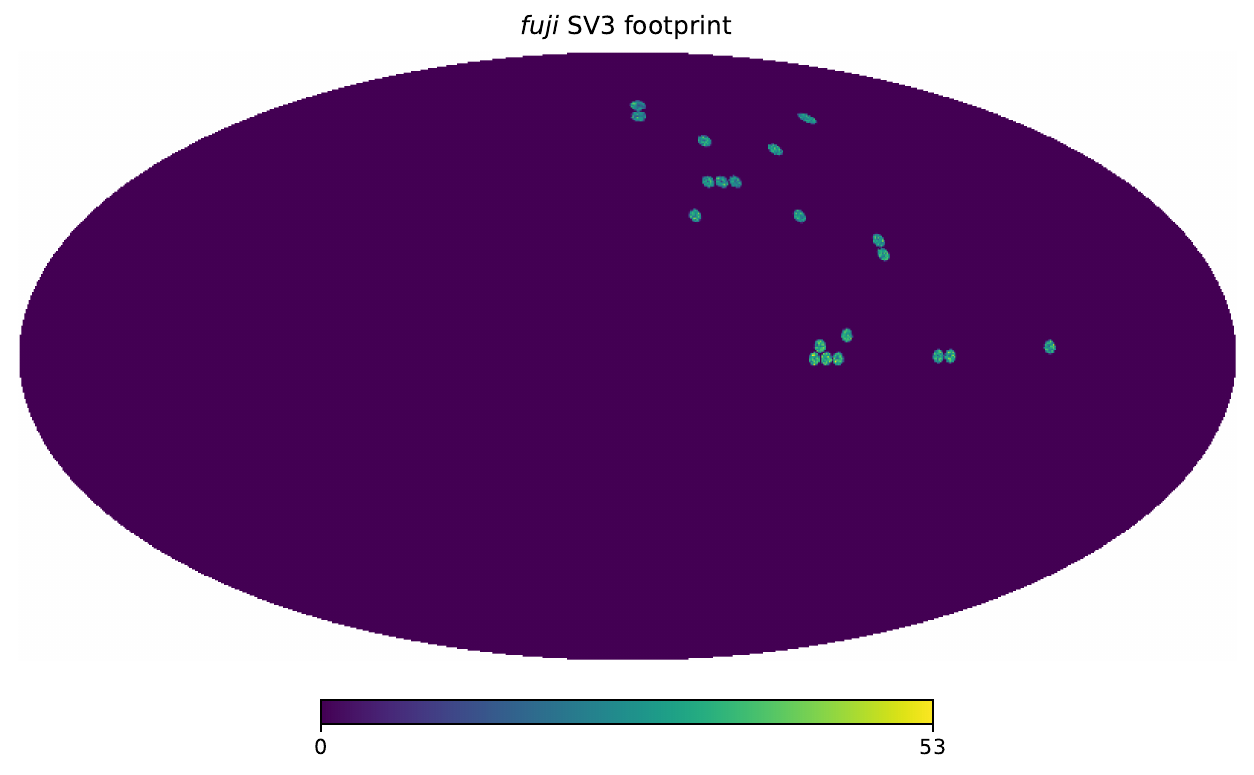}
    \caption{Footprint of the \fuji SV3 catalog in Equatorial projection}
    \label{fig4:sv3}
\end{figure}

The \fuji spectroscopic catalog is an internal data release within the DESI collaboration that contains all the commissioning and survey validation data. The goal of DESI is to determine the nature of dark energy by measuring the expansion of the Universe \cite{levi2013}. The sheer scale of the DESI experiment requires an extensive spectroscopic reduction pipeline \cite{guy2023}, a template-fitting pipeline to derive classifications and redshifts for each targeted source \cite{redrock2024}, a pipeline to tile the survey and to plan and optimize observations as the campaign progresses \cite{schlafly2023}, a pipeline to select targets for spectroscopic follow-up \cite{myers2023}, and Maskbit to generate "clean" photometry \cite{moustakas2023}.

The primary motivation of the survey validation (SV) phase\footnote{``DESI Collaboration Key Paper". Part of Key Project 1, conveners: Kyle Dawson and Christophe Y\`eche} \cite{desisv2024} of the project was to ensure that the DESI instrument met its spectroscopic performance goals, the requirements as a Stage-IV dark energy experiment, fine-tune the target selection algorithms.

The \fuji SV3 catalog contains $397,288$ objects that were photometrically deemed as ELGs, covering about $140$ deg$^2$ of the sky. After rejecting objects whose observed spectra were flagged by the spectral pipeline, we obtained a total of $390,901$ ELGs, giving the SV3 sample a number density of about $2800$ per deg$^2$. Additionally, compared to the ongoing DESI survey, the SV dataset has much deeper spectroscopy, and the ELGs were given equal priority in fiber assignment, making it a suitable dataset for ELG redshift distribution calibration. All the \fuji ELGs are included in the Early DESI Data Release\footnote{``DESI Collaboration Key Paper". Part of Key Project 1, conveners: Anthony Kremin and Stephen Bailey} \cite{desiedr2024}.

Additionally, the SV3 dataset is ideal for dealing with the issue of sample variance. Sample variance in this context refers to the problem of obtaining spectroscopic to photometric redshift calibration from a small patch of the sky. Suppose the galaxy population living in a small patch of the sky does not represent the galaxy population at large. In that case, if the patch traces an underdense or an overdense region in the large-scale structures, the redshift calibration we obtain will be biased. However, as shown in Figure~\ref{fig4:sv3}, the SV3 footprint is spread out in different small patches of the sky, reducing the impact of sample variance. 

\section{Preparation of ELG sample}
\label{sec4:samplesel}

We select the final ELG sample based on photometric features in the DR9 catalog. We initially take the DESI ELG target selection function \citep{raichoor23, raichoor20, karim20} and apply further selection cuts to increase the purity of the final ELG sample. Our selection uses $8$ features -- \textsc{flux\_g}, \textsc{flux\_r} and \textsc{flux\_z} corresponding to the model flux in the $g-$, $r-$ and $z-$ bands, $g-r$ and $r-z$ colors and \textsc{fiberflux\_g}, \textsc{fiberflux\_r} and \textsc{fiberflux\_z} corresponding to the expected fiber magnitude of the target sources assuming a $1.5^{\prime \prime}$  diameter fiber with a $1^{\prime \prime}$ Gaussian seeing. 
The fluxes and the fiber fluxes are converted to magnitudes and corrected for Milky Way transmission using the features \textsc{mw\_transmission\_g}, \textsc{mw\_transmission\_r} and \textsc{mw\_transmission\_z}. We refer to the fiber magnitudes as \gfib, \rfib, and \zfib, respectively. In the following sections, we discuss the various choices we make in preparing the final sample. 

\subsection{Selection of galaxy sample based on photometric and spectroscopic properties}

As we use the \fuji spectroscopic sample to select ELGs from the DR9 catalog, we run several quality checks on the \fuji catalog to define what magnitude and color cuts will define our final sample. These quality checks are necessary to understand whether there are contaminants such as stars, QSOs, and low-redshift interlopers, whose photometric properties are similar to ELGs in the desired redshift range. How these contaminants cluster in the color-magnitude-fiber magnitude space is essential to removing them from our final sample. In addition, these quality checks also help us determine which parts of the data have reliable spectroscopic redshifts to ensure accurate photometric redshift calibration. 

Our quality checks took two forms: 1) for objects based on photometric properties such as the targets' colors, magnitudes or shapes, we apply the same quality cuts on the photometric sample; 2) for objects whose quality cuts depend on spectroscopic properties such as the signal-to-noise of the \oii doublet lines, we look for how galaxies with undesirable properties cluster in the color magnitude space. We train a Random Forest classifier \citep{Kam95} on these clustering properties to delineate where the undesirable galaxies live in this $8$-dimensional space and use the Random Forest on the photometric sample to remove galaxies with similar properties. 

When processing the \fuji catalog, we remove all targets that have been flagged by the DESI spectral pipeline. These quality checks are listed below: 

\begin{itemize}
    \item We select targets with good photometry. Targets with bad photometry are masked in the column \textsc{photsys}.
    \item We select targets that the pipeline considers to have reliable co-added spectra for spectroscopic redshift measurement; we accomplish this task by selecting targets whose \textsc{coadd\_fiberstatus} $== 0$. 
\end{itemize}

\begin{figure}
    \centering
    \includegraphics[width=\columnwidth]{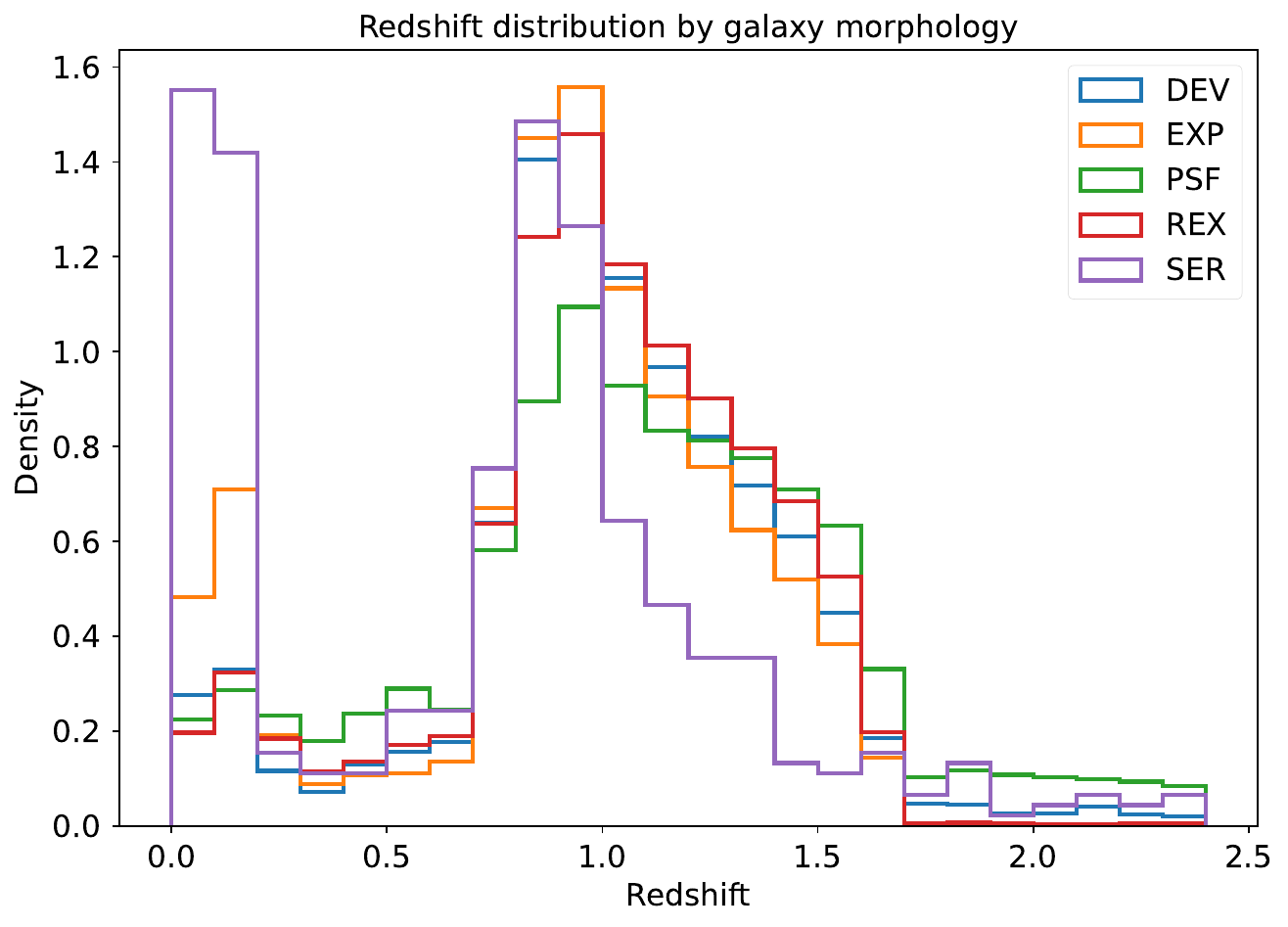}
    \caption{Normalized redshift distribution by galaxy morphology showing Sersic profile galaxies are majority low-redshift interlopers.}
    \label{fig4:sersic}
\end{figure}

With the above selection applied,  a clean catalog is obtained in which further cuts are made on the photometric properties: 

\begin{figure}
    \centering
    \includegraphics[width=\columnwidth]{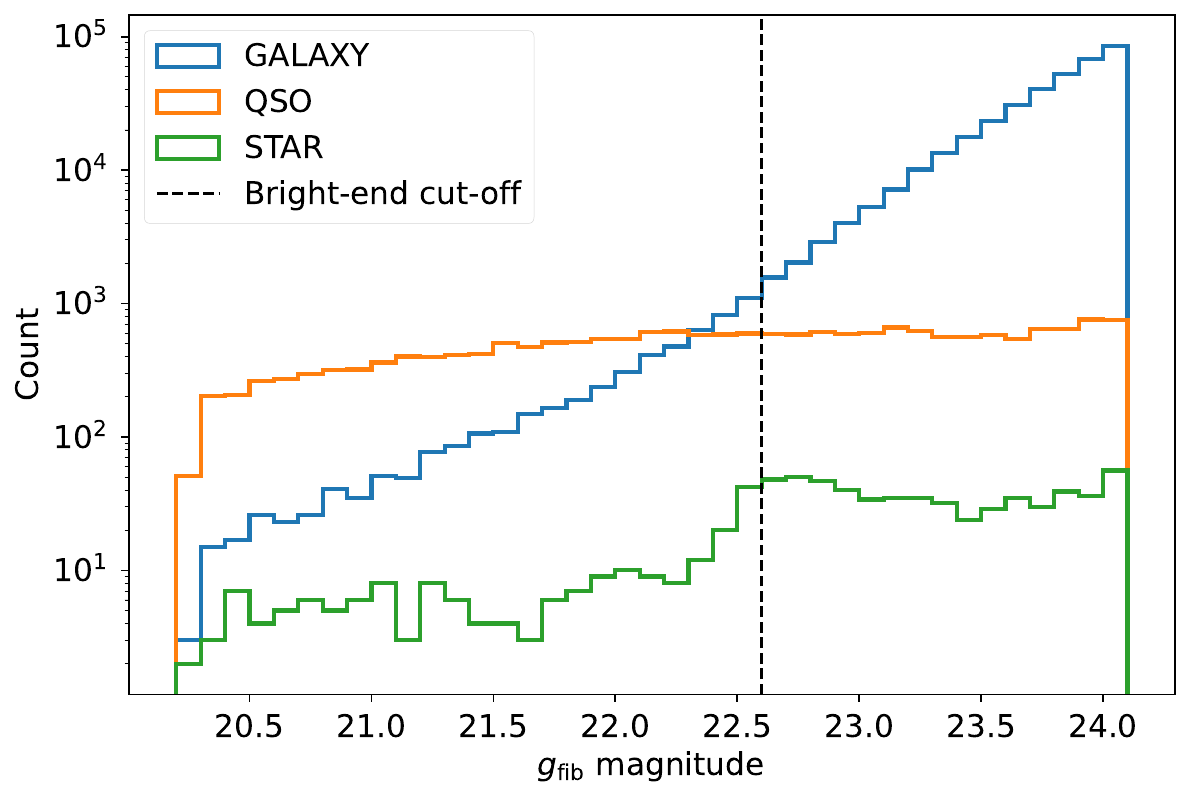}
    \caption{Distribution of ELG targets by spectroscopic types. Above $g_{\rm fib} = 22.6$, the galaxy count is more than half an order of magnitude larger than the QSO counts.}
    \label{fig4:gfib_cut}
\end{figure}

\begin{itemize}
    \item We remove targets whose morphology corresponds to the Sersic profile according to the DR9 catalog pipeline; we accomplish this task by selecting targets with \textsc{morphtype} $!=$ \textsc{ser}. We remove Sersic profile targets because when we look at the spectroscopic redshift distribution of the targets as a function of their morphology, we find that more than half of the Sersic profile targets are low-redshift interlopers (Figure~\ref{fig4:sersic}). In addition, the Sersic profile targets form only $0.12\%$ of the overall catalog. Hence, removing them does not affect the overall number density by any discernible amount.\\
    \item We remove targets with \textsc{shape\_r} $\geq 1.5$ arcsec. \textsc{shape\_r} refers to the half-light radius of galaxies with extended morphologies. We find that targets with larger half-light radii make up only $1\%$ of the total sample. Therefore, calibration concerning these galaxies will be more prone to shot noise. \\ 
    \item We apply a fiber magnitude bright end cut of \gfib $> 22.6$ because below this threshold, a large fraction of ELG targets are deemed as QSOs by the DESI pipeline (Figure~\ref{fig4:gfib_cut}). Specifically, the fraction of QSO contaminants is $65.19\%$ and $2.49\%$ below and above the threshold, respectively. Since the QSOs may have a different linear bias than the ELGs, we apply this cut to obtain a purer sample. As a result, we discard $3.92\%$ of the \fuji sample. \\
    \item We look at the histograms of the different photometric features and apply cuts to select the most sampled region in the color-magnitude-fiber magnitude space with the following: $r \leq 27$, $z \leq 25$, $ - 0.4 < g-r < 0.5$, $0.1 < r-z < 1.25$, \rfib $ \leq 27.5$, and \zfib $ \leq 25.5$.
\end{itemize}

\begin{figure}
    \centering
    \includegraphics[width=\columnwidth]{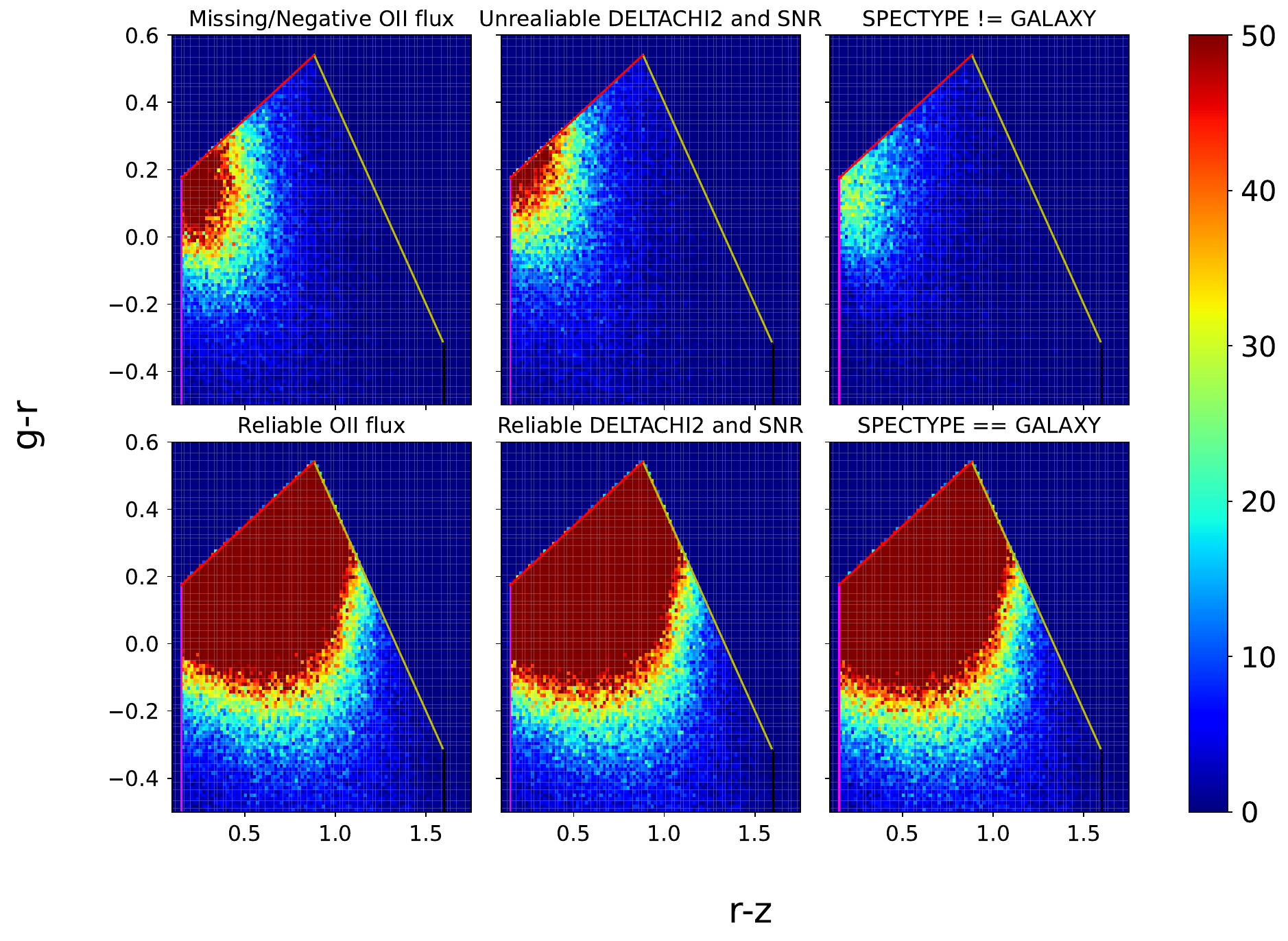}
    \caption{$2$D histogram of ELG targets in the $g-r$ versus $r-z$ space. These plots show the clustering of ELG targets based on their spectroscopic properties. The top row shows objects deemed as Label $2$.}
    \label{fig:hist_clust}
\end{figure}

Next, we look at how the targets that the spectroscopic pipeline flags as contaminants cluster in the color-magnitude-fiber magnitude space. There are three sources of contamination -- (i) ELG targets with spectroscopic redshift $z < 0.6$ because these do not fall into our redshift range sample, (ii) ELG targets that the spectroscopic pipeline flags as not galaxies, (iii) targets present in \fuji whose spectroscopic properties make them unreliable for spectroscopic redshift measurements. 

To isolate these contaminants from the good targets, we train a Random Forest Classifier that draws decision boundaries in the color-magnitude-fiber magnitude space based on labels for each ELG target in the \fuji sample. We denote ELGs that have spectroscopic redshifts in the range $0.6 < z < 1.6$ as Label $1$, low-redshift interlopers belonging to category (i) as label $0$ and contaminants belonging to categories (ii) and (iii) as label $2$. Thus, with each target assigned a label in the \fuji catalog, the Random Forest Classifier learns to identify which parts of the feature space belong to our desired category, i.e., $1$, and which ones belong to the ones we do not want in our sample, i.e., $0$ and $2$. 

We identify these contaminants as follows:

\begin{itemize}
    \item We look at the column \textsc{spectype} in the \fuji catalogue. This column indicates whether the DESI pipeline deemed the target galaxy, a QSO, or a star. We select targets that meet the requirement \textsc{spectype} $==$ \textsc{galaxy}. We find that $4.95\%$ and $0.2\%$ of the ELG targets in \fuji are QSOs and stars, respectively. These latter contaminants are assigned as Label $2$.  
    \item We consider any ELG targets with no or negative \oii flux as bad because \oii doublet is the key signature the DESI pipeline uses to measure spectroscopic redshift. These targets are also assigned as Label $2$.
    \item Reference \cite{raichoor23} showed that there is a relationship between the reliability of spectroscopic redshift measured by the DESI pipeline and the two spectroscopic features, \textsc{foii\_snr} and \textsc{deltachi2}. \textsc{foii\_snr} refers to the signal-to-noise of the measured \oii flux and
    \textsc{deltachi2} refers to the $\chi^2$ difference between the best-fit template and the second-best-fit template. Based on visual inspection of $3500$ ELGs in the SV sample, Reference \cite{raichoor23} shows that ELGs that have reliable spectroscopic redshift obey the following relationship: $\log_{10} \left( \textsc{foii\_snr} \right) > 0.9 - 0.2 \times \log_{10} \textsc{deltachi2}$. We apply this same criterion to the \fuji ELGs and assign Label $2$ to any targets that fail this criterion.
    \item Finally, we look at the spectroscopic redshifts of the galaxies that pass the three criteria mentioned above. Any targets with $z < 0.6$ are assigned the Label $0$, and the rest are assigned the Label $1$. 
\end{itemize}

These contaminants cluster in a specific way on the color-magnitude-fiber magnitude space. We show this clustering on the $g-r$ versus $r-z$ space as an example in Figure~\ref{fig:hist_clust}, where we show $2$D histograms of these contaminants. 

\subsection{Random Forest Classifier training}
Once all the targets in the \fuji catalog are assigned labels, we train a random forest classifier to determine which regions in the color-magnitude-fiber magnitude space are unreliable for spectroscopic redshift calibration. We make a $80-20$ split to prepare the training and test sets. 

We make this split by fields rather than by randomly selecting $80\%$ of the data as the training set. Figure~\ref{fig4:sv3} shows $20$ ELG fields observed in the \fuji catalog; we choose $16$ fields without replacement as training sets, and the rest of $4$ fields are the test sets. We split in this specific way to deal with the so-called \emph{sample variance} of the calibration set; if a particular field by chance lies on a highly fluctuating region, e.g., a void or a galaxy cluster, then calibrating against it would introduce bias \citep{sanchez20}. Hence, the split by fields ensures we are accounting for such variances. 

However, with a bin width of $\Delta z = 0.1$, the total number of bins we are estimating is $30$, while the number of independent fields we have is $16$. Thus, to have the number of training sets in the same order of magnitude as the number of histogram parameters, we further divide the $20$ ELG fields into equal halves to create a total of $40$ fields. As a result, we ultimately use $32$ fields for training and $8$ for testing. Since the Random Forest Classifier has many hyperparameters, we tune the hyperparameters using the \textsc{scikit-learn} \citep{scikit-learn11} function \textsc{RandomSearchCV}. The tuned hyperparameters are listed in Table~\ref{tab4:hypparams}.

\begin{table}
    \centering
    \begin{tabular}{|c|c|}
    \hline
        Hyperparameters & Values \\
        \hline
         \textsc{class\_weight} & [0: 0.325, 1: 0.35, 2: 0.325] \\
         \textsc{max\_features} & sqrt \\
         \textsc{min\_samples\_leaf} & 10 \\
         \textsc{min\_samples\_split} & 10 \\
         \textsc{n\_estimators} & 20 \\
         \hline
    \end{tabular}
    \caption{Hyper-parameters of Random Forest Classifier used to calibrate photometric redshift distribution.}
    \label{tab4:hypparams}
\end{table}

\begin{figure}
    \centering
    \includegraphics[width=0.5\columnwidth]{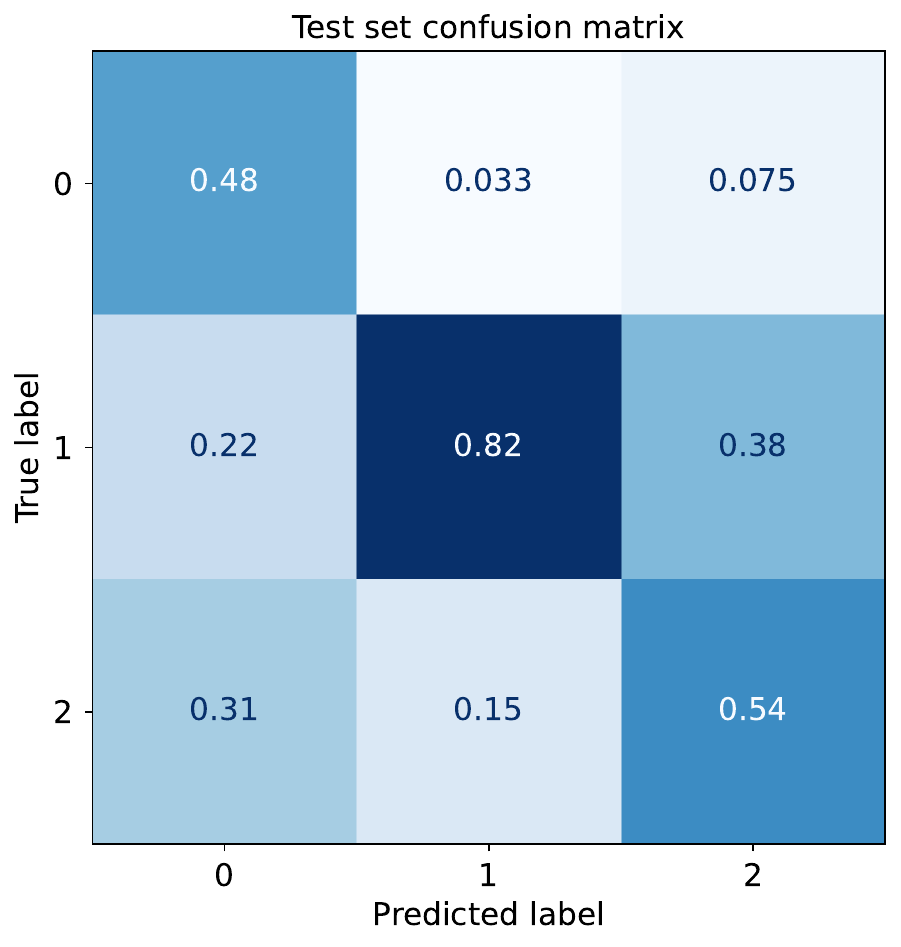}
    \caption{Confusion Matrix of the Random Forest Classifier. The columns are normalized. $15\%$ of the ELGs deemed as label $1$ are label $2$, and the source of unknown redshift contamination.}
    \label{fig4:conf_matrix}
\end{figure}

This procedure selects about $24$ million ELGs from the parent ELG sample in the Legacy Surveys, yielding an overall ELG number density of about $1770$ deg$^{-2}$. 

Overall, the training sample is composed of $4\%$, $74\%$ and $22\%$ of Label $0$, $1$ and $2$ respectively. The Random Forest Classifier training results are shown in Figure~\ref{fig4:conf_matrix}; the columns are normalized to reflect what fraction of the Classifier predicted labels are false positives. Compared to the baseline, we have a $8\%$ improvement in identifying label $1$ objects. Although the Classifier predicts $3.3\%$ of the Label $0$ objects as Label $1$, we can model these objects into our analysis since we know their spectroscopic redshifts. In contrast, the Classifier predicts $15\%$ of the Label $2$ objects as Label $1$ objects. We discuss in-depth how we model these objects in Section~\ref{sec4:model_unknown}. 

Note that given the broadband photometry of the DESI Legacy Surveys and because a large portion of our sample ELGs are PSF-type objects without morphological information, it is not possible to report reliable per object photometric redshift estimations. As such, we resort to using our Random Forest Classifier-based approach to model the overall redshift distribution.

\subsection{Modeling Redshift Distribution of Unknown Galaxies}
\label{sec4:model_unknown}
The confusion matrix plot (Figure~\ref{fig4:conf_matrix}) shows that the Classifier misclassifies $15\%$ of Label $2$ objects as Label $1$ objects, i.e., galaxies that the Classifier considers to be in the redshift range $0.6 < z < 1.6$, but do not have secured spectroscopic redshift measurements. Thus, while these galaxies contribute to the angular clustering measurements, without knowing their redshifts, it is not easy to assess their contribution to the overall power spectra. Therefore, we develop a model to estimate the redshift distribution of these Label $2$ galaxies misclassified as being Label $1$. Note that the integral of the density normalized redshift histogram of the known galaxies and these unknown galaxies should add up to $1$ because these two types of galaxies make up all the samples in the defined color-magnitude-fiber magnitude space. Keeping this in mind, we make some assumptions about their distributions: 

\begin{itemize}
    \item We assume that a fraction of these unknown galaxies are $z > 1.6$; hence, DESI could not locate any \oii doublets to get secured redshift measurements. This assumption is valid because galaxies just above $z \sim 1.6$ have similar color-magnitude-fiber magnitude properties to those just below the redshift threshold. However, since we do not know what this fraction is, we assign a sampling variable, $r$, between $0$ and the failure rate, $f$. 
    \item We further assume that the fraction of objects $1 - r$ is in our desired tomographic range, except for stochastic reasons, we could not estimate their redshifts. Thus, we rescale the known redshift bins $z < 1.6$ by that fraction. 
    \item Finally, we assume that the $z > 1.6$ objects fall off exponentially with redshift, meaning that it is more likely that we have ELGs in the bin $1.6 < z < 1.7$ than in the bin $1.7 < z < 1.8$. Our assumption implies that galaxies from a vastly different redshift would have characteristically different color-magnitude-fiber magnitude properties. Since we know the bin values below $1.6$, and $r$ represents the area under the curve for $z > 1.6$ of the redshift distribution, we have a unique exponential decay curve for each of the $1000$ bootstrapped redshift distributions. We go up to $ z = 3$  and get the final curves in Figure~\ref{fig4:dndz}. We derive the functional form of the exponential tail of the redshift distribution in Appendix~\ref{app4:exp_decay}.
\end{itemize}

As an illustrative example, consider a bootstrap run where the Random Forest Classifier misclassified $15\%$ of Label $2$ objects as being $1$; we have to model the redshift distribution of this $15\%$. First, we uniformly sample a number, $r \sim \mathcal{U} [0, 0.15]$. We rescale the known redshift bins by $1 - r$. Afterwards, since we know the value of the last bin, it serves as the first point of the exponential decay curve. The total area under this curve has to be equal to $r$. The first point of the decay curve and the total area under it uniquely define it. Thus, we obtain the redshift distribution of all the galaxies.

This estimation is, of course, inherently noisy due to the assumptions made. However, in the absence of deep spectroscopic samples that we can use to calibrate the misclassified objects, our method is robust enough to handle this problem. We explicitly account for the uncertainty coming from this method in the covariance matrix; we produce correlated Gaussian mocks with these $1000$ bootstrapped values to estimate the final covariance matrix, following the approach outlined in Reference \cite{Karim23}. 

\subsection{Estimating Redshift Distribution Uncertainty}
\label{sec4:redz_uncertainty}
\begin{figure}
    \centering
    \includegraphics[width=0.6\columnwidth]{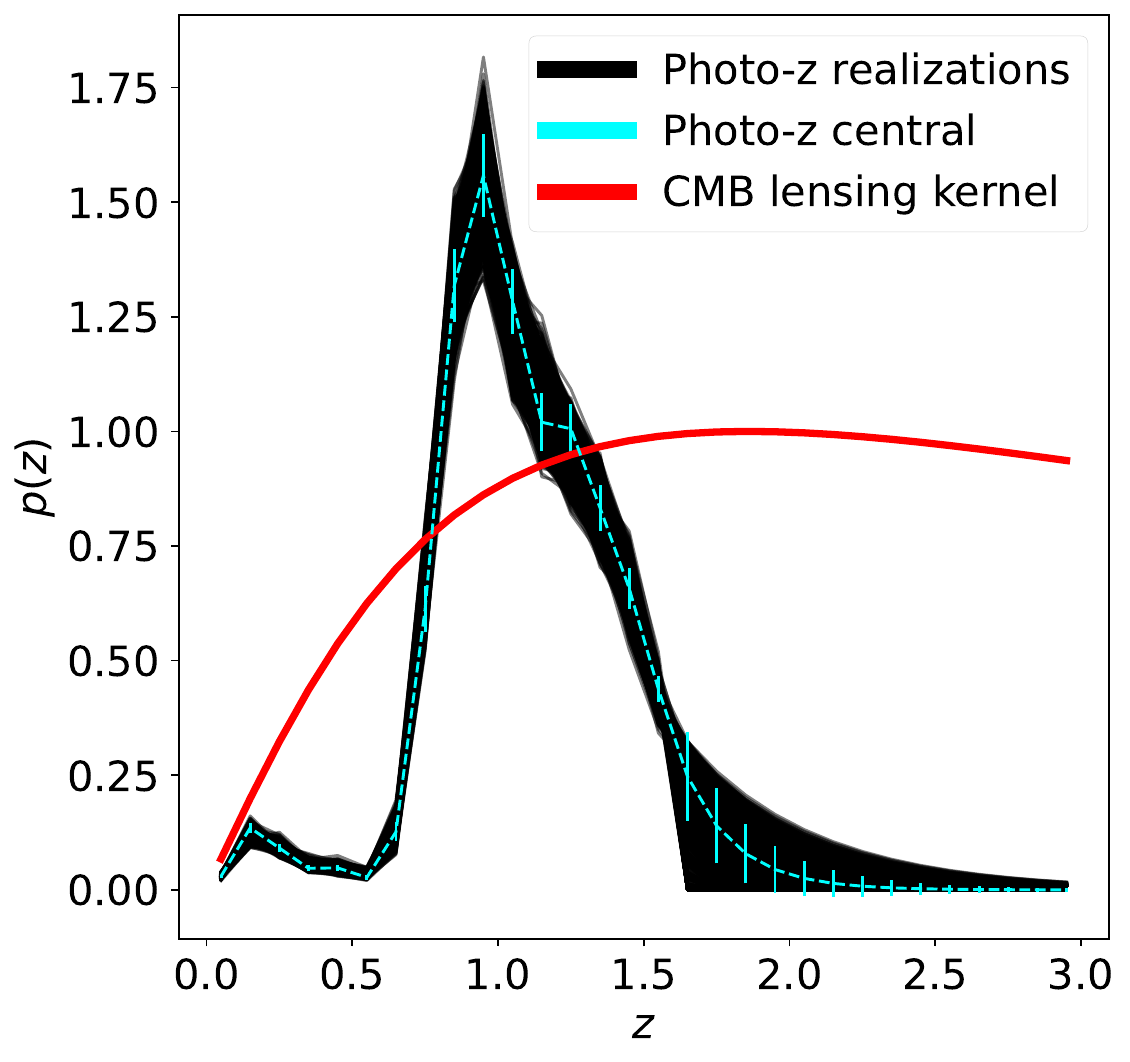}
    \caption{Photometric Redshift Distribution of Legacy Surveys DR$9$ ELGs. The cyan line shows the distribution with $1\sigma$ error bars per bin. The black contour shows $1000$ bootstrapped photometric redshift distributions. The red line shows the CMB lensing kernel with some arbitrary normalization.}
    \label{fig4:dndz}
\end{figure}

We use the bootstrap resampling method \citep{efron92} to estimate the redshift distribution uncertainty. We do this by splitting the $40$ fields into two sets of $32$ and $8$ fields, respectively; these form the basis for the training and the test sets. Then, we resample the indices of the $32$ training fields and the $8$ test fields with replacements. We use the resampled training fields to train the Random Forest Classifier with the same hyperparameters as Table~\ref{tab4:hypparams} to estimate the redshift distribution. We do this $1000$ times shown by the black curves in Figure~\ref{fig4:dndz}, and the cyan curve represents the photometric redshift distribution based on the training of the Random Forest Classifier without replacement. The spread of the black curves indicates the uncertainty. Based on the redshift distribution, the mean redshift, $z_{\rm mean}$, is $1.09$. 

\section{Method}
\label{sec4:method}

Our analysis is based on forward-modelling the matter power spectrum, $P_{mm} (k)$ at redshifts of our interest to the observed $2\times2$ angular pseudo-power spectra $D_{gg} (\ell)$ and $D_{\kappa g} (\ell)$. We use the \textsc{skylens} package introduced in \cite{Singh21} to achieve our goal. \textsc{skylens} is a package that can quickly evaluate angular power spectra of tracers using the Limber approximation \citep{Limber53} by solving the following integral:

\begin{equation}
\label{eq:cell}
    C_{XY} (\ell) = \frac{1}{c} \int_{z_1}^{z_2} \frac{H(z)}{\chi^2 (z)} W_X W_Y P_{mm} \left( k = \frac{\ell}{\chi (z)}, z \right) dz
\end{equation}

\noindent where $C_{XY} (\ell)$ is the angular power spectra between tracers $X$ and $Y$, $z_1$ and $z_2$ denote the redshift range over which the angular power spectra are being measured, $\chi (z)$ is the line-of-sight comoving distance at redshift $z$, $W_X$ is the radial kernel of tracer $X$ and $P_{mm} (k = \ell/\chi(z), z)$ is the matter power spectrum at redshift $z$. We do not take redshift-space distortions (RSD) into account because our redshift integral kernel is broad enough ($\sim 6400$ Mpc in comoving distance) that the impact of RSD is negligible.

Note that for CMB lensing, $W_\kappa$ represents the CMB kernel given by:

\begin{equation}
    W_{\kappa} (z) = \frac{3 \Omega_{\rm M}}{2c} \frac{H^2_0}{H(z)} (1 + z) \chi (z) \frac{\chi_* - \chi(z)}{\chi_*}
\end{equation}

\noindent where $\chi_*$ refers to the line-of-sight co-moving distance to the surface of the last scattering. 

On the other hand, the galaxy radial kernel, $W_g$, consists of both clustering and lensing magnification bias terms, $\mu (z)$, given by:

\begin{align}
\label{eq4:magbias1}
    W_g (z) &= b(z) \frac{dN}{dz} + \mu (z) \\
    \label{eq4:magbias2}
    \mu (z) &= \frac{3 \Omega_M}{2c} \frac{H^{2}_{0}}{H(z)} (1 + z) \chi (z) \notag \\
    &\times \int^{z_*}_{z} dz' \left( 1 - \frac{\chi(z)}{\chi(z')} \right) \left(\alpha (z') - 1 \right) \frac{dN}{dz'}
\end{align}

\noindent where, $b(z)$ is the galaxy bias, $dN/dz$ is the normalized redshift distribution, $z_{*}$ is the redshift of the last scattering surface, and $\alpha (z)$ is the slope of the galaxy number count at the flux density limit of the survey \cite{Bianchini16}.

Furthermore, \textsc{skylens} can consider the effect of survey geometry and quickly evaluate pseudo angular power spectra on fractional sky coverage. Because spherical harmonic signals are defined over the entire sphere, but the observed spherical harmonics are measured from only a fraction of the whole sphere, we must account for mode coupling induced by the survey geometry to make proper inferences. More importantly, rather than just using a binary survey geometry mask (a binary function) that tells us whether a portion of the sky is observed or not, \textsc{skylens} can handle a user-defined survey window function (a continuous function) that represents how to apply weights to different pixelized portions of the sky to account for various systematics effects properly. \textsc{skylens} achieves this by solving the following:

\begin{align*}
    D_\ell &= M_{\ell, \ell'} C_{\ell'} \\
    M_{\ell, \ell'} &= \frac{2 \ell' + 1}{4\pi} \sum_{\ell''} W_{\ell''} (2\ell'' + 1)  \begin{pmatrix}%
    \ell & \ell' & \ell'' \\
    0 & 0 & 0
    \end{pmatrix} \begin{pmatrix}%
    \ell & \ell' & \ell'' \\
    0 & 0 & 0
    \end{pmatrix}
\end{align*}

\noindent where $D_\ell$ is the pseudo angular power spectrum up to multipole $\ell$, $C_\ell'$ is the power spectrum shown in Equation~\ref{eq:cell} up to multipole $\ell'$, $M_{\ell, \ell'}$ is the coupling matrix representing how to transform from the basis of full-sky power spectra to fractional-sky power spectra, $W_{\ell''}$ represents the angular auto power spectrum of the window function up to multipole $\ell''$, and the last matrix represents the Wigner-$3j$ symbol that accounts for the normalization \cite{hivon2002}. 

Thus, given user-provided models of the matter power spectrum, galaxy bias, galaxy redshift distribution, slope of the galaxy number count, and the window function (accounting for properties such as imaging systematics), \textsc{skylens} forward models what the expected observed pseudo angular power spectra should look like. In the rest of this section, we discuss the modeling of these user-defined parameters. 

\subsection{Galaxy bias}
Galaxies are biased tracers of the underlying matter density field. Consequently, galaxy clustering is also a biased tracer of the underlying matter power spectrum. In linear theory, the galaxy auto-power, $C_{gg}$, and the galaxy-CMB lensing cross-power spectra, $C_{\kappa g}$, are related to the matter power spectrum via:

\begin{align}
    C_{gg} (k) &\propto b^2(z) \sigma^2_8  \label{eq:cgg} \\
    C_{\kappa g} &\propto b(z) \sigma_8^2 \label{eq:ckg}
\end{align}
\noindent where, $b(z)$ is the galaxy linear bias and $\sigma_8$ is the variance of the matter density field at $8$ Mpc h$^{-1}$ scale. Thus, we need a model for $b(z)$ for accurate modeling of galaxy clustering. In this work, we use a simple model of the form: 

\begin{equation}
\label{eq4:bias}
    b(z) = \frac{b_0}{D^{*}(z)}
\end{equation}

\noindent where, $b_0$ is the bias constant and $D^{*}(z)$ is the growth function, normalized to $1$ at $z = 1$ (as opposed to the general convention of defining $D(z = 0) = 1$). Note that in this formalism, we implicitly assumed that the galaxy linear bias is scale-independent, but for a more thorough investigation, one should consider scale-dependence.

\subsection{Magnification Bias}
\label{sec:magbias}

In a flux-limited survey, the galaxy number count can be modulated based on the impact of weak gravitational lensing induced by intervening matter between the observer and the farthest galaxies in the sample. The weak lensing effect can reduce the number of galaxies observed by stretching space around lenses and, at the same time, can increase the number of galaxies by amplifying the flux of galaxies outside the flux limit due to magnification. Both of these effects together are known as magnification bias; they must be taken into our modeling of galaxy auto power spectrum and galaxy-CMB lensing cross power spectrum because this effect induces an additional signal in the power spectra, as shown in Equations~\ref{eq:cell} and \ref{eq4:magbias1}. 

We notice in Equation~\ref{eq4:magbias2} that the parameter that determines the impact of magnification bias is the $\alpha$ parameter, which represents the slope of cumulative number count distribution evaluated at the magnitude of the faint-end cut. For a single-band faint-end cut selection, $\alpha$ is given as \citep{Hildebrandt09}:

\begin{equation}
    \alpha = 2.5 \frac{d \log_{10} N(<m)}{dm} \rvert_{{\rm faint-end~cut}} \label{eq4:magbias}
\end{equation}

However, our selection function is complex, depending on both magnitudes and fiber magnitudes in all three bands. The fiber magnitudes cut makes it more complicated because a $1\%$ decrease in total flux does not necessarily correspond to a $1\%$ decrease in fiber flux. Instead, the total and fiber flux relationship depends on the galaxy morphology. Reference \cite{zhou23} calculated a look-up table to convert between total flux and fiber flux depending on DR9 morphology and used this table to define a more robust approach for measuring magnification bias, and we use this look-up table in our analysis. In this procedure, we first apply our trained Random Forest Classifier on the entire DR9 ELG catalog and count how many are assigned label $1$. We then reduce the galaxies' observed total flux by $1\%$ and the corresponding fraction of the observed fiber flux according to the look-up table from \cite{zhou23}. We reapply the Random Forest Classifier and count the number of ELGs with the label $1$. Finally, we use Equation~\ref{eq4:magbias} and a finite-difference method to calculate $\alpha = 2.225 \pm 0.018$. 

\subsection{Survey Geometry}
\label{sec4:survey_geom}

The survey geometry defines the survey footprint of the analysis. The first-order survey geometry is defined by the scope of the survey itself, which may depend on the observatory location (for terrestrial telescopes), target selection, and survey tiling strategies, but also may depend on the closeness to the plane of the Milky Way, and closeness to bright stars. The latter factors are important for cosmological surveys that are detecting objects right around the flux-limit because if a particular patch of the sky has more reddening due to its proximity to the Milky Way, then we may not be able to count all the galaxies up to the flux-limit. This heterogeneity can induce spurious signals in the power spectra and, by proxy, parameter estimation. One strategy to deal with such problematic parts of the survey footprint is to mask them out completely.  

However, liberally masking the footprint comes at the expense of losing higher signal-to-noise measurements. We investigated this question by using $7$ different foreground imaging systematics maps, namely Galactic extinction, point spread function (PSF) size in $g-$, $r-$ and $z-$ bands based on the Legacy Surveys, and the galaxy depth in $g-$, $r-$ and $z-$ bands based on the Legacy Surveys. We found that the following cuts preserve the maximum area of the survey while masking out the most problematic pixels: 

\begin{align*}
    &E(B-V) < 0.11 \\
    &{\rm g-band~Galaxy~Depth} > 23.9\\
    &{\rm r-band~Galaxy~Depth} > 23.1\\
    &{\rm z-band~Galaxy~Depth} > 22.4\\
    &{\rm g-band~PSF~Size} < 2.5^{\prime \prime}\\
    &{\rm r-band~PSF~Size} < 2.5^{\prime \prime}\\
    &{\rm z-band~PSF~Size} < 2.0^{\prime \prime}\\
\end{align*}

As described in detail in Section~\ref{sec:galwin}, we use a regression method to estimate how the foreground affects the observed number density. As long as we have a sufficient number of pixels with similar foreground values, we can model the impact of foregrounds in an unbiased manner. Thus, the aforementioned choices ensure that we cut only the most extreme pixels.

Afterward, we apply another layer of masking taking bright stars and foreground galaxies into account. Our initial assessment showed that the stellar and foreground galaxy masks used in the official DR$9$ catalog were not robust; stray light coming from bright stars and the foreground blue galaxies created artificial ELG targets. As a result, we were seeing spurious spikes of ELG number count closer to certain stars and foreground galaxies than expected. To fix this issue, we apply a more conservative correction by increasing the radius of masks around bright stars and galaxies. This correction results in a more accurate final analysis and we obtain our final survey geometry, shown in non-gray on the right-hand side of Figure~\ref{fig:delta-maps}). A more detailed discussion about this can be found in Appendix~\ref{app4:stellar_gal_masks}. 


\subsection{Window Function or Imaging Systematics Weights}
\label{sec:galwin}

Apart from the survey mask, we must account for the foreground systematics in the footprint because even the reliable pixels inside the survey footprint are affected by large-scale variations of imaging systematics. Such variation affects the number count per pixel, as explained in Section~\ref{sec4:survey_geom}. If this number variation effect is not considered, then the observed power spectra measurement will be biased by the contribution of such foreground imaging systematics. In this case, our fundamental assumption is that what happens in the foreground should \textit{not} affect the intrinstic ELG number density. In other words, we expect a zero correlation between foreground systematics and the ELG number density. 

Thus, if we see an apparent correlation, then we can infer that the foreground is responsible for this correlation. Following \cite{Rezaie20}, we use a neural network-based multivariate regression method to estimate the observed correlation, and also calculate what weights per pixel needs to be applied to null out the effect of the observed correlation; such regression-based technique has been used extensively in many surveys such as the DESI Year 1 analysis \cite{ross2024}, the Dark Energy Survey \cite{elvinpoole2018}, SDSS \cite{alam2017, ross2012}.

As long as we have a sufficient number of pixels that span a certain foreground value range, e.g., $0 \leq E(B-V) \leq 0.11$, the regression method can model the modulation of the galaxy number density as a function of the foreground. Hence, clipping only the most extreme values are sufficient.

In a companion paper \cite{Karim23} we derive and discuss the various choices in estimating the imaging systematics weights. Briefly, we use $11$ foreground maps -- Galactic extinction, stellar number density, point-spread function (PSF) size, PSF depth and galactic depth; the latter three systematics maps measured in $g-$, $r-$ and $z-$bands respectively. If the foreground systematics map $i$ is represented by $s_i (x)$ where $x$ represents the pixel number, then we can model the observed number count as:

\begin{align}
    n_{\rm obs} (x) = n_{\rm true} (x) F(x, s(x))
\end{align}

\noindent where $F(x, s(x))$ represents a function that takes all the systematics maps as an input and outputs a single number at pixel $x$ that can be thought of as an effective weight coming from all the foreground maps. This $F(x, s(x))$ can be modeled as simply as a linear model (shown below) or a more complicated non-linear model represented by a neural network.

\begin{equation*}
        F(x, s(x)) = \sum a_i s_i(x) + b_0
\end{equation*}

Note that our formalism implies for a perfect survey without any foreground systematics, $F(x, s(x))$ would be $1$ at every pixel, and the coefficients of $s_i$ would be $0$ everywhere. Thus, the regression approach tells us whether the coefficients, $a_i$, are non-zero. For non-linear foregrounds, neural-network-based approaches perform better than linear model approaches \cite{ross2024, Karim23, Rezaie20}.

\subsubsection{Sagittarius Stream}

During our analysis, we found excess large-scale power in the ELG auto-power spectrum that could not be explained by the $11$ foreground systematics used in the main DESI analyses. A further investigation revealed that the Sagittarius Stream is also contributing to the ELG foreground systematics. The Sagittarius Stream is a sizeable stellar complex that wraps around the Milky Way and crosses the Galactic poles closely. As such, we consider the possibility of the Stream contaminating our ELG catalog since the blue stars in the Stream could be misclassified as ELGs. The DESI QSO target selection paper \citep{chaussidon23} has already shown that the blue stars in the Stream affect the QSO DR9 catalog. 

\begin{figure}
    \centering
    \includegraphics[width=\columnwidth]{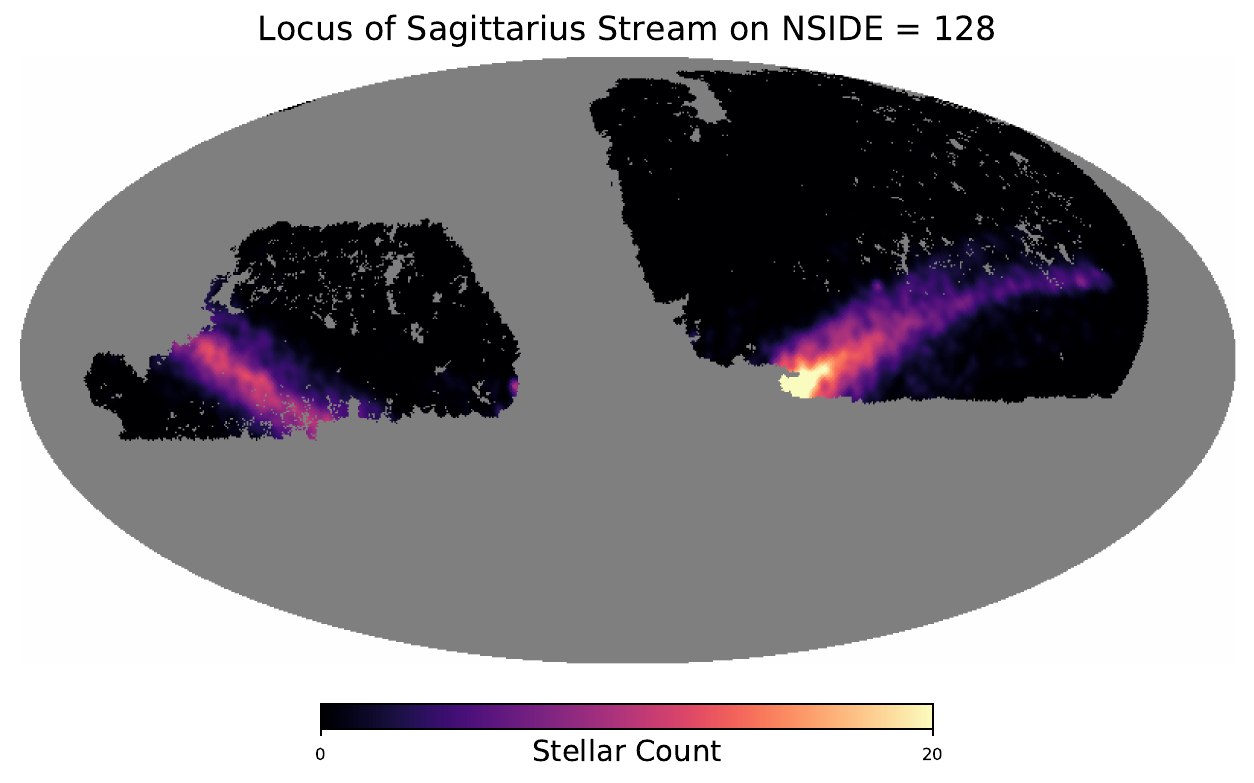}
    \caption{Extent of the Sagittarius Stream on the Legacy Surveys DR9 footprint based on candidate stars selected by Reference \cite{ramos2022}. A Gaussian smoothing of $\sigma = 1^{\circ}$ is applied.}
    \label{fig:sag-stream}
\end{figure}

To best model the extent of the Sagittarius Stream, we consider the catalog produced by \cite{ramos2022}\footnote{Data accessed: \url{http://cdsarc.u-strasbg.fr/viz-bin/cat/J/A+A/666/A64}} where they produce a sample of about $7\times 10^6$ candidate stars using the \textit{Gaia} Early Data Release 3. However, only about $85000$ stars fall within our footprint. As such, the locus of the Stream, as defined by these stars, is rather noisy. As such, we apply a Gaussian smoothing with $\sigma = 1^{\circ}$ on the Sagittarius Stream map and consider the smoothed map as one of our features, as shown in Figure~\ref{fig:sag-stream}.

\begin{figure}
    \centering
    \includegraphics[width=0.75\columnwidth]{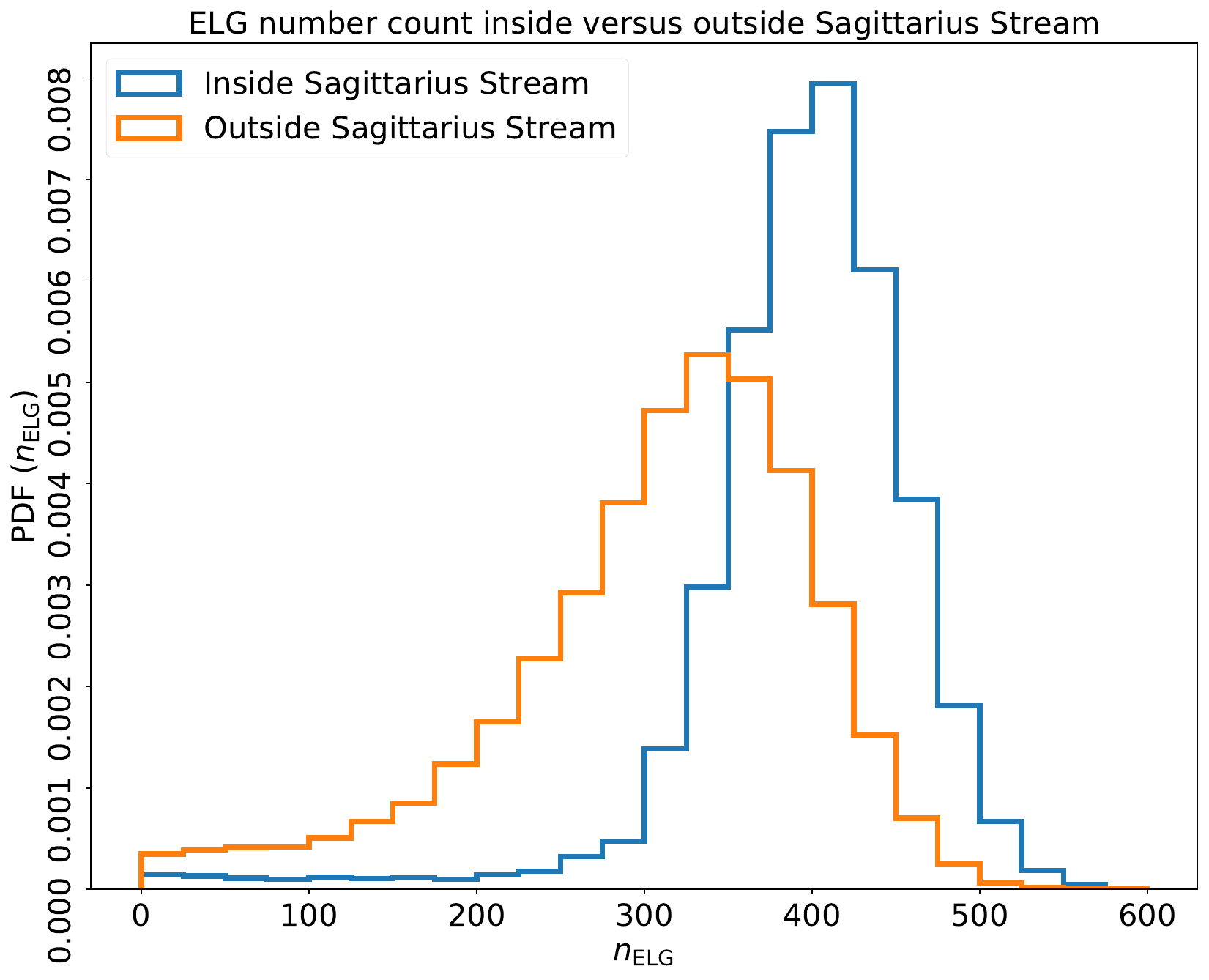}
    \caption{ Normalized ELG number count comparison between DR9 ELGs inside versus outside the Sagittarius Stream. The plot shows that the two distributions are different.}
    \label{fig:sag-elg}
\end{figure}

Figure~\ref{fig:sag-elg} shows that the Stream affects the ELG number count; the ELG number density skews towards the right compared to the number density histogram outside the Stream. We also performed an Anderson-Darling test \citep{anderson-darling} to check whether the two samples were drawn from the same underlying distribution; the null hypothesis was rejected with a p-value $< 0.001$. We use the Sagittarius Stream as one of the features in estimating the ELG galaxy window function. 

\subsection{Multiplicative Bias in \emph{Planck} CMB Lensing}
\label{sec:mc-norm}

The \emph{Planck} 2018 CMB lensing paper \citep{PlanckCMB18} mentions five main steps in lensing reconstruction; the final step refers to applying a multiplicative correction factor — this multiplicative correction factor arises due to some approximations and assumptions made in the prior steps. Below, we discuss the the approximations and assumptions of steps 3 and 4 which are relevant to this analysis.

In step 3, the \emph{Planck} team subtracts the mean field and normalizes the lensing map. Masks and anisotropies bias the lensing reconstruction estimator. This bias, known as the mean field, is the map-level signal expected from the mask, noise, and other anisotropic features of the map in the absence of lensing. It is estimated using the quadratic estimator, also used for lensing reconstruction. Once subtracted, the map is normalized using an \emph{approximate} isotropic normalization; this isotropic normalization is based on an analytic derivation for the whole sky \citep{okamoto2003}. Afterward, in step 4, the \emph{Planck} team measured the lensing potential power spectrum of the reconstructed map and subtracted three separate sources of noise -- $N^{(0)}$ that refers to the Gaussian noise even in the absence of lensing, $N^{(1)}$ that refers to the non-Gaussian noise, and the noise induced by point sources. 

However, the approximate isotropic normalization applied prior to the noise subtraction is suboptimal (For further discussion, refer to Sections 2.1 and 2.3 of \cite{PlanckCMB18}) because the noise results in significant anisotropy. The isotropic approximations made earlier in Step 3 become a significant enough factor downstream. To fix this issue, the \emph{Planck} team uses inhomogeneous filtering on their polarization data during reconstruction, drastically improving the lensing map. However, the inhomogeneous anisotropic filter results in a position-dependent (survey mask-dependent) normalization, or the multiplicative bias factor that we correct for when measuring the $D^{\kappa g}$ power spectrum. 

We calculate the multiplicative renormalization factor using the method outlined in Reference \cite{Farren23}. In this procedure, we use the $300$ pairs of \emph{Planck} lensing simulations that come with the CMB lensing data; each of these pairs has one simulation that is based on the pure CMB lensing convergence signal, while the other is a reconstruction of the lensing convergence signal in the presence of masks, noises, and other systematics. We can then apply the ELG mask on both the pure lensing convergence map and the reconstructed map and then take their ratio to estimate by how much the cross-power spectrum is biased per bin, i.e.,

\begin{equation}
    A (\ell) = \frac{<\delta_{\kappa}^{\kappa \rm{-mask}} \delta_{\kappa}^{g \rm{-mask}}>}{<\delta_{\hat{\kappa}}^{\kappa \rm{-mask}}\delta_{\kappa}^{g \rm{-mask}}>} = \frac{C_{\kappa_{\kappa \rm{-mask}} \kappa_{g \rm{-mask}}} (\ell)}{C_{\hat{\kappa}_{\kappa \rm{-mask}} {\kappa_{g \rm{-mask}}}} (\ell)}
\end{equation}

\noindent where $\kappa$ is a pure lensing convergence field simulation and $\hat{\kappa}$ is the corresponding reconstructed lensing convergence field. 

Since $A_{\ell}$ is a ratio estimator, taking the mean of the ratio from the $300$ simulations is far noisier than taking the ratio of the mean of the $300$ pairs of power spectra. We describe in Section~\ref{app:mc-norm} the result of the multiplicative bias factor, $A_{\ell}$, for different choices of the Galactic extinction map to correct the ELG overdensity. We multiply the measured $C_{\kappa g}$ power spectrum with this correction vector to obtain the debiased cross-power spectra. 

In Appendix~\ref{app:mc-norm}, we show a detailed study to validate that the variance arising from estimating this multiplicative bias factor is subdominant to other systematics, such as photometric redshift uncertainty; as a result, we do not propagate the variance of estimating the multiplicative bias factor into our final covariance matrix. 

Beyond the Monte Carlo norm correction there are two additional biases that may contribute to the cross-power spectrum measurement -- 1) the $N_L^{(3/2)}$ bias and 2) the extragalactic foreground contamination in CMB lensing reconstruction. The $N_L^{(3/2)}$ bias refers to the contribution of the bispectrum induced by the non-linear growth of the large-scale structures and post-Born lensing; however the contribution of this bias at redshift around $1$ and also in the \emph{Planck} lensing reconstruction is sub-percent level, \cite{fabbian2019} which is why we do not consider it. On the other hand, extragalactic foreground such as the cosmic infrared background and Sunyaev-Zeldovich clusters can affect the CMB lensing reconstruction estimator; however these biases are order of magnitudes smaller than the \emph{Planck} CMB lensing primary noise \cite{osborne2014}, which is why we do not consider this bias either in our analysis.

\section{Inference of Cosmological Parameters and The Covariance Matrix}
\label{sec4:cov}

Ultimately, we use the power spectra measurements to infer two cosmological and two galaxy bias parameters. The problem can be expressed in terms of Bayes' Theorem:

\begin{equation}
    P\left( \vec{\theta}|D \right) \propto \mathcal{L} \left( D|\vec{\theta} \right) \Pi \left( \vec{\theta} \right) \label{eq4:posterior}
\end{equation}

\noindent where, $D$ is the data (in our case the two power spectra measurements), $\vec{\theta}$ are the parameters we are interested in inferring ($b_0$, $\Omega_{\rm CDM}$ and $A_s$ or $\sigma_8$), $P$ is the posterior probability, $\mathcal{L}$ is the likelihood function, and $\Pi$ represents the prior probability of the parameters of interest. 

The priors of the parameters are:
 
\begin{align}
    \Pi \left( b_0 \right) &= \mathcal{U} \left(0.5, 3.5 \right) \\
    \Pi \left( A_s \right) &= \mathcal{U} \left(1, 3.5 \right) \\
    \Pi \left( \Omega_{\rm CDM} \right) &= \mathcal{U} \left(0.1, 0.6 \right)
\end{align}

\noindent where $\mathcal{U}$ and $\mathcal{N}$ refer to uniform and Gaussian distributions respectively. We use broad uniform priors for $b_0$, $A_s$, and $\Omega_{\rm CDM}$; we set the bounds so that existing literature values are within the range. 

Note that we fix other cosmological parameters, i.e., $h$, $\Omega_{\rm b}$, $\Omega_{\rm K}$, $\sum m_{\nu}$, $\tau$, $n_s$, $w_0$, $w_a$, $T_{\rm CMB}$, $r$, and constant $N_{\rm eff}$ according to Table~\ref{tab4:cov_params}. The estimated magnification bias parameter, $alpha$, is also fixed based on Section~\ref{sec:magbias}. Finally, the galaxy shot noise is also estimated using Equation~\ref{eq:shot-noise} by taking the ensemble average over the $1000$ window functions. 

\begin{equation}
    \label{eq:shot-noise}
    \delta^2_{N, W} = \widebar{W}_g \frac{1}{\bar{n}_g}
\end{equation}

\noindent where $\widebar{W}_g$ is the average of the imaging systematics weights and $\bar{n}_g$ is the average number of galaxies per solid angle.  

On the other hand, we model the likelihood as Gaussian because, according to the Central Limit Theorem, the likelihood will become Gaussian for scales much smaller than the survey size. As the DESI Legacy Imaging Surveys is about $14,000$ deg$^2$, and the largest mode we use in our analysis is $\ell_{\rm min} = 50$ (or $\theta \sim 3.6$ deg), the Gaussian likelihood is a good approximation. Thus, the logarithm of the likelihood is:

\begin{align}
    \log \mathcal{L} &=  -\frac{1}{2} d^T \left( \frac{n - p -2}{n - 1} \Sigma^{-1} \right) d \\
    d &= D^{\rm Obs}_{\ell} - D^{\rm Model}_{\ell} \notag
\end{align}

\noindent where, $D^{\rm Obs}_{\ell}$ refers to the binned pseudo-$C_{\ell}$ measurements of both galaxy-galaxy and galaxy-CMB lensing power spectra in a single vector, $D^{\rm Model}_{\ell}$ refers to the binned theoretical models and $\Sigma$ refers to the binned covariance matrix of the $D_{\ell}$. The analysis is done in the range $50 \leq \ell \leq 400$ with $9$ equally spaced bins in the logarithmic space. We choose $\ell_{\rm max} = 400$ because that is the recommended small-scale cut used in the official Planck lensing paper. We also restrict our galaxy-galaxy power spectrum analysis to the same $\ell$-space range to not be affected by mismodeling of non-linear effects at high-$\ell$. The models, $D^{\rm Model}_{\ell}$, are calculated using \textsc{skylens}, which in turn uses the power spectrum code \textsc{camb} \citep{camb} to calculate $P_{mm} (k, z)$. For the non-linear correction, we use \textsc{halofit} \citep{takahashi2012} inside \textsc{camb}. Note that \textsc{halofit} agrees at the level of $1\%$ with state-of-the-art emulators and N-body simulations at $k \sim 0.12$ Mpc$^{-1}$ \cite{garrison2018, mead2015}; it is the smallest scale probed by our analysis at the effective redshift of $1$. At this scale, the non-linear correction is about $1\%$ compared to the linear matter power spectrum. 

Note that the prefactor in front of the $\Sigma^{-1}$ is the Hartlap correction \citep{hartlap2007} to obtain an unbiased estimate of the precision matrix (inverse of the covariance matrix), where $n$ refers to the number of simulations and $p$ refers to the number of bins. In this analysis, we set $n = 1000$ and $p = 18$ since we have $9$ bins per power spectra. 

\begin{table}
    \centering
    \begin{tabular}{c|c}
        \hline
        Fiducial Parameter & Value \\
        \hline 
         $h$ &  $0.6776$ \\
         $\Omega_{\rm b}$ & $0.04897$ \\
         $\Omega_{\rm CDM}$ & $0.26069$ \\
         $\Omega_{\rm K}$ & $0.0$ \\
         $A_{s} \times 10^9$ & $2.097$ \\
         $m_{\nu}$ & [$0$, $0$, $0.6$] eV \\
         $\tau$ & $0.06$ \\
         $n_s$ & $0.965$ \\
         $w_0$ & $-1$ \\
         $w_a$ & $0$ \\
         $T_{\rm CMB}$ & $2.7255$ K \\
         $r$ & $0$ \\ 
         $N_{\rm eff}$ & $3.046$ \\
         Magnification Bias $\alpha$ & $2.225$ \\ 
         Shot noise $N_{gg} (\ell)$ & $4.618 \times 10^{-8}$ \\
         $b_0$ & 1.4 
    \end{tabular}
    \caption{Fiducial parameters used for estimating simulation-based covariance matrix.}
    \label{tab4:cov_params}
\end{table}

We estimate the covariance matrix using a simulation-based approach based on the method outlined in Reference \cite{Karim23}. We opt for a simulation-based approach rather than using a Gaussian analytic covariance matrix because we want to account for the variance of the galaxy window function and photometric redshifts so that we can marginalize over them. Briefly, we simulate $1000$ pairs of correlated Gaussian mocks of the galaxy overdensity field and the CMB lensing convergence field and apply the ELG mask described in Section~\ref{sec4:survey_geom} and the CMB lensing mask provided by \emph{Planck}. 

\begin{figure}
    \centering
    \includegraphics[width=0.75\columnwidth]{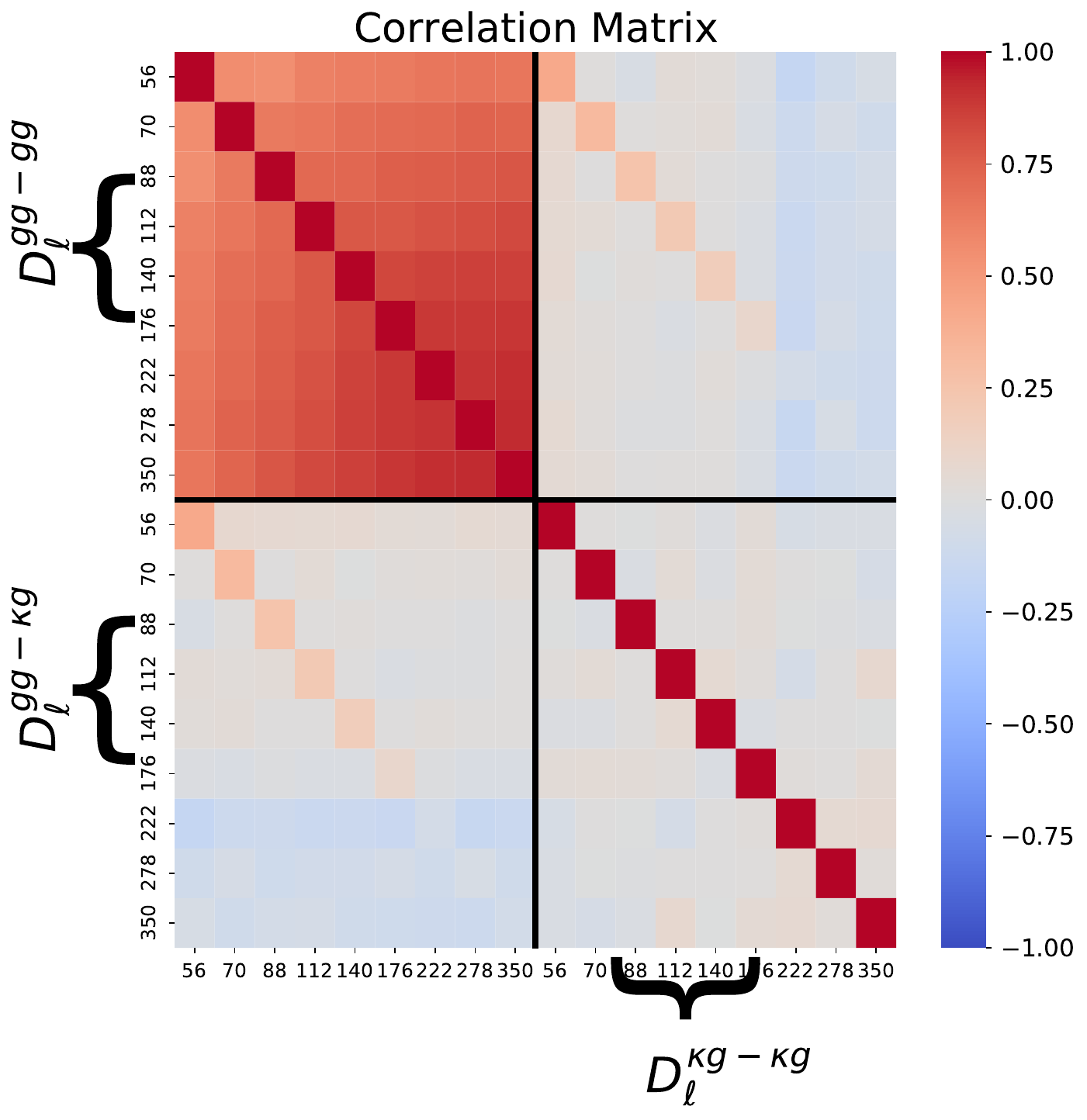}
    \caption{Covariance Matrix used in this analysis expressed as a correlation matrix. The galaxy-galaxy power spectra submatrix show high mode-mode coupling due to photometric redshift uncertainty.}
    \label{fig4:cor_matrix}
\end{figure}

While all of these simulations have the same baseline cosmology and galaxy biases (shown in Table~\ref{tab4:cov_params}), the individual simulations each have a different galaxy window function and redshift distribution, $p(z)$. We sample these window functions by taking snapshots of the neural network per training epoch, and we sample the $p(z)$ as described in Section~\ref{sec4:redz_uncertainty}. Note that shown in Equation~\ref{eq:shot-noise}, a different realization of the galaxy window function yields a different estimate of the galaxy shot noise. 

We then measure the $D_{gg}$ and $D_{\kappa g}$ per simulation and calculate the covariance from these $1000$ pairs of power spectra. Each of the $D_{gg}$ measurements are also corrected for by its corresponding shot noise as discussed above. We show the corresponding correlation matrix of this covariance matrix in Figure~\ref{fig4:cor_matrix} — the galaxy-galaxy power spectra portion of the covariance matrix exhibits high mode-mode coupling due to photometric redshift uncertainty. However, the covariance matrix is well-behaved with a condition number of about $2800$.

\section{Validation with Gaussian Simulations}
\label{sec4:sims}

\begin{figure}
    \centering
    \includegraphics[width=\columnwidth]{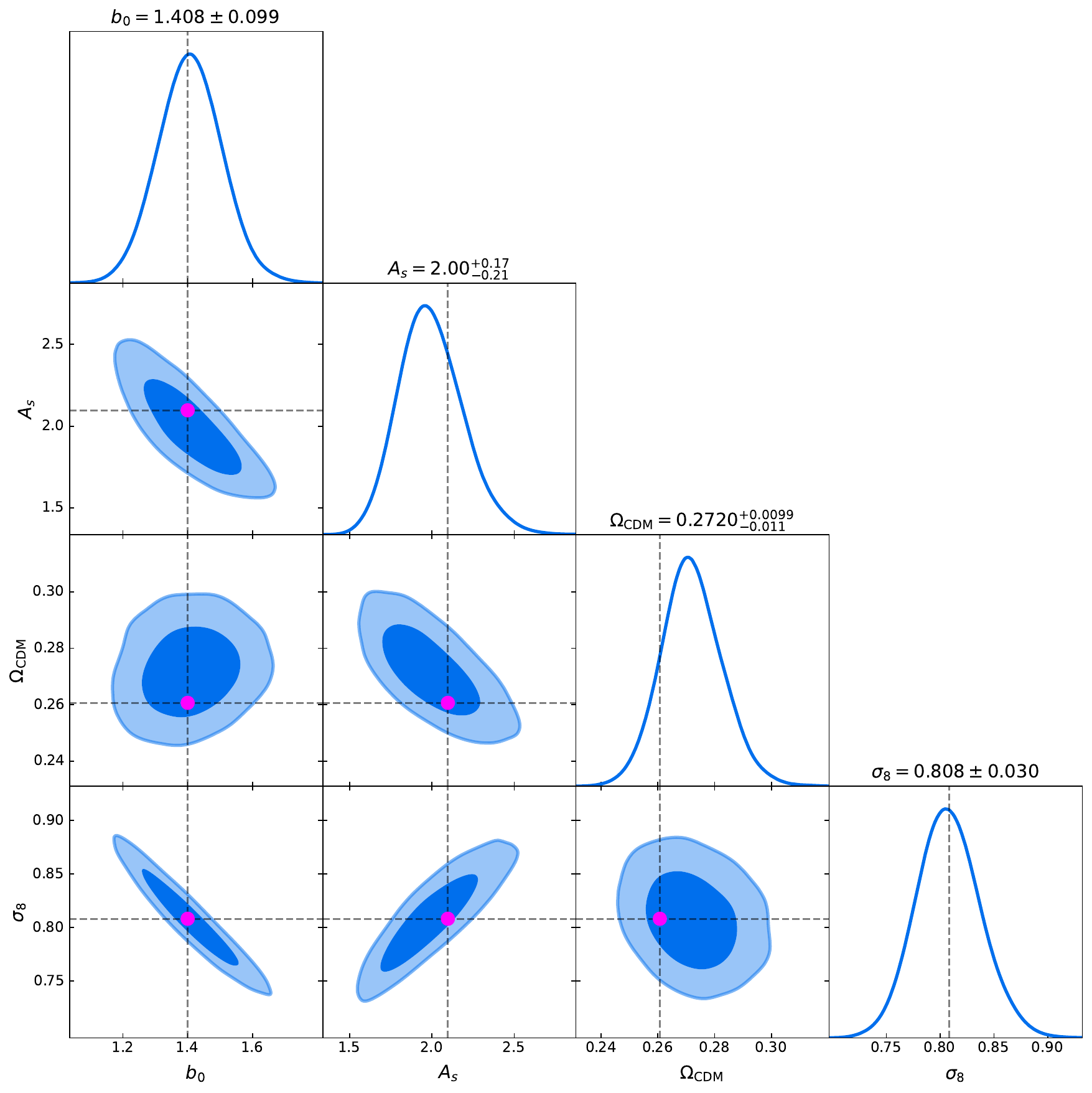}
    \caption{MCMC-based inference of galaxy linear bias, $b_0$, the amplitude of the matter power spectrum, $A_s \times 10^{-9}$, cold dark matter density, $\Omega_{\rm CDM}$, and $\sigma_8$. The red dots and the dashed lines indicate the truth value of the simulation. The contour plots show that the analysis pipeline can recover $b_0$, $\Omega_{\rm CDM}$ and $\sigma_8$ to $7\%$, $4\%$ and $3.7\%$ precision respectively.}
    \label{fig4:mcmc_sim}
\end{figure}

As we saw in Section~\ref{sec4:method}, the analysis pipeline for the pseudo-$C_{\ell}$ measurements of the galaxy-galaxy and galaxy-CMB lensing power spectra are complicated. As a result, we must ensure that the final outputs of our pipeline are reasonable and make sense.

We use the $1000$ correlated Gaussian mocks discussed in Section~\ref{sec4:cov} to do this. These simulations have the same baseline cosmology and galaxy bias properties (shown in Table~\ref{tab4:cov_params}) but vary in galaxy window functions and photometric redshifts. We use the mean of these $1000$ simulations as our validation mock. The posterior of the parameters of interest (Equation~\ref{eq4:posterior}) is sampled using the Markov Chain Monte Carlo sampler \textsc{emcee}. The sampled posterior is shown in Figure~\ref{fig4:mcmc_sim}; the dark-blue and the light-blue regions show the $1-\sigma$ and the $2-\sigma$ contours, respectively, while the dashed black lines and the magenta dots represent the truths. We see that the analysis pipeline can recover all the parameters of interest, with parameters $b_0$, $\Omega_{\rm CDM}$ and $\sigma_8$ having precisions of $7\%$, $4\%$ and $3.7\%$.

\section{Results}
\label{sec4:results}

We measure the galaxy-galaxy and galaxy-CMB lensing power spectra in the $50 \leq \ell \leq 400$ range with $9$ equally-spaced bins in logarithmic space. In our initial analysis, we used the Reference \cite{sfd98} (hereafter SFD) dust map as the tracer of Milky Way extinction in our foreground systematics modeling. However, during this time, we became aware of concerns that the SFD dust map might not be best suited for cosmological analyses, especially with ELGs, because the SFD map uses far-infrared emission to model Galactic extinction. However, the cosmic infrared background (CIB) is also bright in the far infrared. The CIB is a powerful tracer of high-redshift dusty galaxies such as ELGs. Hence, the SFD map may have some intrinsic signal that traces the ELG density field. 

To better understand this effect, we reran our entire analysis using four additional dust maps -- Reference \cite{csfd23} that uses the Sloan Digital Sky Survey to remove large-scale structure signals from the SFD dust map, Reference \cite{planck_dust16} thermal dust emission map, Reference \cite{mudur23} that uses stellar reddening from Pan-STARRS1 and 2MASS surveys, and Reference \cite{zhou24} that uses stellar reddening from Year-$1$ DESI spectroscopic observations of halo stars. Although there are additional Galactic extinction maps, we chose these four because of their wide-area coverage and relatively high resolution. Hence, we report all our results in the following subsections by considering the different dust maps. 

\subsection{Power Spectra Analysis}

\begin{figure*}
    \centering
    \includegraphics[width=\textwidth]{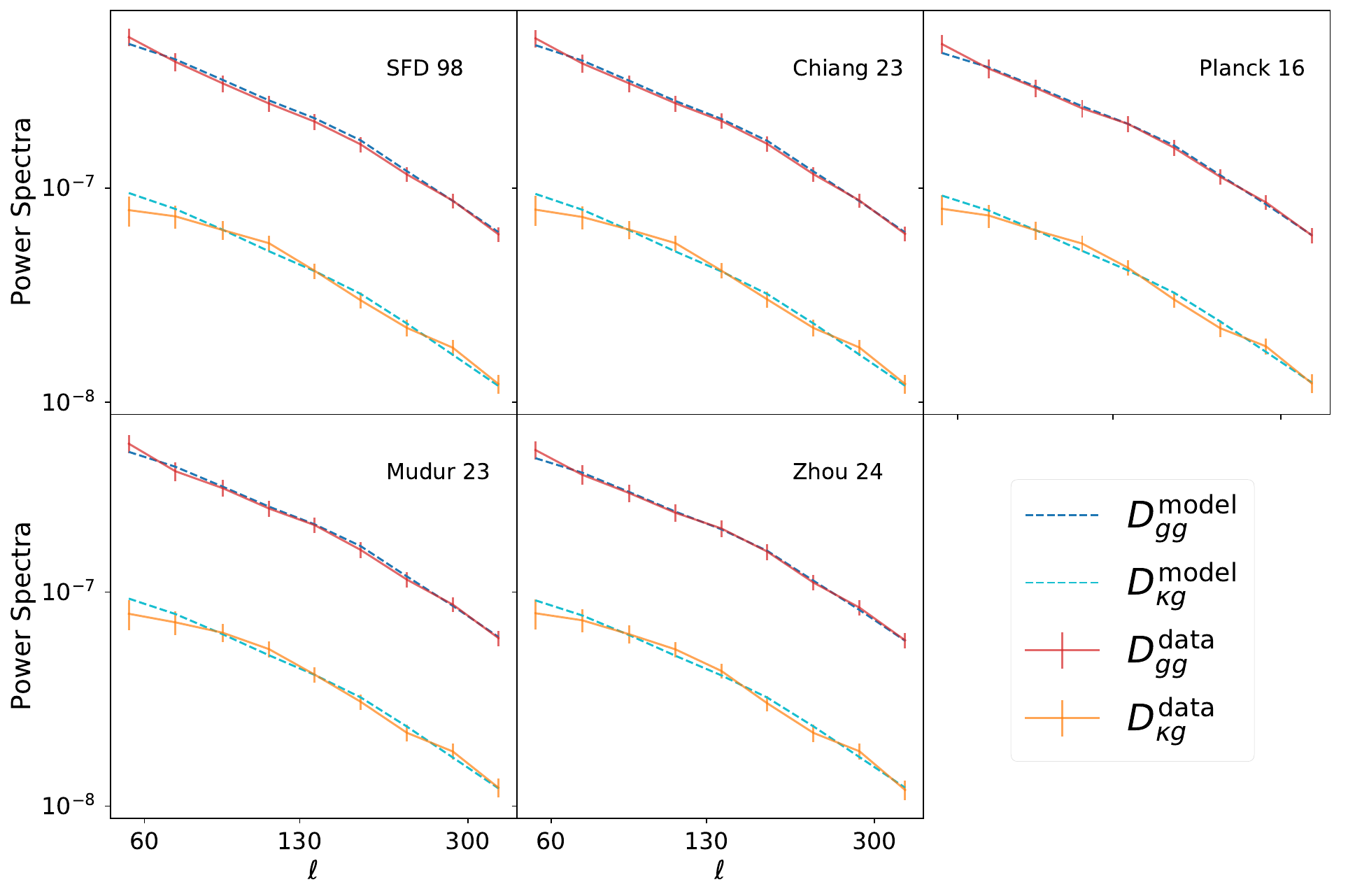}
    \caption{Galaxy-galaxy and galaxy-CMB lensing power spectra as a function of dust maps used to model the galaxy window function. The dashed lines indicate the maximum a posteriori estimate, while the solid lines show the data with error bars estimated from their corresponding covariance matrices.}
    \label{fig4:power_spectra}
\end{figure*}

\begin{figure}
    \centering
\includegraphics[width=0.75\columnwidth]{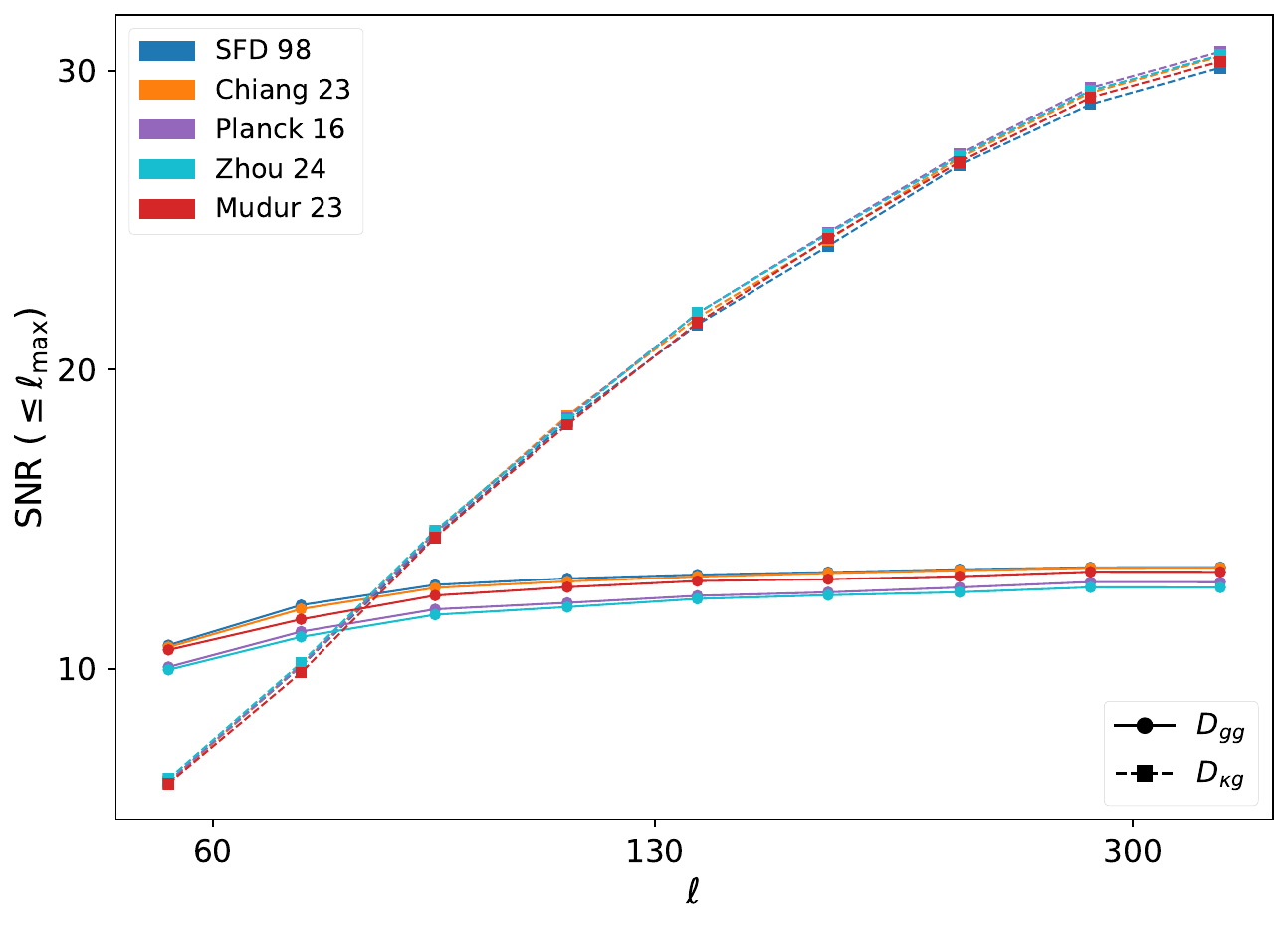}
    \caption{Cumulative signal-to-noise (SNR) of the measured power spectra. The solid lines with circle markers denote the galaxy auto-power spectra, while the dashed lines with square markers denote the cross-power spectra. We find a $30\sigma$ detection of galaxy-CMB lensing cross-correlation and a $13 \sigma$ detection of the galaxy auto power spectra. The SNR of the galaxy-galaxy power spectra is significantly degraded due to photometric redshift uncertainties.}
    \label{fig4:snr}
\end{figure}

The measured power spectra, $D^{gg}_{\ell}$ and $D^{\kappa g}_{\ell}$, are shown in with solid lines in red and orange respectively in Figure~\ref{fig4:power_spectra}, and the corresponding maximum a posteriori (MAP) estimated models are shown with dashed lines in navy and light blue. Each panel shows the MAP estimate as a function of the dust map used to estimate the galaxy window function. We note that the lowest $\ell$ bin, corresponding to the largest scale, generally has a poorer fit to the models compared to the rest of the power spectra. Section~\ref{sec:discussion} offers possible explanations of what may be happening. 

The corresponding cumulative detection significance can be seen in Figure~\ref{fig4:snr}, and we report a $13\sigma$ detection significance of ELG-ELG auto power spectra and a $30\sigma$ detection significance of ELG-CMB lensing cross power spectra. We calculated the detection significance as follows:

\begin{equation}
    {\rm SNR} (\ell_{\rm max})= \sqrt{\sum_{i = 0}^{\ell_{\rm max}} C_{\ell, 0:i} \Sigma^{-1}_{0:i, 0:i} C_{\ell, 0:i}}
\end{equation}

\noindent where $C_{\ell, 0:i}$ refer to the power spectra measurements from the $0^{\rm th}$ bin to the $i^{\rm th}$ bin, and $\Sigma^{-1}_{0:i, 0:i}$ refer to the submatrix of the original covariance matrix (Figure~\ref{fig4:cor_matrix}) running from the $0^{\rm th}$ bin to the $i^{\rm th}$ bin in each direction. 

The galaxy auto-power spectra significantly suffer due to the impact of photometric redshift uncertainty affecting the galaxy-galaxy auto-power spectra part of the covariance matrix. We confirmed this by measuring the cumulative signal-to-noise with a covariance matrix that does not have the photometric redshift uncertainty included in it; we saw that for such a covariance matrix, the galaxy-galaxy signal-to-noise can increase by an order of magnitude. Thus, our analysis shows that a robust marginalization of photometric redshift is paramount to get a high signal-to-noise measurement of the galaxy-galaxy power spectra. 

\begin{figure}
    \centering
    \includegraphics[width=\textwidth]{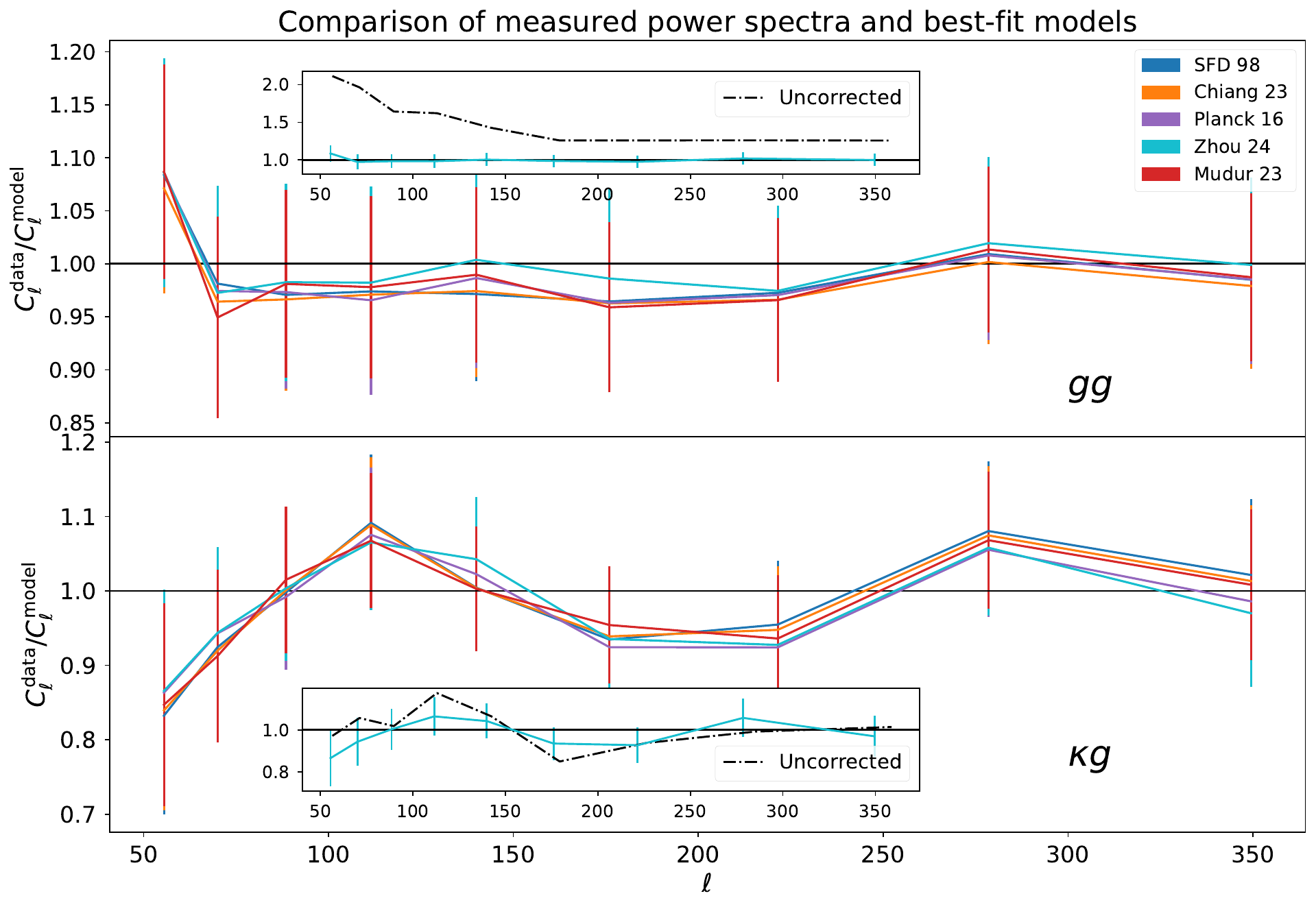}
    \caption{Comparison of galaxy-galaxy and galaxy-CMB lensing power spectra with respect to the best-fit models from Table~\ref{tab:dust-map-constraints}. These plots show that the measured power spectra agree with the best-fit model in all bins. The insets show the power spectra in the case the galaxy overdensity field was not corrected for imaging systematics and their comparison with respect to correction due to Reference \cite{zhou24}. Uncorrected $C_{gg}$ shows significantly higher power at all scales.}
    \label{fig:compare-power-spectra}
\end{figure}

We also note that for galaxy-galaxy clustering, the galaxy window function or imaging systematics correction, were significant. Figure~\ref{fig:compare-power-spectra} shows a comparison of power spectra with respect to the best-fit models for the different dust maps. It also shows in the insets what the power spectra would look like if we had not corrected for the galaxy window function; in this case, the galaxy-galaxy auto-power spectrum is significantly affected at all scales, while the galaxy-CMB lensing cross-power spectrum is unaffected. As we discuss in detail in Section~\ref{sec:discussion-systematics}, not correcting for the galaxy window function affects $C_{gg}$ most substantially. For the same $C_{\kappa g}$, if the measured $C_{gg}$ is higher than some reference value, then the inferred galaxy bias will also be higher and consequently, the inferred $\sigma_8$ will be lower.

\subsection{Parameter Inference}

\begin{figure*}
    \centering
    \includegraphics[width=\textwidth]{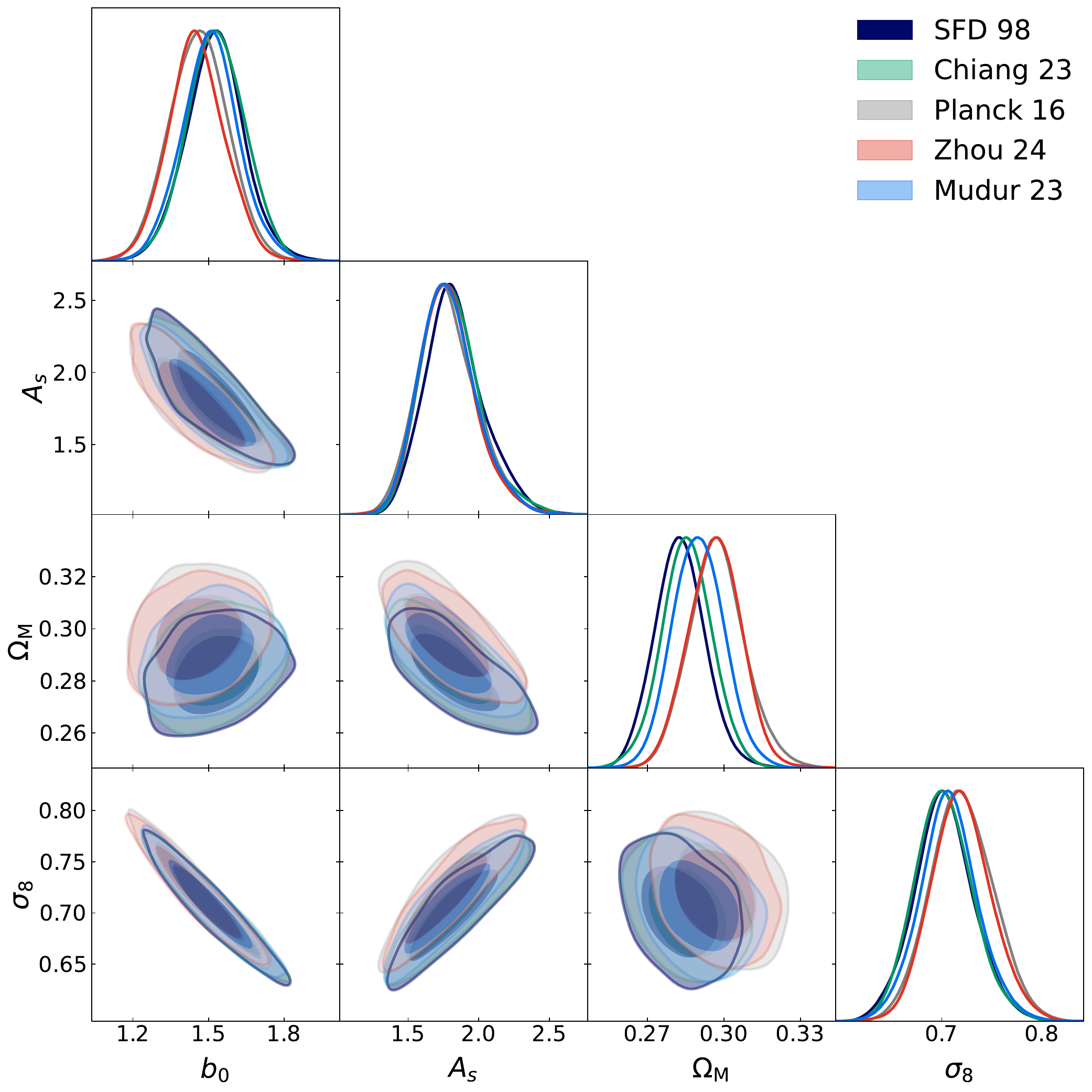}
    \caption{Posterior distribution of galaxy linear bias, $b_0$, amplitude of matter power spectrum, $A_s$ and $\sigma_8$, and matter density, $\Omega_{\rm M}$. The contours show that SFD \cite{sfd98} and Chiang 23 \cite{csfd23} prefer higher $b_0$ and lower $\sigma_8$ and $\Omega_{\rm M}$ compared to Zhou 24 \cite{zhou24} and Planck 16 \cite{planck_dust16}.}
    \label{fig4:mcmc_data}
\end{figure*}

\begin{table*}
\begin{tabular}{|c || c c c c|} 
 \hline
 Dust map & $b_0$ & $A_s$ & $\Omega_{\rm M}$ & $\sigma_8$ \\ [0.5ex] 
 \hline\hline
\\[-1em]
 SFD 98 \cite{sfd98} & $1.53 \pm 0.12$ & $1.83^{+0.17}_{-0.23}$ & $0.2828 \pm 0.0099$ & $0.702 \pm 0.030$ \\[-1em] \\ 
 \hline

\\[-1em]
 Chiang 23 \cite{csfd23} & $1.53 \pm 0.12$ & $1.80^{+0.17}_{-0.22}$ & $0.285 \pm 0.010$ & $0.702 \pm 0.029$ \\[-1em] \\ 
 \hline

 \\[-1em]
 Planck 16 \cite{planck_dust16} & $1.46 \pm 0.12$
 & $1.78^{+0.18}_{-0.23}$ & $0.297 \pm 0.011$ & $0.720 \pm 0.031$ \\[-1em] \\ 
 \hline

\\[-1em]
 Mudur 23 \cite{mudur23} & $1.51 \pm 0.12$
 & $1.78^{+0.17}_{-0.22}$ & $0.290 \pm 0.010$ & $0.707 \pm 0.030$ \\[-1em] \\ 
 \hline

\\[-1em]
 Zhou 24 \cite{zhou24} & $1.45 \pm 0.11$
 & $1.78^{+0.17}_{-0.22}$ & $0.297 \pm 0.010$ & $0.719 \pm 0.030$ \\[-1em] \\ 
 \hline
\end{tabular}
\caption{Cosmological constraints as a function of various dust maps. The errorbars correspond to $1-\sigma$ constraints.}
\label{tab:dust-map-constraints}
\end{table*}

We show the final results and the posterior distributions in Figure~\ref{fig4:mcmc_data}. The sampling of the posterior is done in the {$b_0$, $A_s$, $\Omega_{\rm CDM}$} parameter space while $\sigma_8$ and $\Omega_{\rm M}$ are derived quantities from every MCMC proposal. We derive $\Omega_{\rm M}$ by adding the sampled $\Omega_{\rm CDM}$ to the fixed value of baryon density, $\Omega_{\rm b} = 0.04845$, which is the best-fit value derived from Big Bang Nucleosynthesis measurement of $\Omega_{\rm b} h^2$; here we fixed $h$ to the value shown in Table~\ref{tab4:cov_params}. Additionally, we fix other cosmological parameters constant to the fiducial model in Table~\ref{tab4:cov_params} because the growth of structure measurements are the most sensitive to galaxy biases and a combination of $\Omega_M$ and $\sigma_8$. 

We measure the ELG galaxy bias to be in the range $1.45 - 1.53$ at $z = 1$. On the other hand, we measure cosmological parameters $\sigma_8$ and $\Omega_{\rm M}$ to be in the ranges $0.702 - 0.719$ and $0.2828 - 0.297$ respectively. The uncertainty on the inference of $\sigma_8$ is roughly $4.2\%$. 

\section{Discussions}
\label{sec:discussion}

In this section, we contextualize our findings with a few existing results in the literature and comment on how different systematics may play a role in the apparent $\sigma_8 - \Omega_{\rm M}$ tension that is persistent in the literature. 

\subsection{Comparing Result with Existing Measurements}

\begin{figure}
    \centering
    \includegraphics[width=\columnwidth]{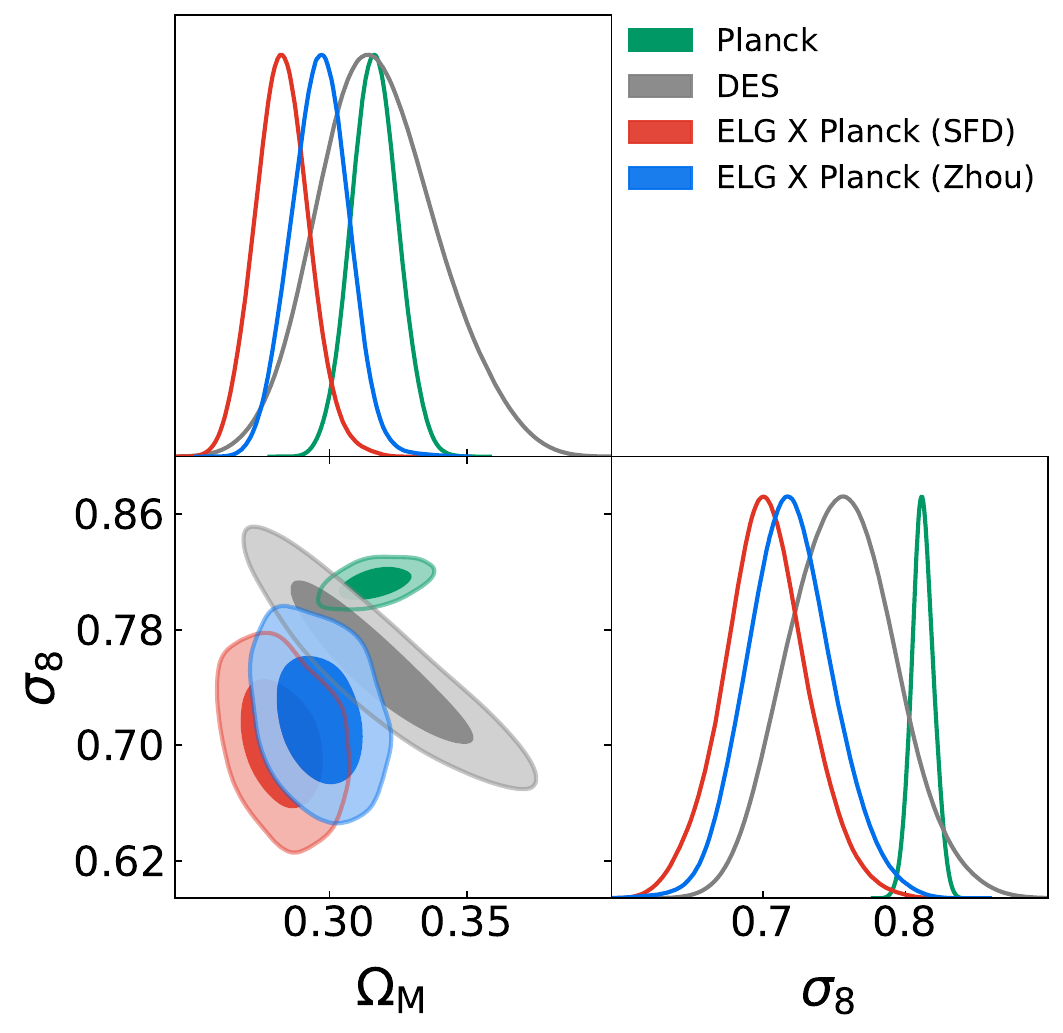}
    \caption{Comparison of DESI-like Emission Line Galaxies from Legacy Surveys DR$9$ X \emph{Planck} CMB lensing with \emph{Planck} CMB anisotropy and DES Year $3$ cosmology. A correction of the observed galaxy overdensity field based on the choice of Galactic extinction maps has a substantial impact on the inferred $\sigma_8 - \Omega_{\rm M}$ contour.}
    \label{fig4:param_compare}
\end{figure}

One of the significant points of contention in the $\sigma_8 - \Omega_{\rm M}$ tension debate has to do with the high $\sigma_8$ values that CMB experiments such as \emph{Planck} and ACT have measured in comparison to low-redshift weak lensing and galaxy clustering surveys such as the Dark Energy Survey. 

For context, the \emph{Planck} 2018 paper reports $\sigma_8 = 0.8083 \pm 0.0076$; this measurement is based on joint analysis of temperature, polarization, temperature cross polarization, and low multipole polarization data (TT, TE, EE + lowE). On the other hand, the DES Year $3$ paper reports $\sigma_8 = 0.756 \pm 0.037$; this measurement is based on a joint analysis of $3\times2$-point correlation function using cosmic shear, galaxy clustering, and cross-correlation of cosmic shear with galaxy-galaxy lensing, BAO, and supernovae. Here we use the DES result as a representative result of other low-redshift surveys. The comparison of these results is shown in Figure~\ref{fig4:param_compare}. While the three surveys closely align on the estimation of $\Omega_{\rm M}$, the picture is different for $\sigma_8$. Additionally, Figure~\ref{fig4:param_compare} includes our measurements, specifically a contour where we used the SFD dust map to model the galaxy window function and another contour where we used the \cite{zhou24} dust map to model the same. 

Two important things stand out from this plot. First, while our measurement of $\sigma_8$ is not competitive with early-Universe experiments such as \emph{Planck} ($ < 1\%$), our precision ($4.3\%$) is competitive with that of the Dark Energy Survey Year $3$ cosmology ($4.8\%$), showcasing that future analyses with better ELG photometric redshift or spectroscopic redshift will have the ability to measure $\sigma_8$ much more precisely.

Second, the choice of dust map has a noticeable effect on pulling the $\sigma_8 - \Omega_{\rm M}$ contour towards the \emph{Planck} results. Since the posterior we are comparing with \emph{Planck} primary anisotropy is two-dimensional ($\Omega_{\rm M} - \sigma_8$ plane), we can use the Mahalanobis distance \cite{mahalanobis2018generalized}, which is analogous to propagating errors for calculating differences in one-dimension: 

\begin{equation}
\label{eq:mahalanobis}
    D^2_{\rm Mahalanobis} = \left( \mu_{\rm Data,A} - \mu_{\rm Data,B} \right)^T \left( \Sigma_{\rm Data,A} + \Sigma_{\rm Data,B} \right)^{-1} \left( \mu_{\rm Data,A} - \mu_{\rm Data,B} \right)
\end{equation}

\noindent where $\mu$ refer to the central values and $\Sigma$ refer to the parameter covariances. Since the problem is multidimensional, $D_{\rm Mahalanobis}$ does not directly translate to the usual $n-\sigma$ convention used in cosmology to quantify tension. Fortunately, $D_{\rm Mahalanobis}$ follows the usual $\chi^2$ distribution, and we can use the numbers of degrees of freedom to calculate the associated $p$-value and the associated $n- \sigma$ value. Note that in this case, the $n-\sigma$ answers the question: if I assume the value to be $\mu_{\rm Data,A}$ with the covariance $\left( \Sigma_{\rm Data,A} + \Sigma_{\rm Data,B} \right)$, then how extreme is the point $\mu_{\rm Data,B}$ along the line that connects the two central values? Unlike the usual one-dimensional $n-\sigma$ that only accounts for variances $D_{\rm Mahalanobis}$ accounts for the often-ignored covariance information too.

\begin{table}
\begin{tabular}{|m{12em}|m{10em}|m{6em}|m{6em}|} 
 \hline
 Data A & Data B & $D_{\rm Mahalanobis}$ & $n - \sigma$ \\ [0.5ex] 
 \hline\hline
\\[-1em]
 ELG $\times$ \emph{Planck} CMB lensing: SFD 98 \cite{sfd98} corrected & Planck Primary Anisotropy \cite{planck2020} & $4.67$ & $4.29$ \\[-1em] \\ 
 \hline

\\[-1em]
ELG $\times$ \emph{Planck} CMB lensing: Zhou 24 \cite{zhou24} corrected & Planck Primary Anisotropy \cite{planck2020} & $3.51$ & $3.07$ \\[-1em] \\ 
 \hline

DES Year $3$ \cite{DES_all_2022} & Planck Primary Anisotropy \cite{planck2020} & $2.36$ & $1.87$ \\[-1em] \\ 
 \hline

ELG $\times$ \emph{Planck} CMB lensing: SFD 98 \cite{sfd98} corrected & DES Year $3$ \cite{DES_all_2022} & $3.41$ & $2.97$ \\[-1em] \\ 
 \hline
 
ELG $\times$ \emph{Planck} CMB lensing: Zhou 24 \cite{zhou24} corrected & DES Year $3$ \cite{DES_all_2022} & $2.14$ & $1.64$ \\[-1em] \\ 
 \hline

ELG $\times$ \emph{Planck} CMB lensing: Zhou 24 \cite{zhou24} corrected & ELG $\times$ \emph{Planck} CMB lensing: SFD 98 \cite{sfd98} corrected & $1.18$ & $0.19$ \\[-1em] \\ 
 \hline

\end{tabular}
\caption{Tension between different cosmological experiments on the $\Omega_{\rm M} - \sigma_8$ plane. The third column refers to the Mahalanobis distance measured using Equation~\ref{eq:mahalanobis}.}
\label{tab:tension-metric}
\end{table}

With this in mind, we report the discordance between parameters inferred using the SFD 98 \cite{sfd98} and the Zhou 24 \cite{zhou24} corrected ELG overdensity maps, and the \emph{Planck} primary anisotropy \cite{planck2020} and DES Year $3$ \cite{DES_all_2022} results in Table~\ref{tab:tension-metric}. We see that the choice of the Galactic extinction correction of the ELG overdensity field has a significant impact on our interpretation of whether there is a cosmological ``tension" on the $\Omega_{\rm M} - \sigma_8$ plane; SFD 98 based correction points towards a significantly higher discrepancy with \emph{Planck} primary anisotropies compared to Zhou 24. Moreover, SFD 98 based correction also shows a noticeable tension with the DES Year $3$ results, while Zhou 24 based correction measurements do not. 

Additional recent papers to compare to are the cross-correlation of the DESI luminous red galaxies photometric sample and ACT CMB lensing \cite{kim2024, sailer2024}, cross-correlation of the unWISE galaxy sample with \emph{Planck} and ACT CMB lensing \citep{Farren23}, the original unWISE cross-correlation with \emph{Planck} CMB lensing \citep{krolewski21} and the cross-correlation of the DESI-like luminous red galaxies (LRGs) sample from the Legacy Surveys with \emph{Planck} CMB lensing \citep{white22}. Philosophically, these papers use ideas similar to ours in that they cross-correlate galaxy overdensity fields with CMB lensing maps. However, we note that the pipeline for this study was developed independently of these other papers. These papers use tomography to measure $\sigma_8$ at different effective redshifts, with unWISE leaning towards a higher redshift than our sample and LRGs leaning towards a lower redshift than ours. 

The most comparable sub-samples from unWISE and LRGs are the "Green" and the \textsc{pz\_bin4} samples. The unWISE Green sample has a broad redshift kernel with a peak around $z \sim 1.1$, and the LRG \textsc{pz\_bin4} sample also has a broad redshift kernel with a peak around $z \sim 0.9$. \cite{Farren23} (ACT DR6 × unWISE + Planck PR4 × unWISE + BAO) and \cite{krolewski21} (fixed $h$) measured $\sigma_8$ using the unWISE Green sample to be $0.818 \pm 0.017$ and $.780 \pm 0.021$ respectively. \cite{white22} measured $\sigma_8$ to be $0.779 \pm 0.044$ (fixed $\Omega_{\rm M} \approx 0.27$). 

In all these examples, they prefer a much higher $\sigma_8$ compared to our ELG-based measurements, and especially Reference \cite{Farren23} finds no tension with \emph{Planck}. While the peculiarity between our results and these earlier works may be attributed to mismodeling galaxy-halo connection, our analysis is restricted to reasonably large scales ($\ell \sim 400$ implies $k_{\rm max} \sim 0.11$ at $z = 1$). The agreement between the data and the best-fit models at smaller scales is strong. 

One of our work's major limitations is that we could not do a tomographic analysis because of sizeable photometric redshift calibration uncertainties. An important step forward to check whether the tension we have found in this paper is cosmological in origin or not, is by re-analyzing the CMB lensing and DESI ELG cross-correlation using spectroscopic redshifts, which we leave for a future paper. 

\subsection{Impact of Systematics on parameter inference}
\label{sec:discussion-systematics}
\begin{table*}
    \centering
    \begin{tabular}{|m{12em}||m{6em}|m{6em}|m{10em}|}
        \hline
        \textbf{Potential Systematics} & \textbf{Impacts $C_{gg}$?} & \textbf{Impacts $C_{\kappa g}$?} & \textbf{Impact on $b_0$ and $\sigma_8$} \\ \hline \hline
        Unaccounted LSS signal in dust map & Yes & Yes & Possible non-linear impact. See Appendix~\ref{app:lss-dust} \\ \hline
        Unaccounted local dust components in dust map & Yes & No & Infer $b_0$ to be higher and $\sigma_8$ to be lower. See Appendix~\ref{app:stellar-dust} \\ \hline
        Unaccounted stellar (stream) contaminants & Yes & No & Infer $b_0$ to be higher and $\sigma_8$ to be lower. See Appendix~\ref{app:stellar-dust} \\ \hline
        Not considering the full distribution of imaging systematics or the galaxy window function & Yes & No & Infer $b_0$ to be higher and $\sigma_8$ to be lower. See \cite{Karim23} \\ \hline
        CMB Lensing Monte Carlo Norm Correction & No & Yes & Infer $b_0$ to be lower and $\sigma_8$ to be higher. See Appendix~\ref{app:unaccounted-mc-norm} \\ \hline
        Mismodeling of magnification bias & Yes & Yes & Possible non-linear impact. See Appendix~\ref{app:mag-bias} \\ \hline
    \end{tabular}
    \caption{Summary of the impact of different systematics on the inference of galaxy bias and $\sigma_8$.}
\label{tab:systematics}
\end{table*}

As we saw in Figure~\ref{fig4:param_compare}, the choice of dust maps to modulate the galaxy window function has a significant impact on the inference of $\sigma_8 - \Omega_{\rm M}$. Our careful studies of various systematics also indicate that different systematics can affect the observed galaxy-galaxy and galaxy-CMB lensing power spectra in specific ways, consequently affecting the inferred values of $b_0$ and $\sigma_8$. 

To first order, we can use Equations~\ref{eq:cgg} and \ref{eq:ckg} to understand how getting one (or both) of the power spectra modeling wrong affects the galaxy bias and $\sigma_8$. At least $3$ out of the $6$ systematics summarized in Table~\ref{tab:systematics} show that they can lead to a lower inferred value of $\sigma_8$. We show derivations of these effects in Appendix~\ref{app:unaccounted}. 

As cosmological surveys get more precise, previously unaccounted-for effects are becoming the dominant source of modeling mismatch, making the determination of cosmological parameters more challenging. First, observational systematics are best modeled using data-driven approaches. As these foreground systematics maps are generated using the maximum likelihood estimator approach, it has become crucial to marginalize over their probability distribution rather than considering only one value. Equally importantly, understanding the mutual information between the foreground systematics map made using different methods or datasets is also becoming necessary for uncertainty quantification. For example, both \cite{csfd23} and \cite{zhou24} are efforts to improve upon the SFD map, but they lead to different estimates of $\Omega_{\rm M}$. Additionally, specific contaminants such as stellar streams are poorly constrained because our best model of them is limited by the number of stars associated with them. Hence, understanding these local, inter-Milky Way contaminants will be significant when we try to constrain cosmological models with galaxies at the flux limit of the telescopes. 

Although we do not have quantitative evidence, Figure~\ref{fig4:power_spectra} hints that the lowest $\ell$ bin has the highest discrepancy with the best-fit models. Imaging systematics such as stellar streams and dust maps are typically large-scale, resulting in steeper observed power spectra at low $\ell$ values. Thus, we will need more comprehensive and robust foreground imaging systematics maps in the future if we want to use information from large scales. We also note that in the pseudo-$C_{\ell}$ framework, the worry is not restricted only to large scales; due to mode-mode coupling signal from large scales, it can also affect small scales.

Second, while many of these systematics have been treated independently of each other in the past, there is now a more pressing need for joint simulations to model them together. As discussed in Appendix~\ref{app:mc-norm}, we found non-zero covariances between the CMB lensing Monte Carlo norm correction factor and the galaxy window function. Although we could avoid modeling the covariance terms exactly since the errors are subdominant to photometric redshift uncertainties, any future survey with smaller photometric redshift uncertainties or spectroscopic redshifts may eventually need to account for this effect. Thus, we encourage the field to consider building joint correlated simulations of surveys to facilitate accurate cross-survey analysis.

\section{Conclusions}
\label{sec4:conc}

In this paper, we have presented the first-ever cross-correlation analysis of CMB lensing with Emission-Line Galaxies to measure $\sigma_8$, $\Omega_{\rm M}$ and the ELG linear bias, $b_0$. We find that the choice of Galactic Extinction map has a noticeable effect on pulling the $\sigma_8 - \Omega_{\rm M}$ contour towards the Planck results. Our analysis based on ELGs from the DESI Legacy Imaging Surveys and \emph{Planck} CMB lensing showcase that ELG-type tracers will allow us to probe cosmological parameters in the high-redshift Universe with high precision. Technological and survey innovations such as the Legacy Surveys, DESI, and \emph{Planck} have finally made this approach of probing the growth of structure viable. 

This paper provides some important methodological developments in dealing with various sources of systematics that affect these kinds of cross-correlation analyses, namely:

\begin{itemize}
    \item We developed a fully forward-model pipeline that can use generative models of systematics to infer cosmological parameters using \textsc{SkyLens}.
    \item We investigated various data-level systematics to obtain the cleanest ELG sample consisting of $22$ million galaxies.
    \item We showcased a thorough analysis of ELG photometric redshift and their uncertainties using deep DESI Survey Validation data.
    \item We showed that the Sagittarius Stream is a significant contaminant source in photometric ELG sample and we correct for its effect by using it as a template to model the galaxy window function. 
    \item We derived a simulation-based covariance matrix that considers various systematics, including marginalizing over the galaxy window function.
    \item We devised a method to jointly estimate the covariance between the Monte Carlo norm correction systematics and the galaxy window functions. 
    \item We showcased how $b_0$, $\Omega_{\rm M}$, and $\sigma_8$ are susceptible to the choice of dust maps used to estimate the galaxy window function.
\end{itemize}

The outcome of these advancements allowed us to measure $\sigma_8 - \Omega_M$ competitively with other major weak lensing cross-correlation surveys such as DES. Our key results are:

\begin{itemize}
    \item The posterior estimate of the $\Omega_{\rm M} - \sigma_8$ plane based on SFD correction indicates a $4.29 \sigma$ tension with \emph{Planck}, which is reduced to $3.07 \sigma$ when we used the DESI stellar-reddening based correction. However, even using these maps results in $>3 \sigma$ tension with \emph{Planck}, indicating that while dust can somewhat alleviate $\sigma_8$ tension, it cannot do so completely. 
    \item The MAP estimate of ELG linear bias, $b_0$, based on the non-SFD dust maps, especially \cite{zhou24} and \cite{planck_dust16}, is consistent with values found in literature, that typically anticipates ELG linear bias to be $1.4$ at $z = 1$ \citep{fdr016, merson2019, tamone2020, zhai2021}. 
    \item The detection significance of the galaxy-galaxy power spectrum was severely affected by photometric redshift uncertainties. 
\end{itemize}

While we tried our best to account for various systematics and uncertainties, our experience has also taught us about things that can be improved upon in the future and can help ELG-based photometric surveys with their analysis design. The main limitations of our work are as follows:

\begin{itemize}
    \item We were ultimately affected the most by photometric redshift uncertainties in our modeling. The DESI Legacy Imaging Surveys have only three broadband filters, so we could do little to improve this. Additionally, we could not account the redshift distribution of about $15\%$ of our sample which led to us using a long-tail redshift distribution. Our work shows the need for spectroscopic analysis for galaxies with a limited number of identifiable lines in the observed frame wavelength. It also shows that narrow-band photometry is paramount to reducing the impact of photometric redshift uncertainty in cosmological analyses.
    \item We also have used a relatively simple galaxy linear bias model for ELGs as we restricted our analysis to relatively large scales. However, recent studies that use more massive galaxies such as luminous red galaxies have shown that one can use Lagrangian Perturbation Theory and the Effective Field Theory of the Large-Scale Structures to extract information from much smaller scales \citep{sailer2024, Farren23, chen2022, white22}. However, initial data from DESI has shown that the small-scale clustering of ELGs are affected by conformity bias and a higher satellite velocity dispersion \cite{rocher2023}. Hence, extending either the Perturbation Theory or the Halo-Occupation Distribution approaches to DESI ELGs will require careful modeling of small-scale clustering, especially if the goal is to use next-generation CMB lensing maps such as the ACT CMB lensing map. We leave this task for a future paper.
    
    \item Computational resources limited our estimation of the window function posterior. On the computation side, it took us a week to sample $1000$ unique window realizations from the neural networks. We would like to use many more realizations to reduce the noise due to a finite number of realizations. Additionally, sampling the posterior of a neural network is still an active area of research, and many different methods are in use. As these methods improve, the galaxy window functions' modeling quality will also improve.
    \item We used Gaussian simulations to estimate the covariance matrix. Although our method is robust because it considers effects such as photometric redshift and window function uncertainties, it does not account for non-Gaussian contributions to the covariance due to the nature of our baseline simulation. A similar covariance estimation based on $n$-body simulations will capture non-Gaussian components we currently fail to capture. 
    \item We used five different dust maps to get cosmological parameter estimates, but we did not make any scientific judgments on which dust map is the ``best" one. 
\end{itemize}

Ultimately, the ELGs have opened up a new era of cosmological analysis. With ongoing and future ELG-based surveys such as DESI, \emph{Euclid} \cite{euclid2024}, \emph{Roman} \cite{roman2021}, SPHEREx \cite{spherex2018} and the Rubin Observatory \cite{lsst2019}, as well as CMB lensing surveys such as the Simons Observatory \cite{simons2019} and CMB-S4 \cite{cmbs42019}, it is only a matter of time before we begin to explore the growth of structure in multiple high redshift epochs tomographically. 

\appendix
\section{Functional Form of the Exponential Decay Function}
\label{app4:exp_decay}

In this section, we will derive the functional form of an exponential decay function whose first point and the overall area under the curve are given. 

Let ${z_r, P_0}$ be the first point (correspond to the midpoint bin value and the histogram value of the last histogram of the redshift distribution curve obtained using Random Forest Classifier), and $r$ be the area under the exponential decay distribution. Thus, we can write the usual form of exponential decay distribution as: 

\begin{equation*}
    P(z) = P_0 e^{-\lambda(z - z_r)}
\end{equation*}

We need to solve for the parameter $\lambda$. If $r$ is the area under the curve of $P(z)$, then we can write the following:

\begin{align}
    \int_{z_r}^{\infty} P(z) dz &= r \nonumber \\
    \int_{z_r}^{\infty} P_0 e^{-\lambda(z - z_r)} dz &= r \label{eq3:exp}
\end{align}

We will now apply a change of variable such that $z' = z - z_r$. This changes the Equation~\ref{eq3:exp} to:

\begin{align*}
    P_0 \int_{0}^{\infty}  e^{-\lambda z'} dz' &= r \\
    P_0 \frac{1}{\lambda} &= r \\
    \lambda &= \frac{P_0}{r}
\end{align*}

Thus, if the last bin value and the area under the exponential curve are known, we can write the functional form of the exponential decay distribution as:

\begin{equation}
    P(z) = P_0 e^{-\frac{P_0}{r} (z - z_r)}
\end{equation}

\section{Stellar and Foreground Galaxy Masks}
\label{app4:stellar_gal_masks}


Large-scale structure surveys have to use masks to block out areas around stars and bright foreground galaxies because they can introduce artificial signals in the power spectrum measurement. Bright foreground objects can preferentially block out fainter background objects, distorting the galaxy number count and number density. As a result, a certain amount of area around these objects is completely masked out when preparing large-scale structure catalogs. A similar step was taken with the DR9 Legacy Surveys ELG catalog when identifying potential DESI targets based on galaxy photometry. 


However, when we ran our regression analysis to model the galaxy window function, we found that specific outlier pixels confused the regression model because there were an unusually large number of ELGs in these pixels. We isolated some of the most extreme pixels and visually inspected these fields using the pixel right ascension and declination. We used the Legacy Survey Viewer\footnote{https://www.legacysurvey.org/viewer}, which is a website built by the Legacy Surveys team containing imaging of the entire Legacy Surveys footprint and also includes value-added catalogs such as all the DESI-like ELG targets. 


\begin{figure*}
\centering
\subfloat[Star: *alf Leo, and Galaxy: Z 64-73, RA = $152.0889$, Dec = $11.9671$]{
  \includegraphics[width=65mm]{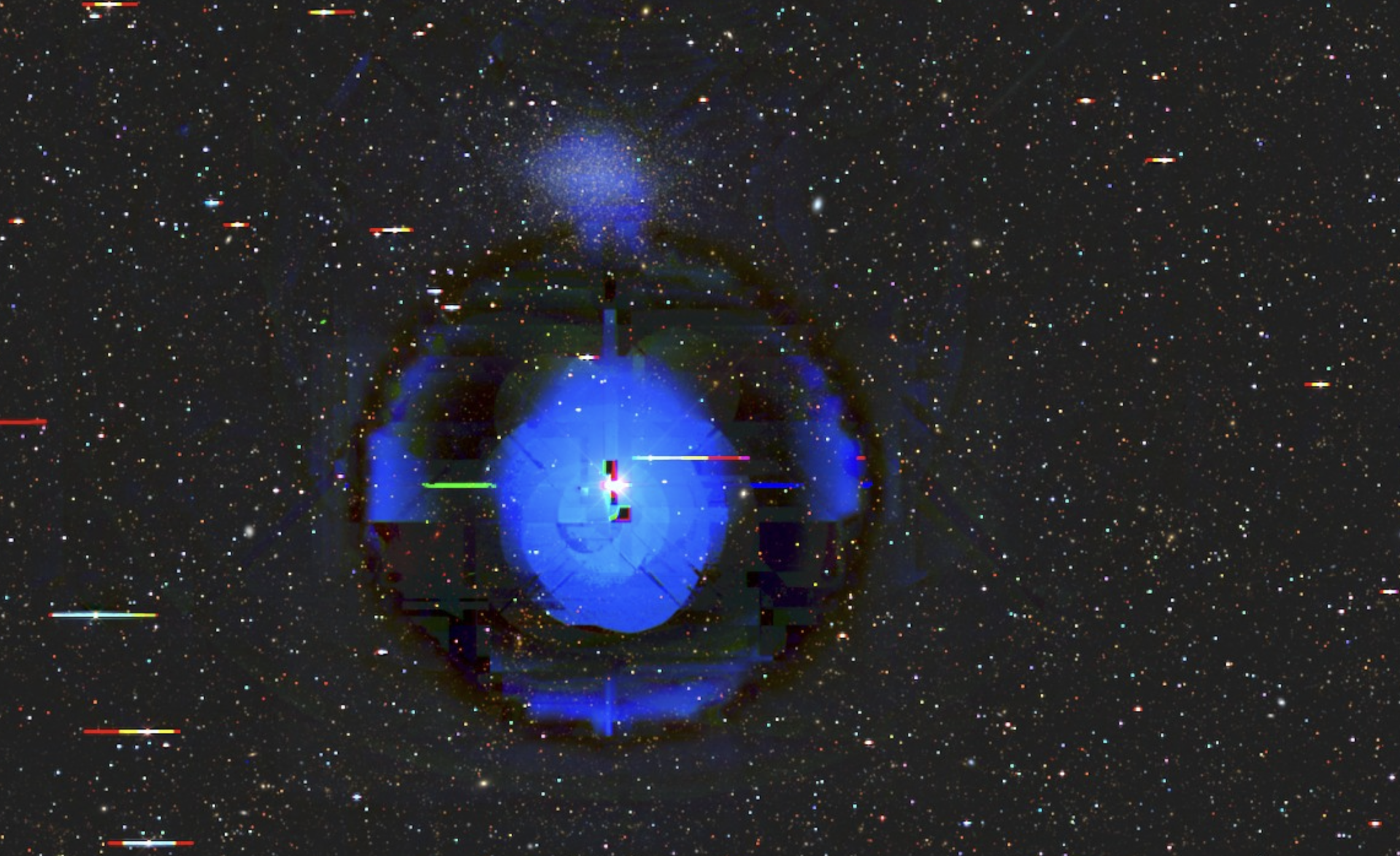}
}
\subfloat{
  \includegraphics[width=65mm]{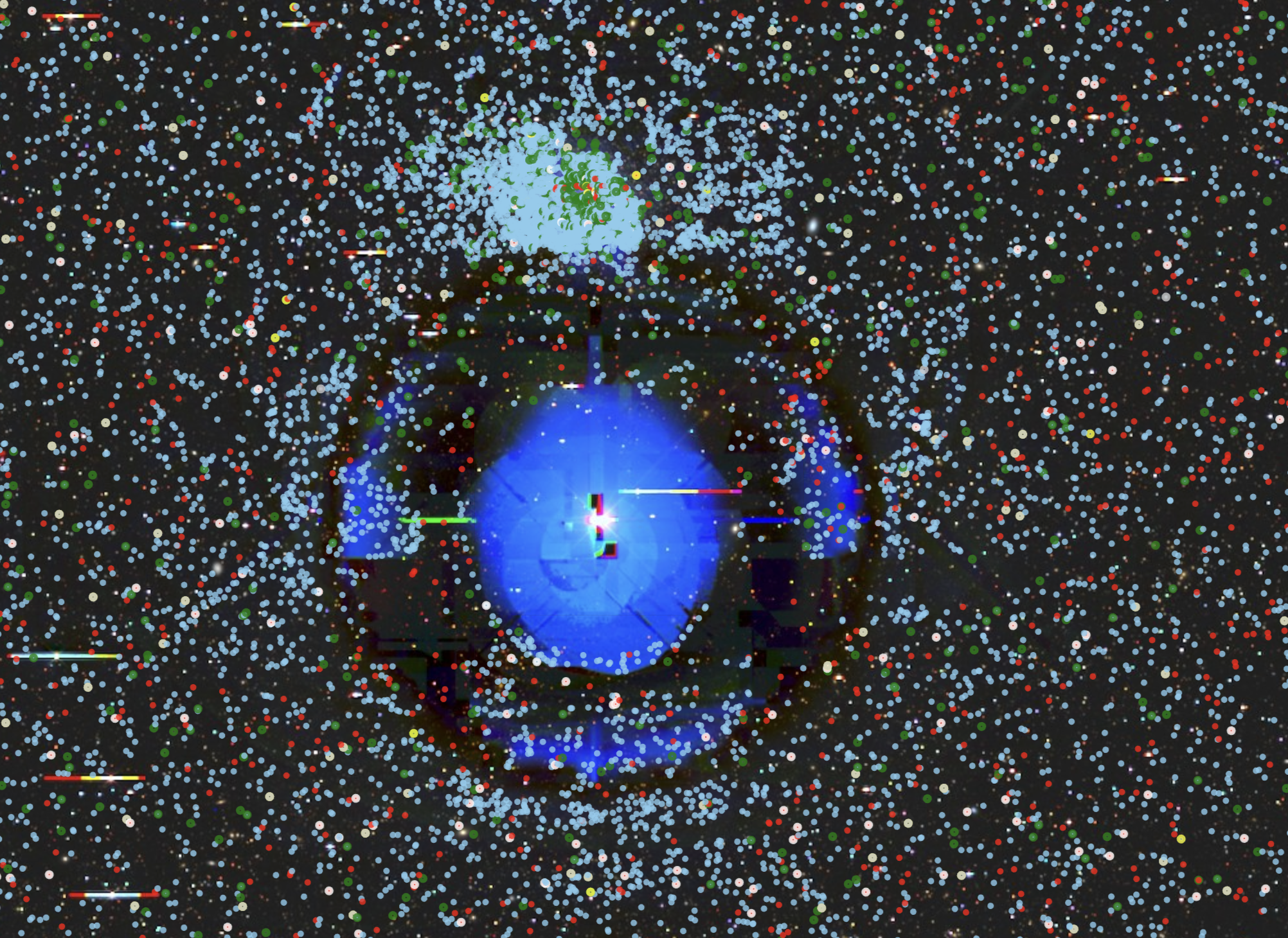}
}
\hspace{0mm}
\subfloat[Galaxy: NGC 2403, RA = $114.2180$, Dec = $65.6022$]{
  \includegraphics[width=65mm]{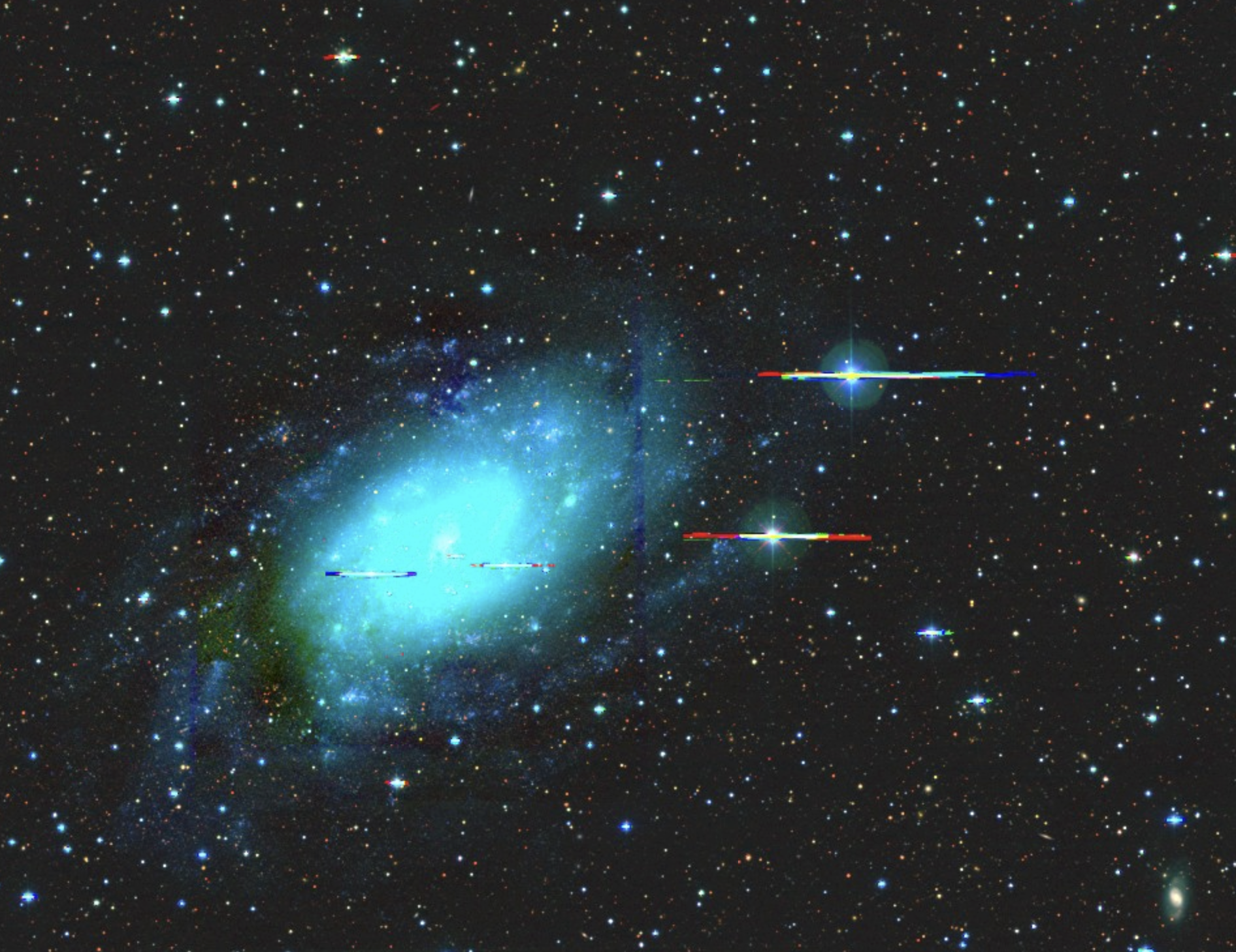}
}
\subfloat{
  \includegraphics[width=65mm]{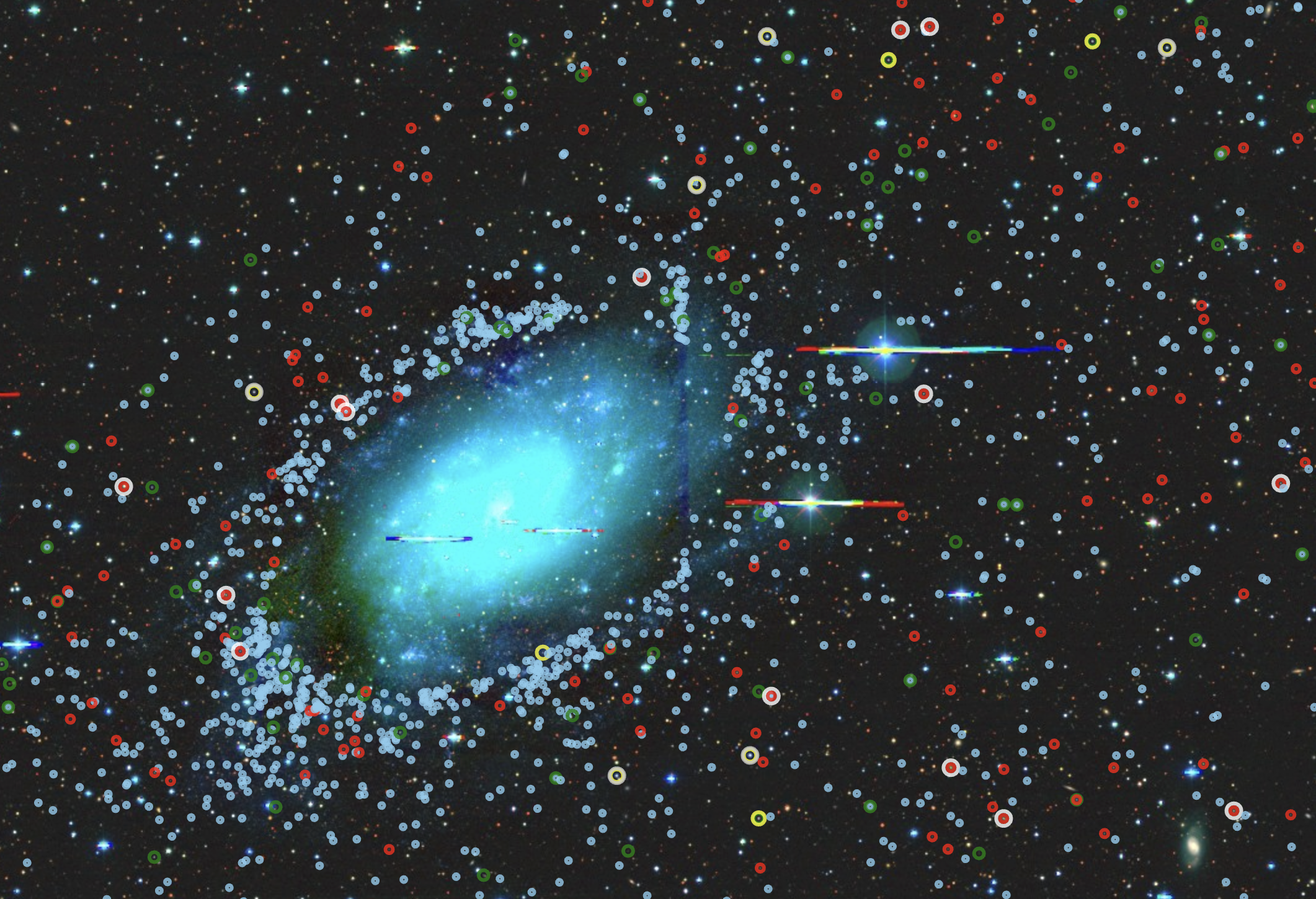}
}
\caption{Legacy Surveys Viewer snapshot of the bright star *alf Leo (top panel) and the bright galaxy NGC 2403 (bottom panel). The left column shows the field in the Legacy Surveys imaging, and the right shows DESI ELG targets (in blue circles) in the same field. The stray lights inside the diffraction arc of *alf Leo, the nearby galaxy on its top, Z 64-73, and the arms of the galaxy NGC 2403 all exhibit similar photometric properties as ELG targets, thereby artificially increasing the number of ELGs.}
\label{fig4:stellar_galactic_examples}
\end{figure*}

Figure~\ref{fig4:stellar_galactic_examples} shows two such examples. Here, we show a bright star, *alf Leo, and an overlapping diffuse galaxy, Z 64-73, in the top panel and a bright foreground galaxy, NGC 2403, in the bottom. Ideally, the DESI target catalog would have masked out these objects properly. However, the right column in the figure clearly shows ELG targets (shown in blue circles) inside the diffraction arc of the star. The diffuse light of both galaxies mimics ELG targets, exemplifying a case where the large-scale structure catalog is affected by artificial ELGs, leading us to investigate this issue statistically and implement more conservative stellar and foreground galaxy masks. 


\begin{figure}
  \centering
  \subfloat[a][ELG fractional overdensity before stellar mask cut]{\includegraphics[width=\columnwidth]{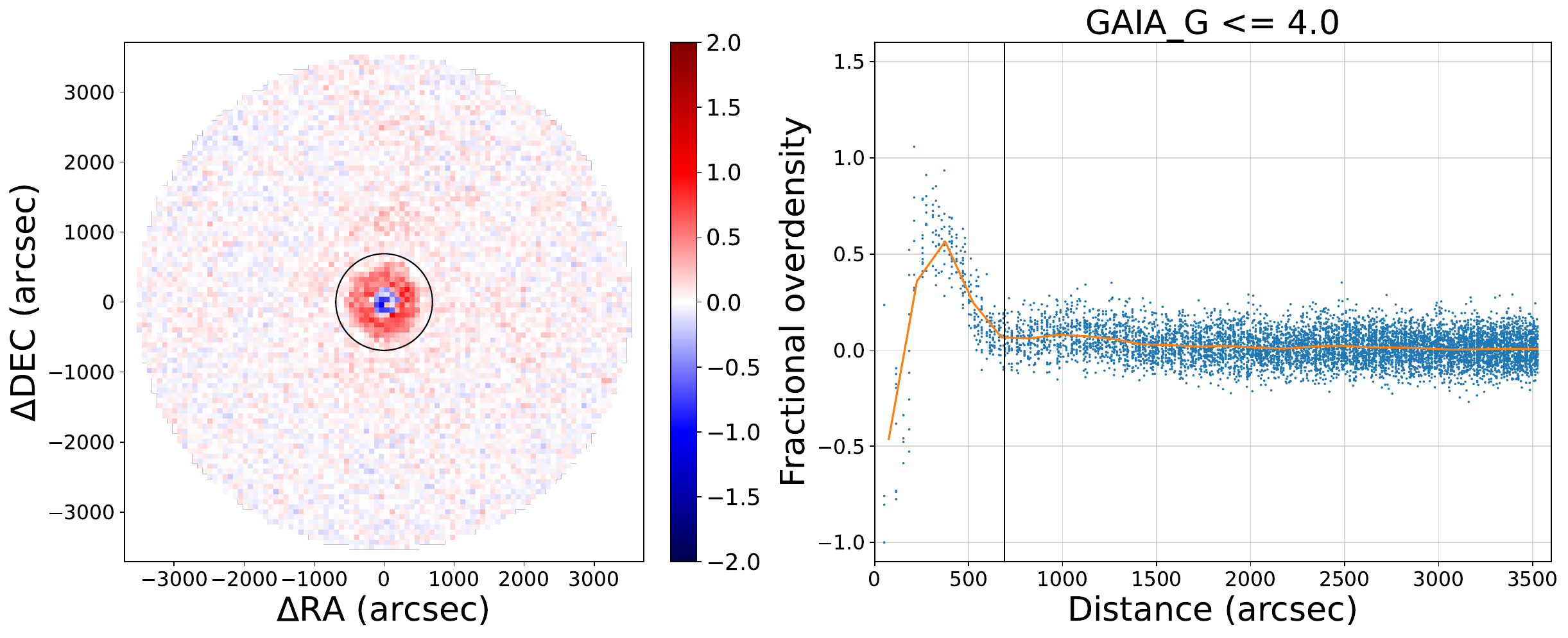} \label{fig:a}} \\
  \subfloat[b][ELG fractional overdensity after stellar mask cut]{\includegraphics[width=\columnwidth]{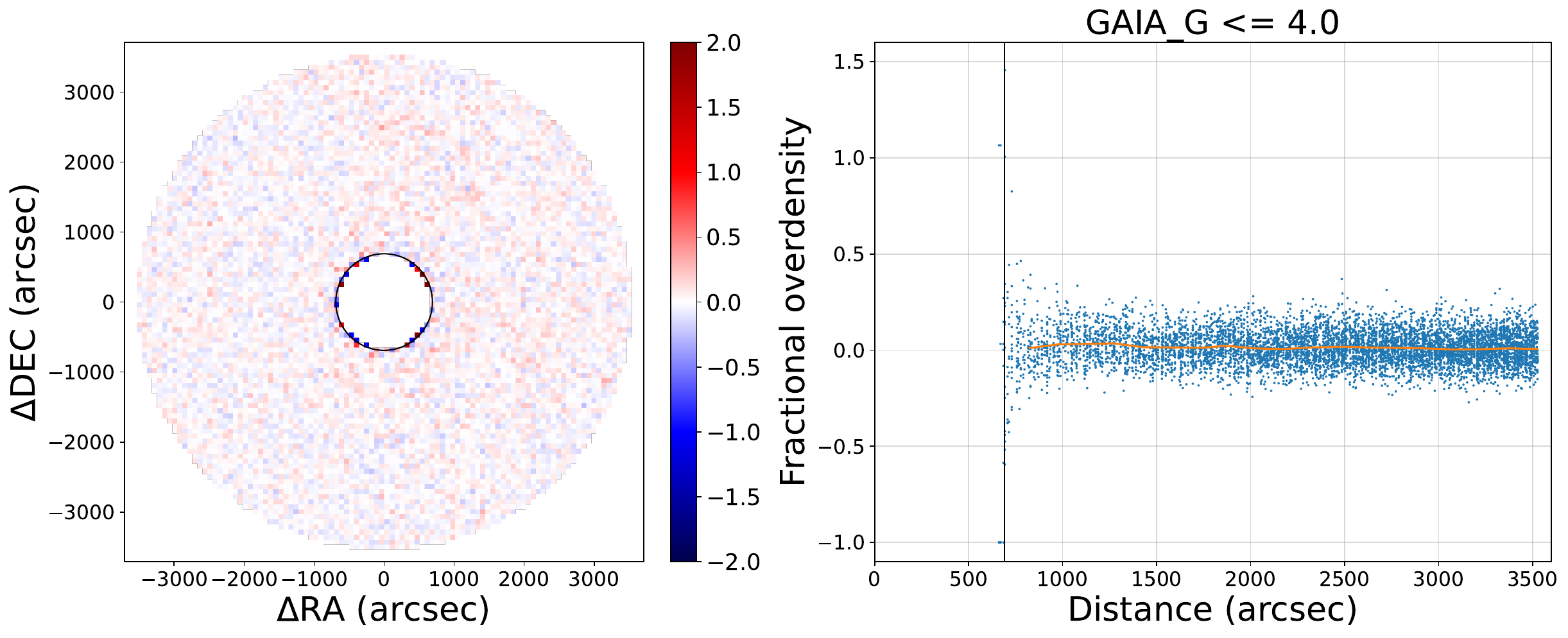} \label{fig:b}}
  \caption{ELG overdensity field around bright \emph{Gaia} stars with magnitudes $G < 4$. We see that the ELG overdensity is affected out to about $600^{\prime \prime}$ in radius. The black line shows the cut we implemented for the new mask since the overdensity field does not fluctuate outside of this line. } \label{fig4:donut}
\end{figure}

We used the \emph{Gaia} \citep{gaia2021} and the Siena Galaxy Atlas\footnote{https://www.legacysurvey.org/sga/sga2020/} \cite{moustakas2023} to identify stars and foreground galaxies in our footprint. We built the new stellar and galactic masks separately, but the analysis procedure was the same. To build the new masks, we binned the objects by their $g-$band magnitude, and the width of the bin was $1$ magnitude. We then centered all of these binned objects and stacked them together. Afterward, we picked radii of $80 - 3000$ arcsec, depending on the brightness of the bin. We then calculate the fractional overdensity of ELG targets inside this radius. The expectation is that the ELG overdensity field around foreground objects should be a function of radius. If we detect a radial dependency, we increase the mask size for that bin until we do not see any radial dependency outside of the mask. Figure~\ref{fig4:donut} shows an example of the brightest stars ($G < 4$) from the \emph{Gaia} catalog. It shows that we must apply to masks as large as $600^{\prime \prime}$ in radius to remove such artificial number density fluctuations.

\begin{figure}
\centering
\subfloat{
  \includegraphics[width=65mm]{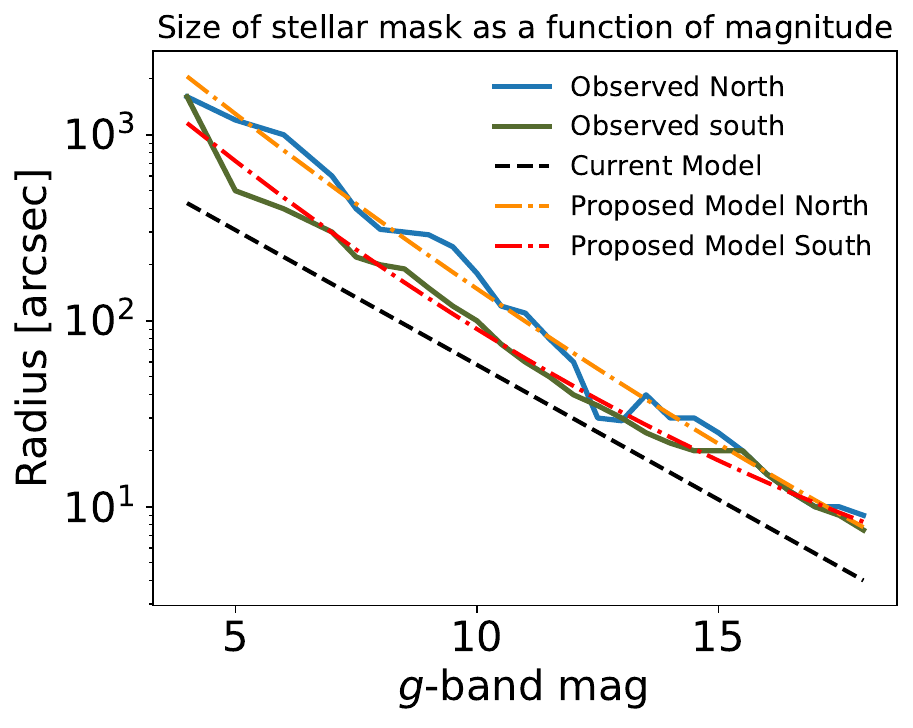}
}
\hspace{0mm}
\subfloat{
  \includegraphics[width=65mm]{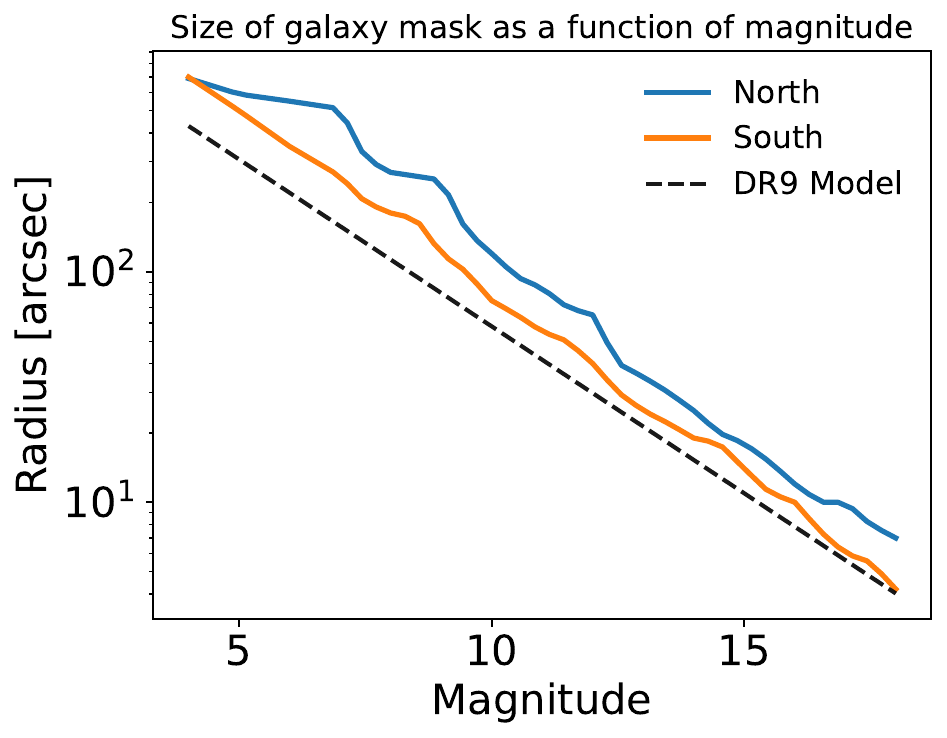}
}
\caption{Size of mask versus $g$-band magnitude of \emph{Gaia} (top panel) and Siena Galaxy Atlas (bottom panel). These plots show that the existing DR9 Legacy Surveys mask size needs to be improved for large-scale structure analysis. The other curves show our models.}
\label{fig4:mask_new}
\end{figure}

We show the summary of this exercise in Figure~\ref{fig4:mask_new}, which shows the relationship between mask size and the magnitude of the stellar and foreground galaxy contaminants. Notably, the DR9 mask size (in the black dashed line) is inadequate for large-scale structure analysis. The resulting radius vs. g-band mag curves fit the data better, and we use these models to construct our survey mask. 

\section{Uncertainty of the Multiplicative Bias Estimator due to \emph{Planck} CMB Lensing}
\label{app:mc-norm}

As explained in Section~\ref{sec:mc-norm}, the \emph{Planck} Collaboration (also other experiments such as the Atacama Cosmology Telescope \citep{Farren23}) make some assumptions about the CMB lensing noise properties which makes cross-power spectrum unbiased as long as the galaxy mask is the same as the CMB lensing mask. However, this is not true for almost all cases, including ours, so we must estimate the correction factor by considering a new galaxy mask. Because the noise property that the lensing reconstruction is optimized for is a function of sky position, we change the optimality of the default \emph{Planck} lensing reconstruction by masking additional parts of the sky. 

Suppose we know our CMB lensing and galaxy masks. In that case, we can use the $300$ pairs \emph{Planck} CMB lensing simulations where each pair consists of one simulation with a pure CMB lensing convergence signal, $\delta_{\kappa}$, and one simulation of the associated reconstruction of the lensing signal by taking into account various noise properties, denoted as $\delta_{\hat{\kappa}}$. The correction factor, $A_{\ell}$ is then given by:

\begin{equation}
    A_{\ell} = \frac{C^{\kappa_{\kappa \rm{-mask}} \kappa_{g \rm{-mask}}}_{\ell}}{C^{\hat{\kappa}_{\kappa \rm{-mask}} \kappa_{g \rm{-mask}}}_{\ell}}
\end{equation}

\noindent Since the actual \emph{Planck} CMB lensing map is a lensing reconstruction, we have to multiply the correction factor, $A_{\ell}$, to get an estimation of what the true cross-power spectra, $C_{\kappa g}$, ought to be. Note that both the numerator and the denominator are estimators since the $\delta_{\kappa}$ field and the associated reconstructed $\delta_{\hat{\kappa}}$ field are estimated using the $300$ simulations. As such, $A_{\ell}$ is a ratio estimator, and estimating $A_{\ell}$ by taking ratios over all $300$ simulations and then taking the mean has a high variance. Instead, we choose to estimate $A_{\ell}$ by taking the ratio of the means of all the simulations, i.e.,

\begin{equation}
\label{eq:mc-norm-avg}
    A_{\ell} = \frac{\sum_{i = 1}^{300} C^{\kappa_{\kappa \rm{-mask}} \kappa_{g \rm{-mask}}}_{\ell}}{\sum_{i = 1}^{300} C^{\hat{\kappa}_{\kappa \rm{-mask}} \kappa_{g \rm{-mask}}}_{\ell}} 
\end{equation} 

However, beyond this issue, the subtlety in our paper is that we treat the galaxy window function, $W_g$, or the imaging weights as the effective mask such that our galaxy mask is no longer binary but has a continuous value in the range $[0, \infty)$. In Section~\ref{sec:galwin} and in our companion paper \cite{Karim23}, we show how the estimation of $W_g$ can have biases and the uncertainty in estimating $W_g$ must be marginalized over to get the proper covariance matrix. As a result, the numerators and the denominators in Equation~\ref{eq:mc-norm-avg}, and consequently $A_{\ell}$, vary not only due to the variance in $\delta_{\kappa}$ and $\delta_{\hat{\kappa}}$, but also due to the variance in the galaxy mask. Thus, we have to properly account for the covariance between realizations of $\delta_{\kappa}$ and the galaxy window function to understand the noise properties of the $A_{\ell}$ estimator. 

\begin{figure}
    \centering
    \includegraphics[width=0.75\columnwidth]{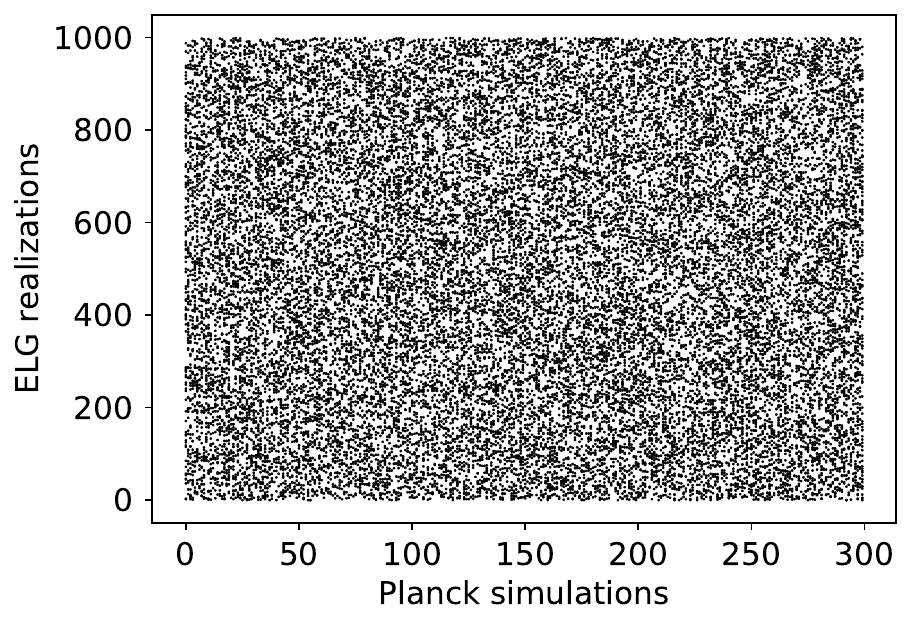}
    \caption{Coverage of \emph{Planck} CMB lensing simulations and ELG window function realizations. The plot shows that the bootstrapping points cover the ranges well.}
    \label{fig:sim-coverage}
\end{figure}

To explore this problem, we consider $100$ bootstrapped realizations of $A_{\ell}$; in each bootstrap realization, we randomly sample with replacements $300$ \emph{Planck} $\delta_{\kappa}$ and $\delta_{\hat{\kappa}}$ simulations with replacement as well as $300$ realizations of the galaxy window function, $W_g$. We then use Equation~\ref{eq:mc-norm-avg} to estimate $A_{\ell}$. The $100$ bootstrapped realizations have enough coverage along both the \emph{Planck} simulations of the ELG window function realizations to yield a robust answer, as shown in Figure~\ref{fig:sim-coverage}.

\begin{figure}
    \centering
    \includegraphics[width=0.75\columnwidth]{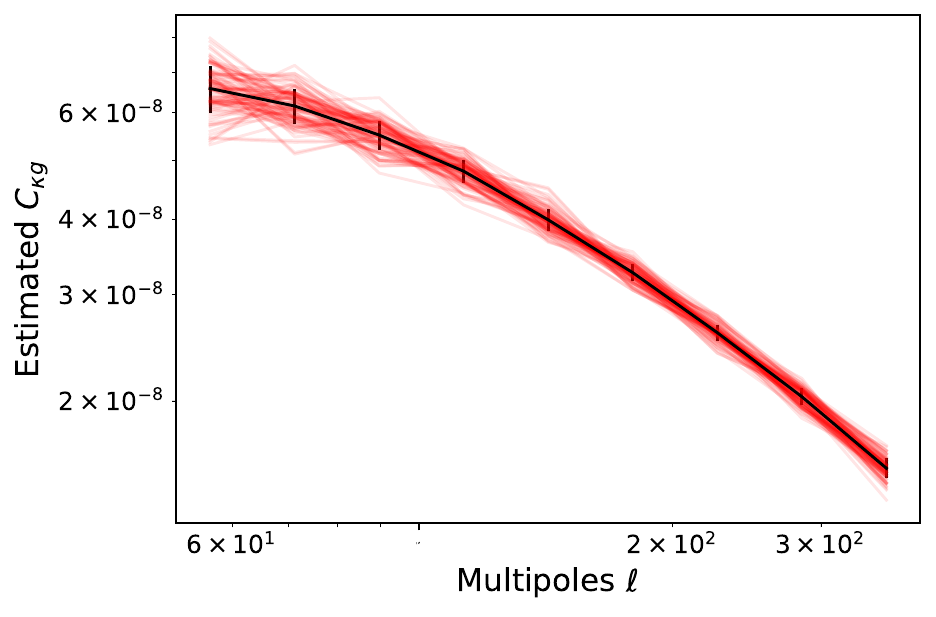}
    \caption{Distribution of the corrected $C_{\kappa g}$ estimator based on the Monte Carlo multiplicative biased discussed in Section~\ref{sec:mc-norm}. This plot shows $100$ samples based on the bootstrapping resampling technique.}
    \label{fig:ckg-dist}
\end{figure}

\begin{figure}
    \centering
    \includegraphics[width=0.75\columnwidth]{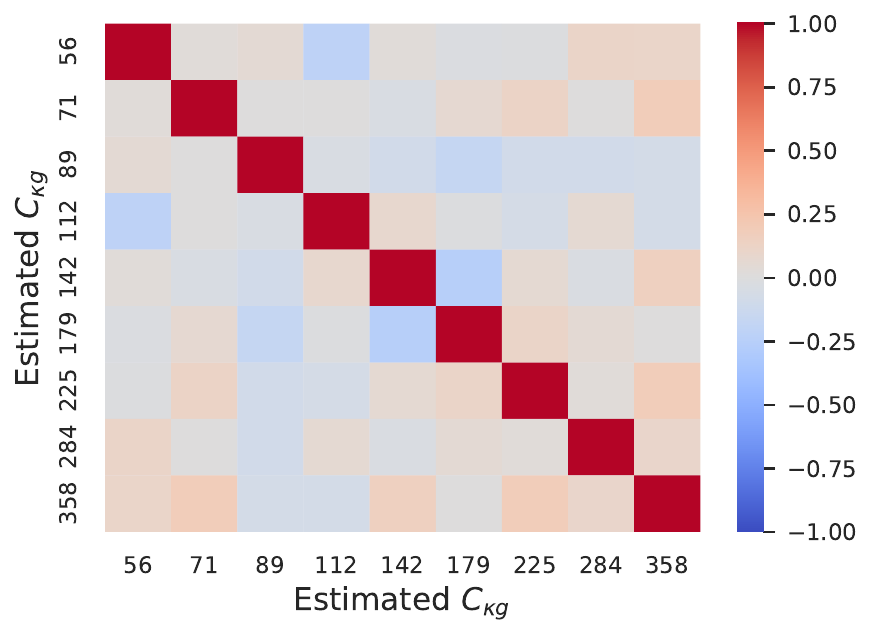}
    \caption{Covariance matrix of the corrected $C_{\kappa g}$ estimator based on the Monte Carlo multiplicative biased discussed in Section~\ref{sec:mc-norm}. This plot shows non-negligible off-diagonal terms, especially at high $\ell$.}
    \label{fig:ckg-cov-mc-norm}
\end{figure}

Finally, we need to convert the estimations of $A_{\ell}$ to estimations of $C_{\kappa g}$ to get a sense of how varying the \emph{Planck} CMB lensing simulations and the ELG galaxy window functions jointly affect the $C_{\kappa g}$ estimator. For each $A_{\ell}$ realization, we randomly select one cross-power spectrum between the reconstructed lensing field, $\delta_{\hat{\kappa}}$ and the galaxy window function and estimate the corrected $C_{\kappa g}$. Figure~\ref{fig:ckg-dist} shows the distribution of the estimated $C_{\kappa g}$, and the corresponding covariance matrix is shown in Figure~\ref{fig:ckg-cov-mc-norm}. The covariance matrix shows non-negligible off-diagonal terms, especially at high $\ell$. This finding validates the concern that we should consider the joint covariance of the variance in CMB lensing simulations and the galaxy window function. 

However, it is difficult to combine this covariance matrix properly with the covariance matrix we estimate in Figure~\ref{fig4:cor_matrix} because while in Figure~\ref{fig4:cor_matrix}, we vary both $n(z)$ and $W_g$, in Figure~\ref{fig:ckg-cov-mc-norm} we vary the \emph{Planck} CMB lensing simulations and $W_g$; for the latter $n(z)$ does not directly enter in the analysis. Hence, combining these two covariance matrices becomes an intractable problem without a joint cross-survey simulation. As a side note, this kind of joint systematics analysis will become more critical in future cosmological papers. So, it merits the field to develop joint, correlated cross-survey simulations for proper forward modeling. 

\begin{figure}
    \centering
    \includegraphics[width=.75\columnwidth]{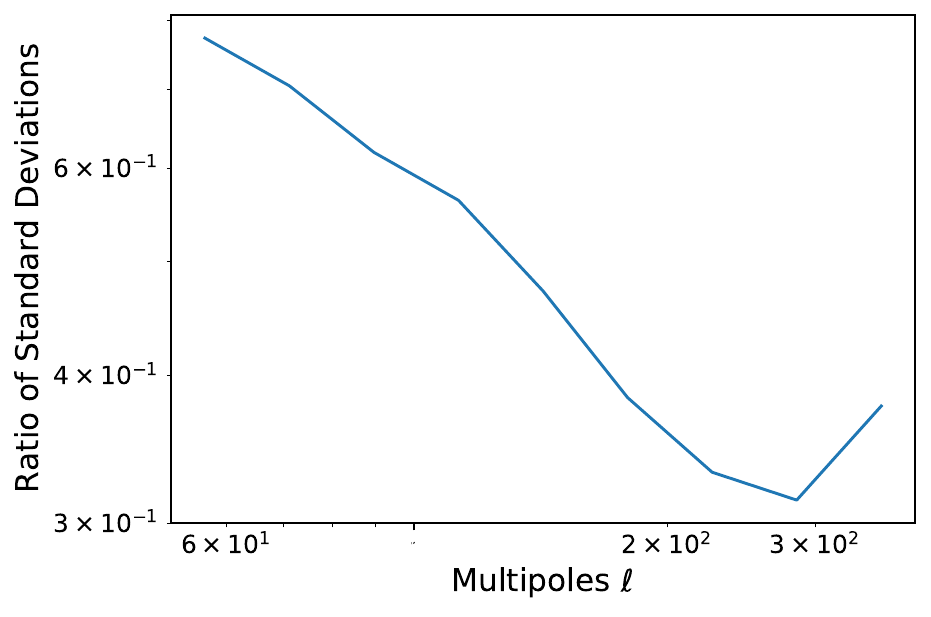}
    \caption{Ratio of the standard deviation of samples where the \emph{Planck} CMB lensing simulations and the galaxy window function were jointly varied to the standard deviation of the sample where the galaxy window function and the galaxy redshift distributions were jointly varied. The plot shows that the uncertainty of $A_{\ell}$ is subdominant to the uncertainty of the galaxy redshift distribution. Hence, we do not propagate the uncertainty of $A_{\ell}$ in the likelihood covariance matrix.}
    \label{fig:ratio-stds}
\end{figure}

Nonetheless, the variance (and the covariance) due to the multiplicative bias can be avoided as in our analysis because Figure~\ref{fig:ratio-stds} shows that the error bars due to the multiplicative biases are only $\mathcal{O} (10\%)$ that of the error bars due to photometric redshift uncertainties. Hence, we do not propagate the error and consider only the average value. We do caution that for spectroscopic surveys where the redshift uncertainty is orders of magnitude smaller, this effect should be reassessed to ensure that it is subdominant; otherwise, one may have artificially smaller error bars than what they ought to be. 

\begin{figure}
    \centering
    \includegraphics[width=0.75\columnwidth]{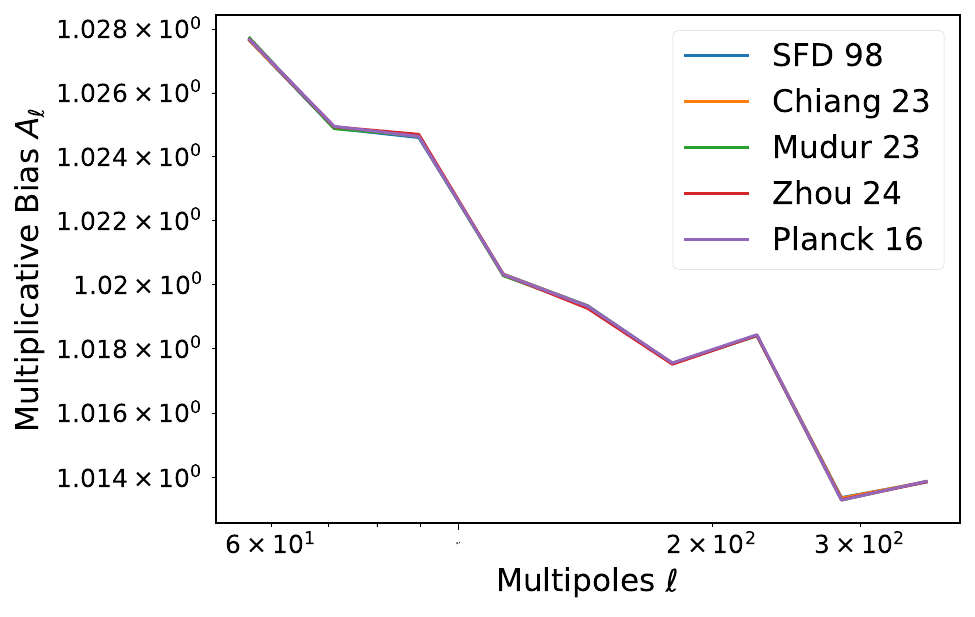}
    \caption{Monte Carlo Norm Correction Factor $A_{\ell}$ used to correct the observed galaxy-CMB lensing cross-power spectra. Each line represents a different dust map.}
    \label{fig:Aells-dust}
\end{figure}

Finally, Figure~\ref{fig:Aells-dust} shows that no matter our choices of the Galactic Extinction map to estimate the galaxy window function, the multiplicative correction factor $A_{\ell}$ remains almost the same.

\section{Impact of Unaccounted Systematics on galaxy bias and $\sigma_8$}
\label{app:unaccounted}

This appendix shows the theoretical calculations that helped us interpret the various systematics, primarily those described in Table~\ref{tab:systematics}.

\subsection{Unaccounted LSS signal in foreground dust maps}
\label{app:lss-dust}

In this situation, we consider a dust map that may have unaccounted large-scale structures (LSS) signal, e.g., we may have not accounted for large-scale structure due to cosmic infrared background contaminating the SFD dust map. 

Here, we assume that our best estimate of the galaxy window function, $W_g$, is composed of the pure, strictly foreground systematics ($W_g^{u}$), and a part that is correlated with the LSS ($W_g^{c}$). Using this information, we calculate observed power spectra and note from Equations~\ref{eq:cgg} and \ref{eq:ckg} that to the first order, the linear bias is a ratio of them. 

\begin{align*}
    \delta_g^{\rm obs} &= \widehat{W}_g \delta_g \\
    &= (W_g^{u} + W_g^{c}) \delta_g \\
    &= W_g^{u} \delta_g + W_g^{c} \delta_g \\
    C_{gg}^{\rm obs} &= <\delta_g^{\rm obs} \delta_g^{' \rm obs}> \\
    &= <W_g^{u} \delta_g W_g^{'u} \delta_g'> + <W_g^{c} \delta_g W_g^{'c} \delta_g'> + 2 <W_g^{u} \delta_g W_g^{'c} \delta_g'> \\
    &= C^{uu}_{gg} + C^{cc}_{gg} + 2 C^{uc}_{gg}\\
    C_{\kappa g}^{\rm obs} &= <W_g^{u} \delta_g \delta_{\kappa}> + <W_g^{c} \delta_g \delta_{\kappa}> \\
    &= C^{u}_{\kappa g} + C^{c}_{\kappa g} \\
\end{align*}

Hence, the inferred linear bias is:

\begin{align*}
    b_{\rm inferred} &= \frac{C_{gg}^{\rm obs}}{C_{\kappa g}^{\rm obs}} \\
    &= \frac{C^{uu}_{gg} + C^{cc}_{gg} + 2 C^{uc}_{gg}}{C^{u}_{\kappa g} + C^{c}_{\kappa g}}
\end{align*}

If we expect the correlated part of the galaxy window function, $W_g^{c}$ to be small, we can then Taylor expand $(C_{\kappa g}^{u} + C_{\kappa g}^{c})^{-1}$ around $C_{\kappa g}^{c} = 0$:

\begin{equation*}
    \frac{1}{C_{\kappa g}^{u} + C_{\kappa g}^{c}} \approx \frac{1}{C_{\kappa g}^{u}} - \frac{C_{\kappa g}^{c}}{(C_{\kappa g}^{u})^2}
\end{equation*}

Plugging this expansion into $b_{\rm inferred}$ we get:

\begin{align}
    b_{\rm inferred} &= \frac{C^{uu}_{gg} + C^{cc}_{gg} + 2 C^{uc}_{gg}}{C^{u}_{\kappa g} + C^{c}_{\kappa g}} \\
    &= \frac{C_{gg}^{uu}}{C_{\kappa g}^u} + \frac{C_{gg}^{cc}}{C_{\kappa g}^u} + 2\frac{C_{gg}^{uc}}{C_{\kappa g}^u} - \frac{C_{\kappa g}^c C^{uu}_{gg}}{(C_{\kappa g}^u)^2} - \frac{C_{\kappa g}^c C^{cc}_{gg}}{(C_{\kappa g}^u)^2}  - 2\cancelto{0}{\frac{C_{\kappa g}^c C^{uc}_{gg}}{(C_{\kappa g}^u)^2}} \\ 
    &= \frac{C_{gg}^{uu}}{C_{\kappa g}^u} + \boxed{\frac{C_{gg}^{cc}}{C_{\kappa g}^u} + 2\frac{C_{gg}^{uc}}{C_{\kappa g}^u} - \frac{C_{\kappa g}^c C^{uu}_{gg}}{(C_{\kappa g}^u)^2} - \frac{C_{\kappa g}^c C^{cc}_{gg}}{(C_{\kappa g}^u)^2}} \\ 
    &= b_{\rm true} + {\rm boxed}
\end{align}

\noindent we cancel the final term in the second line because the numerator is a minimal term. In the final expression, we see that our inferred linear bias is composed of the linear bias we are interested in measuring plus some linear and non-linear terms of various powers. Depending on the strength of the contamination, the inferred bias could either be higher or lower than the actual bias. 

\subsection{Unaccounted Magnification Bias signal in $\delta_g$ field}
\label{app:mag-bias}

Next, we consider a situation in which we have not adequately modeled the magnification bias. As a result, our observed galaxy overdensity field is composed of the galaxy overdensity field we modeled properly and an additional term representing the unmodeled galaxies that are still contributing to our density field. 

We similarly calculate the power spectra under these assumptions: 

\begin{align*}
    \delta_g^{\rm obs} &= W_g \hat{\delta}_g \\
    &= W_g (\delta_g^m + \delta_g^u) \\
    &= W_g \delta_g^m + W_g \delta_g^u \\
    C_{gg}^{\rm obs} &= <\delta_g^{\rm obs} \delta_g^{' \rm obs}> \\
    &= <W_g \delta_g^m W'_g \delta_g^{'m}> + <W_g \delta_g^{u} W'_g \delta_g^{'u}> + 2 <W_g \delta_g^{m} W'_g \delta_g^{'u}> \\
    &= C_{gg}^{mm} + C_{gg}^{uu} + 2 C_{gg}^{mu} \\
    C_{\kappa g}^{\rm obs} &= <W_g \delta_g^m \delta_{\kappa}> + <W_g \delta_g^u \delta_{\kappa}> \\
    &= C^{m}_{\kappa g} + C^{u}_{\kappa g}
\end{align*}

Hence, the inferred linear bias is:

\begin{align*}
    b_{\rm inferred} &= \frac{C_{gg}^{\rm obs}}{C_{\kappa g}^{\rm obs}} \\
    &= \frac{C^{mm}_{gg} + C^{uu}_{gg} + 2 C^{mu}_{gg}}{C^{m}_{\kappa g} + C^{u}_{\kappa g}}
\end{align*}

As we assumed in the first subsection, if we expected the unmodeled part, $\delta_g^m$ to be small, then we can Taylor expand the denominator around $\delta_g^m = 0$:

\begin{equation*}
    \frac{1}{C_{\kappa g}^{m} + C_{\kappa g}^{u}} \approx \frac{1}{C_{\kappa g}^{m}} - \frac{C_{\kappa g}^{u}}{(C_{\kappa g}^{m})^2}
\end{equation*}

Plugging this expansion into $b_{\rm inferred}$ we get: 

\begin{align*}
    b_{\rm inferred} &= \frac{C^{mm}_{gg} + C^{uu}_{gg} + 2 C^{mu}_{gg}}{C^{m}_{\kappa g} + C^{u}_{\kappa g}} \\
    &= \frac{C_{gg}^{mm}}{C_{\kappa g}^m} + \frac{C_{gg}^{uu}}{C_{\kappa g}^m} + 2\frac{C_{gg}^{mu}}{C_{\kappa g}^m} - \frac{C_{\kappa g}^u C^{mm}_{gg}}{(C_{\kappa g}^m)^2} - \frac{C_{\kappa g}^u C^{uu}_{gg}}{(C_{\kappa g}^m)^2} - 2\cancelto{0}{\frac{C_{\kappa g}^u C^{mu}_{gg}}{(C_{\kappa g}^m)^2}} \\
    &= \frac{C_{gg}^{mm}}{C_{\kappa g}^m} + \boxed{\frac{C_{gg}^{uu}}{C_{\kappa g}^m} + 2\frac{C_{gg}^{mu}}{C_{\kappa g}^m} - \frac{C_{\kappa g}^u C^{mm}_{gg}}{(C_{\kappa g}^m)^2} - \frac{C_{\kappa g}^u C^{uu}_{gg}}{(C_{\kappa g}^m)^2}} \\
    &= b_{\rm true} + boxed
\end{align*}

\noindent we cancel the final term in the second line because the numerator is very small. In the final expression, we see that the inferred bias is composed of the bias we are interested in measuring plus additional terms that can either increase or decrease the inferred value of the bias. 

\subsection{Unaccounted stellar (stream) or local dust contaminants}
\label{app:stellar-dust}

In this case, we consider a situation in which we have not adequately modeled stellar contaminants, and as a result, stars with similar colors as our galaxy sample masquerading as galaxies. We also consider a situation in which our dust map does not have leakage from the LSS but has unmodeled parts. For example, a poor modeling of the zodiacal light will leave an imprint of the residual zodiacal light. 

As a result, our estimated galaxy overdensity field will be composed of the actual galaxy overdensity field and some additional contaminant field, $\delta_g^{c}$. 

We calculate the observed power spectra:

\begin{align*}
    \delta_g^{\rm obs} &= W_g \widehat{\delta}_g \\
    &= W_g (\delta_g + \delta_g^c) \\
    &= W_g \delta_g + W_g \delta_g^c \\
    C_{gg}^{\rm obs} &= <\delta_g^{\rm obs} \delta_g^{' \rm obs}> \\
    &= <W_g \delta_g W'_g \delta_g> + <W_g \delta_g^c W'_g \delta_g^{'c}> + 2 \cancelto{0}{<W_g \delta_g W'_g \delta_g^{'c}>} \\
    &= C_{gg} + C_{gg}^{cc} \\
    C_{\kappa g}^{\rm obs} &= <W_g \delta_g \delta_{\kappa}> + \cancelto{0}{<W_g \delta_g^c \delta_{\kappa}>} \\
    &= C_{\kappa g} \\
\end{align*}

Hence, the inferred linear bias is: 

\begin{align*}
    b_{\rm inferred} &= \frac{C^{\rm obs}}{C^{\rm obs}_{\kappa g}} \\ 
    &= \frac{C_{gg} + C_{gg}^{cc}}{C_{\kappa g}} \\
    &= b_{\rm true} + \boxed{\frac{C_{gg}^{cc}}{C_{\kappa g}}}
\end{align*}

\noindent Here, we see a clear excess signal that makes the inferred linear bias greater than the actual bias. Moreover, if the inferred bias is larger, the inferred $\sigma_8$ will correspondingly be lower. 

\subsection{Unaccounted MC Norm Correction}
\label{app:unaccounted-mc-norm}

In this case, we consider the situation where the Monte Carlo Norm correction is not accounted for when debiasing the observed galaxy-CMB lensing power spectra. As discussed in Appendix~\ref{app:mc-norm}, not accounting for the norm correction bias leads to a slightly lower value of the observed $C_{\kappa g}$. If we consider the observed $C_{\kappa g}$ to be composed of the true cross-power spectrum and the correction factor, then:

\begin{align*}
    \delta_{g}^{\rm obs} &= W_g \delta_g \\
    C_{gg}^{\rm obs} &= C_{gg} \\
    C_{\kappa g}^{\rm obs} &= A_{\rm norm} C_{\kappa g}
\end{align*}

Hence, the inferred bias is: 

\begin{align*}
    b_{\rm inferred} &= \frac{C_{gg}^{\rm obs}}{C_{\kappa g}^{\rm obs}} \\
    &= \frac{C_{gg}}{A_{\rm norm} C_{\kappa g}} \\
    &=  b_{\rm true} \boxed{\frac{1}{A_{\rm norm}}}
\end{align*}

\noindent Thus, the inferred bias will be scaled by the norm correction factor.

\section{Impact of Photometric Redshift Distribution Uncertainty on Parameter Inference}

As shown in Figure~\ref{fig4:cor_matrix}, our photometric redshift uncertainty causes high mode-mode coupling and this in turn severely degrades our detection significance of the galaxy-galaxy power spectrum. 

\begin{figure}
    \centering
    \includegraphics[width=0.8\linewidth]{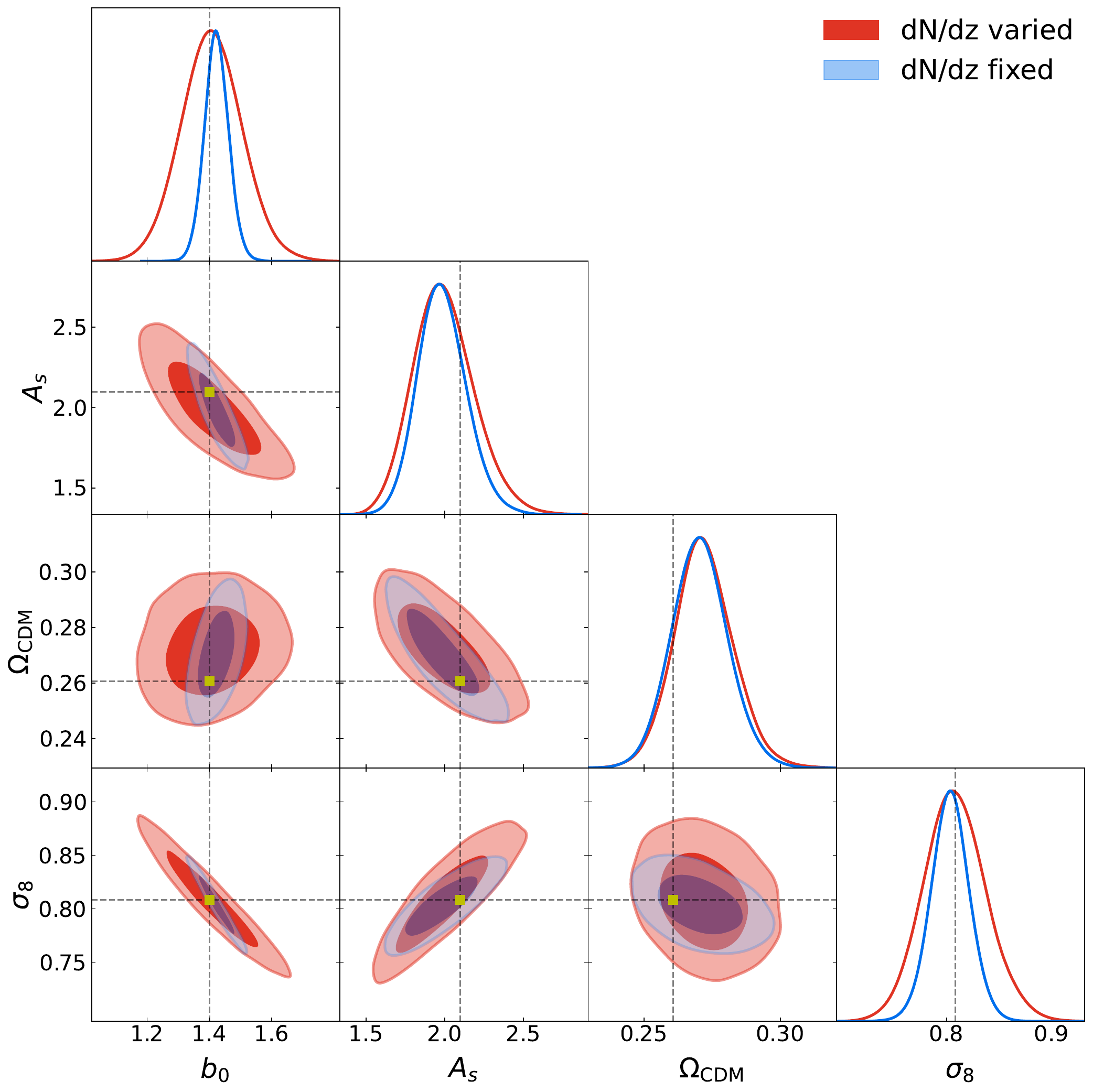}
    \caption{The impact of photometric redshift uncertainty on parameter inference using simulation. The dashed lines and the yellow squares denote the actual values at which the simulation was generated. The red contour shows the case where we marginalize over photometric redshift uncertainty at the covariance matrix level, matching the ELG sample; the blue contour shows the case that mimics a spectroscopic survey with the same number density and mean redshift distribution. The photometric redshift uncertainty results in the $3-\sigma$ contour volume of the red case to increase by a factor of $26$ compared to the blue case when considering the $b_0$-$\Omega_{\rm CDM}$-$\sigma_8$ space.}
    \label{fig:dndz-varied}
\end{figure}

A natural question that may arise is how our relatively large photometric redshift uncertainty, and consequently the lower detection significance of the galaxy-galaxy power spectrum, impact the final parameter inference. 

To test this, we run a suite of Gaussian simulation following the same prescription as Section~\ref{sec4:sims}, and keeping all the parameter fixed; the only change we make is we pick the mean of the measured redshift distributions in Figure~\ref{fig4:dndz} as our redshift distribution and do not marginalize over the uncertainty at the covariance matrix level, unlike our main analysis. This choice is equivalent to exploring what a spectroscopic survey with the same areal coverage and number density would measure. A spectroscopic survey such as DESI reports the typical ELG spectroscopic redshift uncertainty at worst to be $\approx 3\times 10^{-4}$ \cite{yu2024} per object, making the redshift distribution uncertainty almost negligible. 

The comparison between the photometric redshift distribution and the spectroscopic redshift distribution is shown in Figure~\ref{fig:dndz-varied}. We see that the photometric redshift uncertainty leads to a wider posterior than a fixed redshift distribution (equivalent to spectroscopic surveys). In fact, if we compare the $3-\sigma$ error ellipse volume of the two cases considering the $b_0$-$\Omega_{\rm CDM}$-$\sigma_8$ axes, we find that photometric uncertainty increases the volume by a factor of $26$. Note that most of this change is coming from $b_0$ and $\sigma_8$ as photometric redshift distribution does not appear to affect $\Omega_{\rm CDM}$ noticeably.  

\acknowledgments
The authors thank Rongpu Zhao, Gerrit Farren, Lukas Wenzl and Ren\'{e}e Hlo\u{z}ek for their valuable discussions. The authors also thank Noah Sailer, Abhishek Maniyar, James Rohlf and Antonella Palmese for providing valuable feedback during the DESI internal review process. 

TK is supported by the Arts \& Science Postdoctoral Fellow at the University of Toronto. TK was supported by the National Science Foundation Graduate Research Fellowship under Grant No. DGE - 1745303 and the Barbara Bell Dissertation Fellowship at the Harvard University during part of this project. SS is supported by the McWilliams Fellowship at the Carnegie Mellon University. MR is supported by the U.S. Department of Energy grants DE-SC0021165 and DE-SC0011840. 

This material is based upon work supported by the U.S. Department of Energy (DOE), Office of Science, Office of High-Energy Physics, under Contract No. DE–AC02–05CH11231, and by the National Energy Research Scientific Computing Center, a DOE Office of Science User Facility under the same contract. Additional support for DESI was provided by the U.S. National Science Foundation (NSF), Division of Astronomical Sciences under Contract No. AST-0950945 to the NSF’s National Optical-Infrared Astronomy Research Laboratory; the Science and Technology Facilities Council of the United Kingdom; the Gordon and Betty Moore Foundation; the Heising-Simons Foundation; the French Alternative Energies and Atomic Energy Commission (CEA); the National Council of Humanities, Science and Technology of Mexico (CONAHCYT); the Ministry of Science, Innovation and Universities of Spain (MICIU/AEI/10.13039/501100011033), and by the DESI Member Institutions: \url{https://www.desi.lbl.gov/collaborating-institutions}. Any opinions, findings, and conclusions or recommendations expressed in this material are those of the author(s) and do not necessarily reflect the views of the U. S. National Science Foundation, the U. S. Department of Energy, or any of the listed funding agencies.

The authors are honored to be permitted to conduct scientific research on Iolkam Du’ag (Kitt Peak), a mountain with particular significance to the Tohono O’odham Nation.

The DESI Legacy Imaging Surveys consist of three individual and complementary projects: the Dark Energy Camera Legacy Survey (DECaLS), the Beijing-Arizona Sky Survey (BASS), and the Mayall z-band Legacy Survey (MzLS). DECaLS, BASS and MzLS together include data obtained, respectively, at the Blanco telescope, Cerro Tololo Inter-American Observatory, NSF’s NOIRLab; the Bok telescope, Steward Observatory, University of Arizona; and the Mayall telescope, Kitt Peak National Observatory, NOIRLab. NOIRLab is operated by the Association of Universities for Research in Astronomy (AURA) under a cooperative agreement with the National Science Foundation. Pipeline processing and analyses of the data were supported by NOIRLab and the Lawrence Berkeley National Laboratory. Legacy Surveys also uses data products from the Near-Earth Object Wide-field Infrared Survey Explorer (NEOWISE), a project of the Jet Propulsion Laboratory/California Institute of Technology, funded by the National Aeronautics and Space Administration. Legacy Surveys was supported by: the Director, Office of Science, Office of High Energy Physics of the U.S. Department of Energy; the National Energy Research Scientific Computing Center, a DOE Office of Science User Facility; the U.S. National Science Foundation, Division of Astronomical Sciences; the National Astronomical Observatories of China, the Chinese Academy of Sciences and the Chinese National Natural Science Foundation. LBNL is managed by the Regents of the University of California under contract to the U.S. Department of Energy. The complete acknowledgments can be found at https://www.legacysurvey.org/.

This research made use of the following packages: \textsc{Numpy} \citep{numpy}, 
\textsc{AstroPy} \citep{astropy:2013, astropy:2018, astropy:2022}, \textsc{Matplotlib} \cite{matplotlib},
\textsc{Healpy} and \textsc{HEALPix} \citep{healpy, healpix}, \textsc{SkyLens} \citep{Singh21}, \textsc{CAMB} \citep{camb},  \textsc{emcee} \citep{emcee} and \textsc{GetDist} \citep{getdist}.






\bibliographystyle{JHEP}
\bibliography{biblio.bib}





\end{document}

%% file: bangla_commands.tex
\def\bng{\bngx}

\font\bngx=bang10

\def\*#1*#2{o\null{#2}{#1}}

\def\sh#1{\setbox0=\hbox{#1}%
     \kern-.02em\copy0\kern-\wd0
     \kern.04em\copy0\kern-\wd0
     \kern-.02em\raise.0433em\box0 }